\documentclass[10pt]{iopart}
\usepackage{graphicx}
\usepackage{color}
\usepackage{amssymb}
\usepackage{iopams}

\usepackage[breaklinks=true,colorlinks=true,linkcolor=blue,urlcolor=blue,citecolor=blue]{hyperref}

\begin{document}
\title{Lateral heterostructures and one-dimensional interfaces in 2D transition metal dichalcogenides}

\author{O \'Avalos-Ovando$^{1,*}$, D Mastrogiuseppe$^{2}$, and S E Ulloa$^{1}$}

\address{$^1$ Department of Physics and Astronomy, and Nanoscale and Quantum
	Phenomena Institute, Ohio University, Athens, Ohio 45701--2979, USA}
\address{$^2$ Instituto de F\'isica Rosario (CONICET), 2000 Rosario, Argentina}

\ead{oa237913@ohio.edu}

\begin{abstract}
The growth and exfoliation of two-dimensional (2D) materials have led to the creation of edges and novel interfacial states at the juncture between crystals with different composition or phases. These hybrid heterostructures (HSs) can be built as vertical van der Waals \textit{stacks}, resulting in a 2D interface, or as \textit{stitched} adjacent monolayer crystals, resulting in  one-dimensional (1D) interfaces. Although most attention has been focused on vertical HSs, increasing theoretical and experimental interest in 1D interfaces is evident. In-plane interfacial states between different 2D materials inherit properties from both crystals, giving rise to robust states with unique 1D non-parabolic dispersion and strong spin-orbit effects. With such unique characteristics, these states provide an exciting platform for realizing 1D physics.
Here, we review and discuss advances in 1D heterojunctions, with emphasis on theoretical approaches for describing those between semiconducting transition metal dichalcogenides $MX_{2}$ (with $M$=Mo, W and $X$= S, Se, Te), and how the interfacial states can be characterized and utilized. We also address how the interfaces depend on edge geometries (such as zigzag and armchair) or strain, as lattice parameters differ across the interface, and how these features affect excitonic/optical response. This review is intended to serve as a resource for promoting theoretical and experimental studies in this rapidly evolving field. 

\end{abstract}

\submitto{\JPCM}
\noindent Orcid O \'Avalos-Ovando: \href{http://orcid.org/0000-0003-3572-7675}{http://orcid.org/0000-0003-3572-7675}\\
Orcid D Mastrogiuseppe: \href{http://orcid.org/0000-0003-4606-2883}{http://orcid.org/0000-0003-4606-2883}\\
Orcid S E Ulloa: \href{http://orcid.org/0000-0002-3091-4984}{http://orcid.org/0000-0002-3091-4984}\\

\ioptwocol
\maketitle

\tableofcontents

\section{Introduction}
\label{sec:Introduction}

Research on two-dimensional (2D) materials has grown enormously over the last few years \cite{Bhimanapati2015}, since the first isolation of monolayer graphene in 2004 by Novoselov \emph{et al.} \cite{Novoselov2004}. Despite graphene flexibility, strength, and high conductivity, with promising applications in electronics and spintronics, it does not have a bandgap, severely restricting its use in optoelectronics and digital electronics \cite{Gibney2015}. Although interest on graphene is far from over, the attention has been drawn recently to other 2D materials, such as hexagonal boron nitride (hBN) \cite{Zhang2017hBN}, the various -enes in group IVA and VA (silicine, arsenene, antimonene, stanene, germanene, bismuthene, etc.) \cite{Balendhran2015,Zhang2015ArseneneAntimonene,Reis2017}, black phosphorus \cite{Balendhran2015,Carvalho2016phosphorene}, and the extensive transition-metal dichalcogenide (TMD) family \cite{Manzeli20172d}.

Naturally, different 2D materials can be combined within the same device and form diverse heterostructures (HSs). As each material has different electronic structure and properties, HSs are capable of enhancing or, better yet, creating new tailored features, which are rather weak or nonexistent in their pristine counterparts. Prominent recent examples of HSs between 2D materials include enhancement of valley splitting by magnetic proximity  \cite{Zhao2017enhanced,Seyler2018}, the appearance of spatially indirect excitons \cite{Calman2018}, and superconductivity in graphene bilayers rotated by a \emph{magic angle} \cite{Cao2018unconventional}.

In particular, group VIB semiconducting TMDs \cite{Novoselov2005} have been suggested as novel components for spintronics devices \cite{Zibouche2014}. They are receiving a great deal of attention due to their unique electronic \cite{Wang2012,Xu2014NatPhys,Liu2015} and optical properties \cite{Castellanos2016}, including strong spin-orbit coupling (SOC) \cite{Zhu2011,Cheiwchanchamnangij2012,Xiao2012} and either direct or indirect bandgap depending on the number of layers \cite{Splendiani2010,Mak2010}. TMDs can be exfoliated down to a `monolayer unit', a stack of three atomic layers ($MX_{2}$), in which transition metal atoms ($M$=Mo, W) are sandwiched between two layers of chalcogen species ($X$=S, Se, Te),  resulting in molybdenum disulphide (MoS$_{2}$), tungsten diselenide (WSe$_{2}$), and molybdenum ditelluride (MoTe$_{2}$), among others.

Two (or more) different TMD monolayer units can either be stacked together to form a vertical HS \cite{Geim2013}, or they can be `stitched' together to build a lateral HS \cite{Ling2016} (also called planar or in-plane).  Given that different TMDs show different bandgaps, work functions, SOC and excitonic spectra, they offer a wide variety of tunable properties when combined. In 2016, Kolobov \emph{et al.} \cite{Kolobov2016}, listed 1 theory and 7 experimental papers in their section 13.2 on lateral HSs; just two years later, those numbers have increased up to 35+ and 40+, respectively, highlighting the rapidly growing interest and activity in these novel structures.

Lateral HSs have been achieved, and the current emphasis is on improving the quality of the interfaces. Experiments on these systems include graphene-hBN \cite{Levendorf2012graphenehBN,Drost2015graphenehBN}, graphene-TMDs \cite{Ling2016}, hBN-TMDs \cite{Ling2016}, and different TMD-TMD combinations \cite{Ling2016,Huang2014NatMat,Gong2014NatMat,Duan2014NatNano,Zhang2015,Li2015,Zhang2016naturecommunications,Zhang2018strain,Sahoo2018Nature,Xie2018ParkGroup}, with many suggested applications as in-plane transistors, diodes, \emph{p-n} photodiodes and complementary metal-oxide-semiconductor (CMOS) inverters. Chemical vapor deposition (CVD) growth techniques have focused on successfully improving the lateral atomic connection between materials \cite{Huang2014NatMat,Gong2014NatMat,Duan2014NatNano,Zhang2018strain,Sahoo2018Nature,Xie2018ParkGroup}, in order to build clean, sharp and well oriented borders. This progress is clearly reflected in the description of HSs systems, which has changed from \emph{alloys} to \emph{interfaces}.  Recent experiments have shown remarkable control on sharpness \cite{Sahoo2018Nature} and strain at the interface \cite{Zhang2018strain,Xie2018ParkGroup}. Moreover, atomically sharp interfaces between crystalline phases of the same TMD, 1T'-WSe$_{2}$ and 1H-WSe$_{2}$, have been studied in the search for topologically protected helical edge states \cite{Ugeda2018arxiv}.
As the quality and diverse composition of lateral HSs continue to improve, these interfaces can serve as  exciting new platforms for the study of 1D physical phenomena.

As control of lateral HSs in TMDs is increasingly achieved in experiments, understanding the structural and electronic properties of the interfaces and their general behavior is important for future progress. Experiments have shown that the optoelectronic behavior has strong 1D character, and theoretical approaches have started to appear, suggesting effective applications for these novel 1D systems. Several interesting reviews on this topic \cite{Voiry2015,Lv2015,Li2016,Novoselov2016,Lin2016Review,Yuan2017Review,Zhao2018Review,Frisenda2018Review,Chen2018Review,Zeng2018Review,Hu2018Review} have focused mainly on experimental advances. In contrast, in this review we  address current theoretical advances on 1D lateral interfaces between  group VIB semiconducting TMDs (with a few exceptions), together with an overview of the experimental efforts to produce these interesting interfaces and associated device geometries.

The review is organized as follows: In section\ \ref{sec:Experiments} we briefly summarize experimental advances, focusing on growth and characterization of the interfacial region. In section\ \ref{sec:TheoreticalAndNumerical} we develop the main scope of this review, the theoretical and numerical descriptions of lateral HSs: We describe structural and electronic properties, and analyze proposals for using the interface as a unique stage to explore 1D physics. In section\ \ref{sec:Applications} we summarize the already available and proposed applications. In section\ \ref{sec:ProspectiveDirections} we give an outlook on the evolution of the field, and lastly, in section\ \ref{sec:SummaryFinal} we provide a concluding summary.

\section{Experiments}
\label{sec:Experiments}

Clean and sharp lateral interfaces in TMD HSs were reported  in 2014 by three groups \cite{Huang2014NatMat,Gong2014NatMat,Duan2014NatNano}, improving on previous alloy growth $M^{(1)}X^{(1)}_x-M^{(2)}X^{(2)}_{1-x}$ \cite{Zhang20142DAlloys,Li2015Lateral,Wang2015SpinOrbit,Xie20152Dalloys,Yoshida2015microscopic,Zheng2015Monolayers,Duan2016Synthesis} with different metal $M^{(j)}$ and chalcogen $X^{(j)}$ combinations. Recent experiments have shown successful growth of longer interfacial sections with great strain, geometry, and/or electronic band alignment tunability \cite{Sahoo2018Nature,Xie2018ParkGroup,Zhang2018strain}.

This section is intended to serve as a brief resource to be considered in theoretical proposals for effective uses of lateral HSs between TMDs. It is divided into three subsections as follows: \ref{subsec:growth} reviews growth techniques, \ref{subsec:ExperimentalInterfaces}  lists important interfacial parameters, and  \ref{subsec:PhaseInterfaces} reviews advances on interfaces between different phases of the same TMD.

\subsection{Growth and characterization techniques}
\label{subsec:growth}

\begin{table*}[t]
\caption{\label{tab:table1} Atomic parameters for lateral TMD HS, including the HS interface, interfacial features (sharpness, strain, length, orientation, stitch or atomic interface), growth technique, source (Ref.), and substrate used. The strain column lists values only if characterized in the study. Length corresponds to the maximum pristine interface presented in the study. On the orientation column, ac* indicates that armchair domains are only sporadic instead of extended. Growth: while most procedures are chemical vapor deposition (CVD), additional techniques are also required, shown with symbols, $\dag$: e-beam lithography, $\ddag$: PTAS seeding, $\S$: assisted NaCl, and $\P$: self aligned. The table is in chronological order, with early work at the top.}
\begin{indented}
\item[]\begin{tabular}{@{}ccccccccc}
\br
&\multicolumn{5}{c}{Interface}&&&\\ \cline{2-6}\\
HS interface & Sharpness & Strain & Length & Orientation & Stitch & Growth & Ref. & Substrate \\
\mr
MoSe$_2$-WSe$_2$ & smooth (16 nm) & - & 30 nm & - & - & 1-step CVD & \cite{Huang2014NatMat} & SiO$_2$/Si \\
MoS$_2$-MoSe$_2$ & smooth (30 nm) & - & - & - & Se-W & 1-step CVD & \cite{Duan2014NatNano} & SiO$_2$/Si \\
WS$_2$-WSe$_2$ & smooth (40 nm)& - & - & - &   &  \\
MoS$_2$-WS$_2$ & sharp & - & 7 nm & zz \& ac* & S-Mo & 1-step CVD & \cite{Gong2014NatMat} & SiO$_2$/Si \\
MoS$_2$-WS$_2$ & sharp & - & 16 u.c. & zz & S-W & 2-step CVD$^\ddag$ & \cite{Zhang2015} & SiO$_2$/Si, sapphire, quartz \\
MoSe$_2$-WSe$_2$  & &  &  &  & Se-W &  &  &  \\
MoSe$_2$-MoS$_2$ & sharp & - & 1 nm & zz & - & - & \cite{Tizei2015Esciton} & \\
WS$_2$-MoS$_2$ & sharp & - & 10 nm & zz & S-W & 2-step CVD & \cite{Heo2015RotationMisfitFree} & SiO$_2$ \\
MoSe$_2$-MoS$_2$ & smooth (5 nm) & - & - & zz \& ac & - & 2-step CVD$^\dag$ & \cite{Mahjouri2015patterned} & SiO$_2$ \\
WSe$_2$-MoS$_2$ & sharp & 1.5\% &  &  & S-W & 2-step CVD & \cite{Li2015}  & sapphire \\
MoSe$_2$-WSe$_2$ & sharp & - &  &  & S-W & 2-step CVD & \cite{Gong2015TwoStep} & SiO$_2$/Si \\
MoS$_2$-WS$_2$ & sharp & - & - & - & - & 2-step CVD & \cite{Chen2015Electronic} & SiO$_2$/Si \\
MoS$_2$-WS$_2$ & sharp & - & 6 nm & - & zz & 1-step CVD & \cite{Chen2015} &  \\
MoS$_2$-WS$_2$ & sharp & - & 8 nm & zz & - & 2-step CVD & \cite{Yoo2015} & sapphire \\
MoS$_2$-graphene & smooth & - & - & - & - & 2-step CVD$^\ddag$ & \cite{Ling2016} & SiO$_2$/Si \\
MoS$_2$-WS$_2$   &  &  &  &  &  &  &  &  \\
MoS$_2$-hBN      &  &  &  &  &  &  &  &  \\
WSe$_2$-MoS$_2$ & smooth (120 nm) & - & - & zz & - & 2-step CVD & \cite{Son2016Observation} & sapphire \& ITO \\
MoS$_2$-WS$_2$ & sharp & - & - & - & - & 2-step CVD & \cite{Bogaert2016Diffusion} & SiO$_2$/Si \\
WS$_2$-MoS$_2$ & sharp & - & - & - & - & 2-step CVD & \cite{Kobayashi2016} & Graphite \\
WSe$_2$-WS$_2$ & sharp & - & 4 nm & zz & S-W & 2-step CVD & \cite{Chen2016Lateral} & SiO$_2$/Si \\
WSe$_2$-WS$_2$ & sharp & - & - & zz \& ac & - & 2-step CVD$^\dag$ & \cite{Li2016Laterally} & sapphire \\
MoS$_2$–MoSe$_2$ & smooth & - & - & - & - & 2-step CVD & \cite{Chen2017InPlaneMosaic} & SiO$_2$/Si \\
WSe$_2$-MoS$_2$ & sharp & - & 5 nm & zz & - & 2-step CVD & \cite{Tsai2017SingleAtomically} & SiO$_2$/Si \\
WS$_2$-MoS$_2$ & sharp & - & 4 nm & zz & S-Mo & 1-step CVD & \cite{Shi2017cascaded} & SiO$_2$/Si \& Al$_2$O$_3$/Ag\\
 & smooth & - & - & - & - & 2-step CVD$^\ddag$ &  & \\
MoS$_2$-WS$_2$ & sharp (0.85 nm) & - & 6 nm & - & - & 1-step CVD$^\S$ & \cite{Wang2017NaClAssisted} & SiO$_2$/Si \\
MX$_2$ combinations & sharp & - & 6 nm & zz \& ac* & - & many-step CVD & \cite{Zhang2017Robust}  & SiO$_2$/Si \\
MoS$_2$-WS$_2$ & sharp \& smooth & - & - & - & - & 1-pot CVD & \cite{Liu2017ARXIVNanoscale} &  SiO$_2$ \\
MoSe$_2$-WSe$_2$, & sharp \& smooth & - & - & zz & X-Mo & 1-pot CVD & \cite{Sahoo2018Nature}  & Si \\
MoS$_2$-WS$_2$ &  & &  &  &  &  &  &   \\
WSe$_2$-MoS$_2$ & sharp & 2.2\% & 5-15 nm & zz \& ac* & Se-Mo & 2-step CVD & \cite{Zhang2018strain} & HOPG \\
 &  & 1.76\% &  &  &  &  &  & WSe$_2$\\
WSe$_2$-MoS$_2$ & sharp & - & - & irregular & - & 2-step CVD$^\P$ & \cite{Li2018SelfAligned} & sapphire \\
WSe$_2$-WS$_2$ & sharp & 1.2\% & 160 u.c. & zz &  & MOCVD & \cite{Xie2018ParkGroup} & SiO$_2$ \\
MoS$_2$-WS$_2$ & - & - & - & - & - & 2-step CVD$\dag$ & \cite{Murthy2018Intrinsic} & SiO$_2$/Si \& hBN\\
MoS$_2$-WS$_2$ & sharp & - & 3 nm & zz & - & 1-step CVD & \cite{Zhou2018Morphology} & SiO$_2$/Si \\
MoS$_2$-WS$_2$ & sharp & - & - & - & - & 1-step CVD & \cite{Wu2018SelfPowered} & SiO$_2$/Si \\
MoSe$_2$-WSe$_2$ & sharp \& smooth & - & - & - & - & 1-pot CVD & \cite{Xue2018NanoOptical} & SiO$_2$/Si \\
\br
\end{tabular}
\end{indented}
\end{table*}

Controlled synthesis of TMD HSs remains challenging due to the difficulty of growth conditions and their tunability. Furthermore, the HSs obtained with most methods are still relatively small,  which restricts possible studies and applications. Different TMDs have similar thicknesses (monolayer height), so that the planar connection depends mostly on lattice constant mismatch. They also have nearly the same lattice constant if the chalcogen is the same ($\approx$0.3\% lattice mismatch) for different transition metals  (e.g., MoS$_{2}$-WS$_{2}$). On the other hand, the lattice constant is very different ($\approx$4\% lattice mismatch) if only the chalcogen changes (e.g., MoS$_{2}$-MoSe$_{2}$). This large difference may lead to dislocations or wrinkles at the interface, with large built-in strains, which have important consequences on the electronic structure, as we will see later.

The growth of 2D TMDs HSs usually involves CVD, where vapor phase reactants are generated by thermally evaporating solid sources, usually powders. A classification scheme considers the degree of growth process manoeuvrability, where fewer changes in the growth conditions (such as sources or reactors) reduce degradation and promote cleaner and sharper interfaces. This \emph{the-fewer-steps-the-better} theme is increasingly mentioned in the literature. On the other hand, more steps allow the construction of more complex HSs, such as quantum wells and periodic HSs. A 1-step CVD process uses \emph{in situ} modulation of the vapor-phase reactants during growth, changing the chalcogen precursor just once in the middle of the growth run (see for example,  Duan \emph{et al.} \cite{Duan2014NatNano}). A 2-step CVD involves the synthesis of one TMD, followed by epitaxial growth of the second one off the edges of the first growth, as reported in \cite{Heo2015RotationMisfitFree} and  \cite{Gong2015TwoStep}. The advantage of this process is that it allows larger and sharper interfaces, avoiding cross-contamination. Multi-step CVD typically consists of modulating the chemical vapor source sequentially, to grow block-by-block multi-HSs \cite{Zhang2017Robust}.

CVD techniques have been used to grow heterotriangles, composed of a central TMD and an outer triangular ring of another TMD \cite{Huang2014NatMat,Gong2014NatMat,Duan2014NatNano}. Truncated triangles, hexagons, and hexagrams \cite{Zhang2015} have been also seen in experiments. These are built by changing the growth conditions,  keeping the same chalcogen and changing the metal to build MoSe$_{2}$-WSe$_{2}$ \cite{Huang2014NatMat} or MoS$_{2}$-WS$_{2}$ \cite{Gong2014NatMat,Zhang2015}, or by keeping the metal and changing the chalcogen, building MoS$_{2}$-MoSe$_{2}$ or WS$_{2}$-WSe$_{2}$ \cite{Duan2014NatNano}. Other works change both species, such as WSe$_{2}$-MoS$_{2}$ \cite{Li2015}.

More complex patterned structures have also been reported \cite{Mahjouri2015patterned}. MoSe$_2$ pristine triangular flakes are coated with SiO$_2$, for subsequent sulfurization of the uncovered parts, obtaining \emph{crosswalk} patterned lateral arrays of MoSe$_2$-MoS$_2$ within the initial triangular flake. Similar approaches allow cutting triangular flakes with electron-beam lithography resulting in armchair interfaces or even irregular logos \cite{Li2016Laterally}. Large area mosaics of lateral HSs of triangular MoS$_2$ sections embedded in a monolayer MoSe$_2$ have been achieved, in a `cheetah spots' configuration \cite{Chen2017InPlaneMosaic}. Ring interfaces between MoS$_2$ and WS$_2$ have been observed as well \cite{Chen2015Electronic}.

Recent work has shown that nearly perfect interfaces can be grown in a `one-pot' synthesis process \cite{Sahoo2018Nature,Xie2018ParkGroup}. This is achieved by changing the composition of the reactive gas environment in the presence of water vapor, allowing for great control and flexibility. TMD controlled growth in the carrier gas N$_2$+H$_2$O(g) promotes growth of MoX$_2$, while Ar+H$_2$(5\%) suppresses Mo and promotes growth of WX$_2$. The approach appears versatile and scalable, as continuous planar multi-interfaces can be grown by controlled sequential edge-epitaxy. Sahoo \emph{et al.} report several MoSe$_2$-WSe$_2$ and MoS$_2$-WS$_2$ lateral HSs, with long and controllable 1D interfaces \cite{Sahoo2018Nature}. Their Se-based HSs are concentric triangles, while S-based HSs have one triangular central section with trapezoidal sections growing off the central edges. A similar one-pot process creates coherent WSe$_2$-WS$_2$ lateral HSs (also WSe$_2$-MoS$_2$-WS$_2$) \cite{Xie2018ParkGroup}. The coherence would allow one in principle to tune optical properties, strain-engineering the HS photoluminescence. The growth modulation uses metal-organic CVD (MOCVD), controlling each precursor individually and precisely,  with linear dependence of transverse width vs growth time. Coherence was shown using different scanning transmission electron (STEM) microscopy techniques.
Similarly, sophisticated CVD growth techniques have allowed the characterization of strain, as discussed by Zhang \emph{et al.} \cite{Zhang2018strain}, that directly determine strain in WSe$_2$-MoS$_2$ and in the coherent and sharp WSe$_2$-WS$_2$ HSs \cite{Xie2018ParkGroup}. Table \ref{tab:table1} summarizes grown techniques, substrates an other interfacial parameters for experiments with lateral TMD HSs.

In-plane lateral HSs have been also achieved between materials with different thicknesses, such as those composed by bilayer-monolayer combinations (also called \emph{terrace} structures) of either MoSe$_{2}$ or WSe$_{2}$ \cite{Zhang2016naturecommunications}.

Other improvements on growth processes are also being considered. For example, temperature control is essential,  promoting mixing at high temperatures and compositional segregation at lower temperatures, so that HSs with sharp interface are achieved at low growth temperatures, and alloying occurs at higher temperatures \cite{Bogaert2016Diffusion}. CVD assisted by sodium chloride (NaCl) requires lower growth temperatures, as Na precursors condensate on the substrate and reduce reaction energies \cite{Wang2017NaClAssisted}. Given that the properties of 2D materials are susceptible to external environments, the encapsulation of HSs between hBN sheets has been recently obtained \cite{Murthy2018Intrinsic}, showing that both photovoltaic and hot electron generation lead to photocurrents that depend on the biasing conditions.

Control over location and size of the CVD flakes is not as well developed, although efforts are underway.  Ling \emph{et al.} developed a \emph{parallel stitching} method for connecting MoS$_2$ to several materials, such as WS$_2$, graphene, and hBN \cite{Ling2016}. This consisted in sowing perylene-3,4,9,10-tetracarboxylic acid tetrapotassium salt (PTAS) molecules on the growth substrate. These serve as seeds to facilitate growth off the edges of a previously deposited 2D material, depending on the wettability of seeds and surfaces \cite{Ling2016,Shi2017cascaded}.
The interface between MoS$_2$-graphene appears more terrace-like than lateral stitching, with overlapping edges extending for 2-30 nm. No lattice distortion is seen at the interface, but atomic defects associated with MoS$_2$ edges, such as Mo-Mo bonds, and S bridge defects were found.

Also, a new scalable 2-step CVD method for lateral growth has been developed, allowing the fabrication of heteroribbons \cite{Li2015,Chen2016Lateral} with long interfaces in a non-triangular structure.
Most recently, WSe$_2$-MoS$_2$ \cite{Li2018SelfAligned} and WS$_2$-MoS$_2$ HSs \cite{Aleithan2018unpublished} are grown starting from both distinct metallic samples. This 2-step process promotes growth from distinct patterned metal contacts in a \emph{position-selective} manner, as the interface is created at the meeting point between both flakes, as shown in figure\ \ref{FigInterfaces}(d). This method allows for control of the geometrical distribution of the interfaces, tailored by pre-growth lithographically patterned electrodes, as done for precontacted monolayer systems \cite{Khadka2017,Aleithan2018unpublished}.

One of the most common and interesting characterization tools of lateral TMD HSs is the excitonic photoluminescence (PL) near the interface \cite{Huang2014NatMat,Gong2014NatMat,Duan2014NatNano,Sahoo2018Nature}.  Most interestingly, sub-wavelength scale resolution reported by Tizei {\em et al.} \cite{Tizei2015Esciton} has measured the spatial variation of excitons in a MoS$_2$-MoSe$_2$ interface using spatially resolved electron energy loss spectroscopy (EELS) with a monochromatic beam size of 1 nm. The exciton maps allow measurements of  optical features with nanometer-scale resolution, and excitonic peaks are seen broader at interfaces, probably due to interfacial roughness.

A different technique of photocurrent spectral atomic force microscopy allowed imaging of currents and photocurrents generated between a PtIr tip and the monolayer WSe$_2$-MoS$_2$ HS \cite{Son2016Observation}.  Changing tip polarity and magnitude showed that the photoresponse can be switched on and off.

Second harmonic generation (SHG) and atomic-resolution STEM have also been used to characterize HS symmetries \cite{Zhang2015,Li2015,Zhou2018Morphology,Wu2018SelfPowered}. A recent study has quantitatively characterized the built-in potential at the interface by scanning Kelvin probe force microscopy (SKPFM) along with SHG at the interface \cite{Wu2018SelfPowered}. SHG measures the angle between the crystal orientation and axis of a linearly polarized pump laser normally incident on the HS.\@  When the incident laser polarization is perpendicular (parallel) to the zigzag (armchair) direction, intensity maxima appear. This allows one to determine the growth direction, and if the interfaces are zigzag or armchair \cite{Li2015}.

Lateral MoSe$_2$-WSe$_2$ \cite{Xue2018NanoOptical} and MoS$_2$-WS$_2$ \cite{Liu2017ARXIVNanoscale} HSs have been used for mapping spatially confined carriers with nanoscale resolution around the interfaces. Near-field plasmonic tip-enhanced photoluminescence has been able to distinguish distinct crystal boundaries with high resolution, showing enhanced PL at the interfaces.

\subsection{Experimental interfacial parameters}
\label{subsec:ExperimentalInterfaces}

\begin{figure*}[tbph]
\centering
\includegraphics[width=1.0\textwidth]{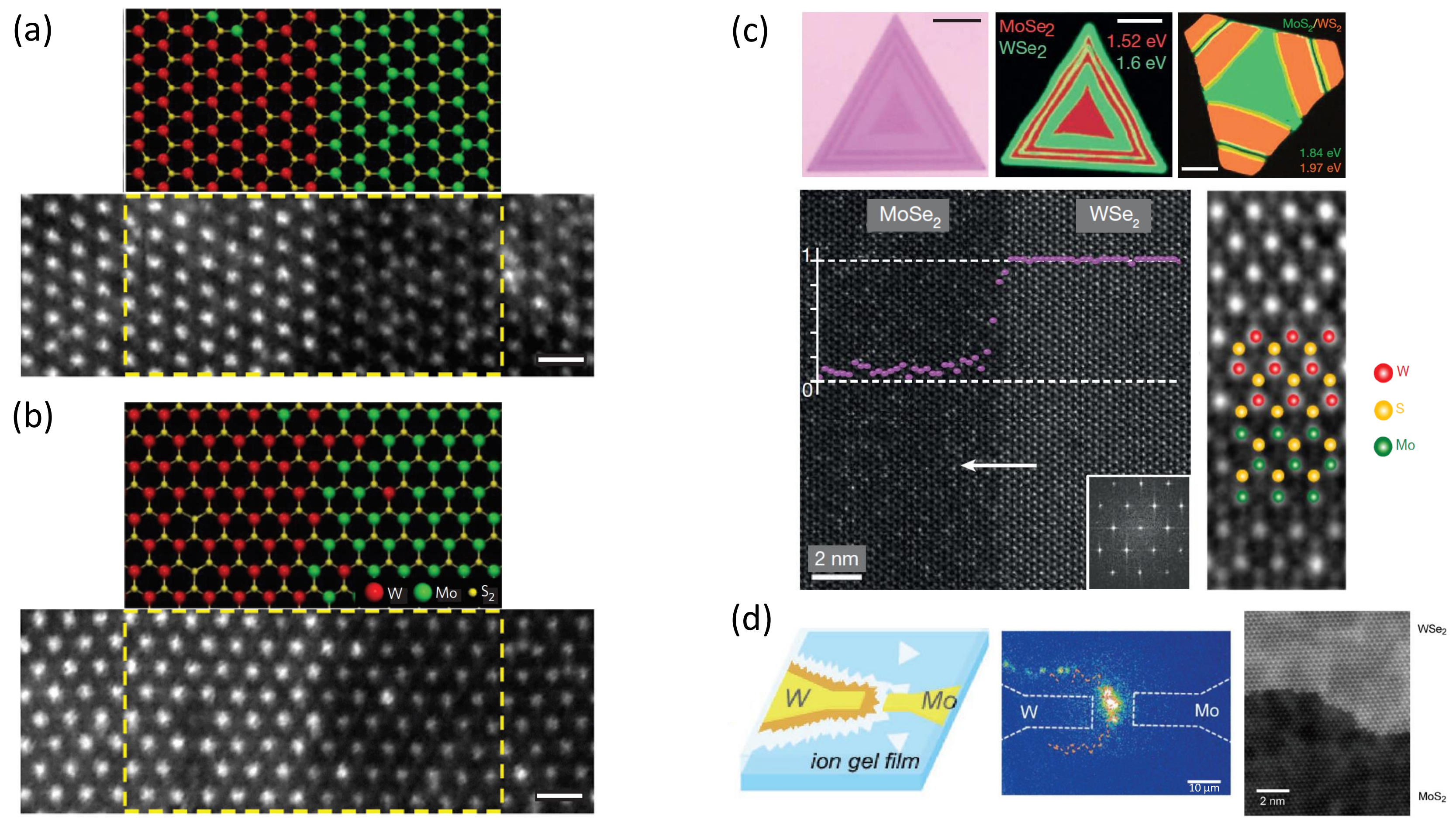}
\caption{(a) Zigzag and (b) armchair interfaces with atomic resolution Z-contrast images, between WS$_2$ and MoS$_2$, with their respective ball-stick models. Scale bar is 0.5 nm \cite{Gong2014NatMat}. (c) One-pot growth of larger and sharper interfaces to date. Upper panels show the optical image of a multi HS, and composite PL maps with TMD and strongest excitonic single peak as indicated. Lower panels show atomic resolution images of a MoSe$_2$-WSe$_2$ sharp zigzag interface, as indicated in the zoomed model \cite{Sahoo2018Nature}. (d) Schematics for the metal deposition and ion-gel film coating process (left panel); electroluminescence image where  white dashed lines show electrode shape and orange dashed line is the interface (middle panel); zoomed sharp interface with atomic-resolution (right panel) \cite{Li2018SelfAligned}. (a) and (b) Reprinted with permission from \cite{Gong2014NatMat}. Copyright 2014 Springer Nature, Nature Materials. (c) Reprinted with permission from \cite{Sahoo2018Nature}. Copyright 2018 Springer Nature, Nature. (d) Reprinted with permission from \cite{Li2018SelfAligned}. Copyright 2018 John Wiley and Sons, Advanced Functional Materials.}
\label{FigInterfaces}
\end{figure*}

Lateral HSs can be grown along both the zigzag and armchair directions, see figure\ \ref{FigInterfaces}(a)-(b). Although zigzag is the most common, armchair interfaces are also  seen often with atomic-resolution STEM  \cite{Gong2014NatMat,Zhang2015}. PL spectroscopy can probe the clean and sharp interface as shown in figure\ \ref{FigInterfaces}(c). The localized excitonic signal is due to the strong built-in electric field at the atomically sharp interface, originating from a type-II band alignment, as will be explained later. This built-in field leads to preferential recombination at the interface. In bulk monolayer regions, radiative recombination of excitons may be suppressed by non-radiative channels \cite{Gong2014NatMat}.

In this section we discuss experimental techniques and parameters that are important for theoretical modeling and characterization. In \ref{subsubsec:InterfacialGeometry} we describe the geometry observed in commensurate HSs, while \ref{subsubsec:InterfacialStrain} describes incommensurate HSs and how strain affects the interface. In  \ref{subsubsec:BandalignmentExperiment} we show measurements in band alignment between both TMD semiconductors forming the HS. Lastly, \ref{subsubsec:PlasmonicsEffectsEXPERIMENT} highlights plasmonic effects observed at the interfaces.

\subsubsection{Interfacial geometry in commensurate HSs}
\label{subsubsec:InterfacialGeometry}

As previously mentioned, when the chalcogen across a HS is the same, the strain is less than 1\%, so that relaxed commensurate interfaces can be achieved [Table \ref{tab:table1} summarizes results for HSs]. One of the 2014 reports \cite{Gong2014NatMat} characterizes the atomic connections between MoS$_2$ and WS$_2$, finding zigzag and armchair interfaces, as shown in figures\ \ref{FigInterfaces}(a)-(b). These were found to be sharp, with 4 unit cells of overall roughness (about 15 nm). The armchair domains were seen to have inter-diffusion over 1-3 unit cells. The longest defect-free zigzag lengths are seen to be about 7 nm, while the armchair are about 2 nm, suggesting the relatively low stability of fresh armchair MoS$_2$ edges during epitaxial growth.

Sahoo \emph{et al.} \cite{Sahoo2018Nature} achieved one-pot CVD growth of either MoSe$_2$-WSe$_2$ or MoS$_2$-WS$_2$, shown in figure\ \ref{FigInterfaces}(c), one of the sharper and longer HSs obtained to date. Their MoSe$_2$-WSe$_2$ [concentric triangles in figure\ \ref{FigInterfaces}(c)] exhibit both atomically sharp and smooth interfaces  just 4 (1 nm) or 21 atomic lines (6 nm) wide in the two different HSs.  This difference is attributed to different oxidation and reduction rates of Mo and W as well as to the gas switching mechanism. Further optimization is anticipated to lead to even sharper interfaces. The MoS$_2$-WS$_2$ trapezoids around a central triangle in figure\ \ref{FigInterfaces}(c) show also sharp interfaces and modulation of the optical bandgap. Inner MoS$_2$ shows two kinds of terminations: Mo- and S-zigzag, depending on the gas environment: chalcogen-deficiency promotes the formation of M-zigzag edges.

Planar and vdWs combinations, terrace interfaces, where the edge of a first monolayer TMD is on top of another, also exhibit zigzag orientations that may act as quantum wires \cite{Zhang2016naturecommunications}.

As most CVD procedures yield zigzag terminations, with sporadic armchair domains, the best approach for obtaining armchair interfaces is perhaps e-beam lithography.  Such cutting of a TMD pristine monolayer followed by deposition of another TMD, achieves a `crosswalk' pattern of lateral MoSe$_2$-MoS$_2$ ribbons \cite{Mahjouri2015patterned} or a bisector strip of the second TMD in a WSe$_2$-WS$_2$ HS \cite{Li2016Laterally}, both with interfaces along the armchair direction.

\subsubsection{Interfacial strain and incommensurate HSs}
\label{subsubsec:InterfacialStrain}

When chalcogens are different at either side of the interface (or the other side is another material altogether), strain plays a large role on the HS properties, which could be used in strain engineering. Although many of these structures have been grown, no detailed analysis of strain distribution had been performed \cite{Duan2014NatNano,Tizei2015Esciton,Ling2016}. Early attempts found a 1.59\% tensile strain and 1.1\% compressive strain in a WSe$_2$-MoS$_2$ HS, as estimated from a PL energy shift rate of 45 meV per \% of strain \cite{Li2015}. Strain effects have now started to be systematically characterized and even tailored in experiments \cite{Xie2018ParkGroup,Zhang2018strain}, allowing coherent HSs. Different interfacial parameters for incommensurate HSs are also listed in table\ \ref{tab:table1}.

\begin{figure*}[tbph]
\centering
\includegraphics[width=1.0\textwidth]{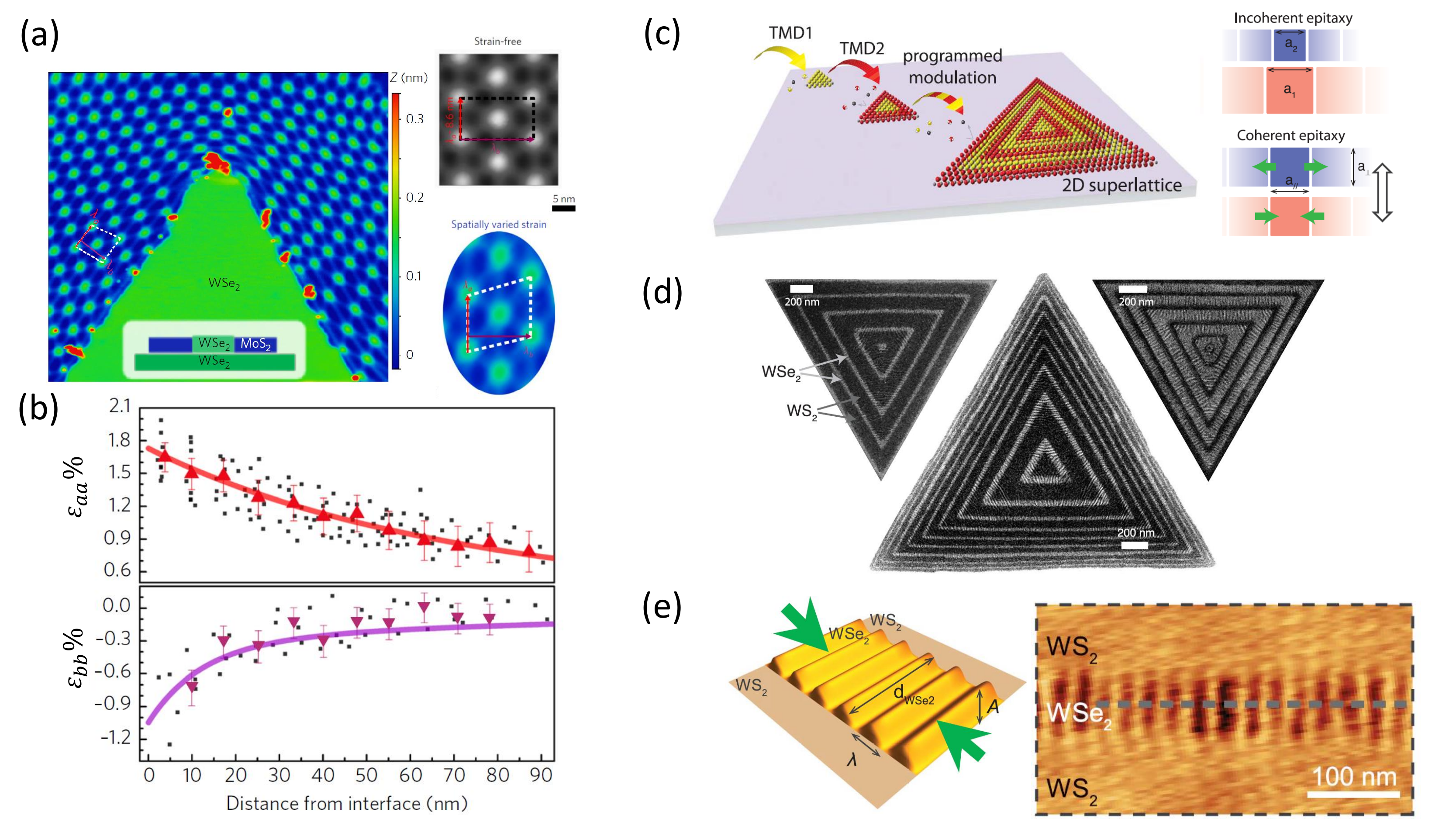}
\caption{(a) Left panel shows STM image of lateral WSe$_2$-MoS$_2$ HS on a WSe$_2$ substrate (inset), where red regions are kink spots due to adsorbates, separating straight interfacial sections. Right panels are moir\'e patterns for both the strain-free and strained regions within the HS, used as \emph{magnification glasses} for changes in lattice constant, given by 3 and 20 respectively \cite{Zhang2018strain}. (b) Strain tensor decays as a function of the distance away from the interface, with experimental results for multiple regions on the WSe$_2$ (black dots), averages along representative zigzag lines (triangles), and fittings (lines) \cite{Zhang2018strain}. (c) Schematic representation of coherent WS$_2$-WSe$_2$ HS (left panel), and its growth epitaxy (right), where $a_{\parallel}$ and $a_{\perp}$ are lattice constants parallel and perpendicular to the interface \cite{Xie2018ParkGroup}. (d) SEM images of achieved coherent planar WS$_2$-WSe$_2$ HSs \cite{Xie2018ParkGroup}. (e) Thermally induced ripples in WSe$_2$ (owing to its larger lattice constant) with respect to WS$_2$ \cite{Xie2018ParkGroup}. (a) and (b) Reprinted with permission from \cite{Zhang2018strain}. Copyright 2018 Springer Nature, Nature Nanotechnology. (c)-(e) Reprinted with permission from \cite{Xie2018ParkGroup}. Copyright 2018 The American Association for the Advancement of Science, Science.}
\label{FigInterfaceStrain}
\end{figure*}

Zhang \emph{et al.} \cite{Zhang2018strain} directly map the anisotropic strain tensor in WSe$_2$-MoS$_2$ using scanning tunneling microscopy/spectroscopy (STM/STS) techniques.  Unlike previous optical techniques, such as Raman and PL, STM/STS is not diffraction limited.  They further use the hexagonal moir\'e pattern ($\sim1$ nm period) as a `magnifying glass' for observing changes in lattice constants, as seen in figure\ \ref{FigInterfaceStrain}(a). When a highly oriented pyrolytic graphite (HOPG) substrate is used, the magnification is found to be 3$\times$, driven by the large lattice mismatch ($>$30\%) between TMD and HOPG substrate, and the nearly zero rotation between them. When a WSe$_2$ substrate is used instead, the magnification factor increases ($>$20$\times$), since there is basically no moir\'e pattern observed due to the small mismatch between HS and the substrate. The strain distribution is characterized by the 2D strain tensor parameters $\epsilon_{aa}$, $\epsilon_{bb}$, and $\epsilon_{ab}=\epsilon_{ba}$, with $a$ and $b$ defined along the zigzag and armchair directions, as shown in figure\ \ref{FigInterfaceStrain}(b). It is seen that $\epsilon_{aa}$ decays much faster than $\epsilon_{bb}$, by a 2 to 1 ratio over a 50 nm length.  This difference is probably due to the fact that there is a free edge during growth, allowing stress relaxation normal to the edge. Analytical modelling for these $\epsilon$ components is discussed later in section \ref{subsubsec:IncommensurabilityAndStrain} below.

Xie \emph{et al.} \cite{Xie2018ParkGroup}, are able to control strain effects in coherent WSe$_2$-WS$_2$ and (WSe$_2$-MoS$_2$-WS$_2$) lateral HSs, as shown in figure\ \ref{FigInterfaceStrain}(c). The interface was repeated without dislocations, matching lattice constants at the interfaces (even though they are $\sim$4\% off), and maintaining structure and triangular symmetry, as seen in figure\ \ref{FigInterfaceStrain}(d). Since one misfit dislocation is expected every 25 unit cells on average, the 160 unit cells ($\sim$50 nm) average length observed is definite evidence of coherent HSs. This data agrees with coarse-grained simulations (see section \ref{subsubsec:IncommensurabilityAndStrain})  which account for bonding and angle interactions. Rippled regions where the lattice constant is larger (WSe$_2$) can also be achieved perturbing the coherent 2D flat HS with thermal cool-down immediately after growth, as shown in figure\ \ref{FigInterfaceStrain}(e). These ripples show characteristic wavelengths of about 30 nm.

\subsubsection{Band alignment}
\label{subsubsec:BandalignmentExperiment}

A key feature of HSs is that bandgaps and Fermi levels of both materials are usually different, leading to polarization dipoles and even charge transfer across the HS, driven by differences in the bulk conduction and valence bands. These can be seen to arise from differences in electronegativity and/or work function of the materials across the HS.
A major question to be addressed is to determine the relative band alignment of the conduction and valence bands across the HS. Borrowing from bulk semiconductors, one identifies three usual types of alignments: type-I, when the bandgap of one material is contained (nested) inside the bandgap of the other  (also called symmetric alignment); type-II, when the conduction band maximum (CBM) of one material is inside the gap of the other (also called staggered alignment); and type-III, when the CBM of one material is lower than the valence band minimum (VBM) of the other material (also called broken alignment). Related useful quantities to measure are the conduction and valence band offsets, CBO and VBO respectively, defined as CBM$_1-$CBM$_2 \equiv$ CBO, and VBM$_1-$VBM$_2 \equiv$ VBO.

For HSs between different TMDs, band alignments have been calculated with DFT \cite{Kang2013,Gong2013,Kosmider2013,Wei2014,Guo2016,OngunOzcelik2016} (see below for details),
and measured experimentally for vertical \cite{Chiu2015,Hill2016} and lateral \cite{Gong2014NatMat,Zhang2018strain} HSs in different works. Vertical HS band alignment has been experimentally measured by STM/STS, scanning photocurrent microscopy and X-ray photoelectron spectroscopy (XPS), in MoS$_{2}$ \cite{Howell2015}, MoSe$_{2}$ \cite{Zhang2016naturecommunications} and WSe$_{2}$ \cite{Zhang2016naturecommunications} terraces, and both MoS$_{2}$-WSe$_{2}$ \cite{Chiu2015} and WS$_{2}$-MoS$_{2}$ \cite{Hill2016} vertical HSs.  Although most of these experimental works consider vertical HSs, Zhang \emph{et al.} \cite{Zhang2018strain} have recently studied band alignment in WSe$_{2}$-MoS$_{2}$ lateral HSs.

\begin{figure*}[tbph]
\centering
\includegraphics[width=1.0\textwidth]{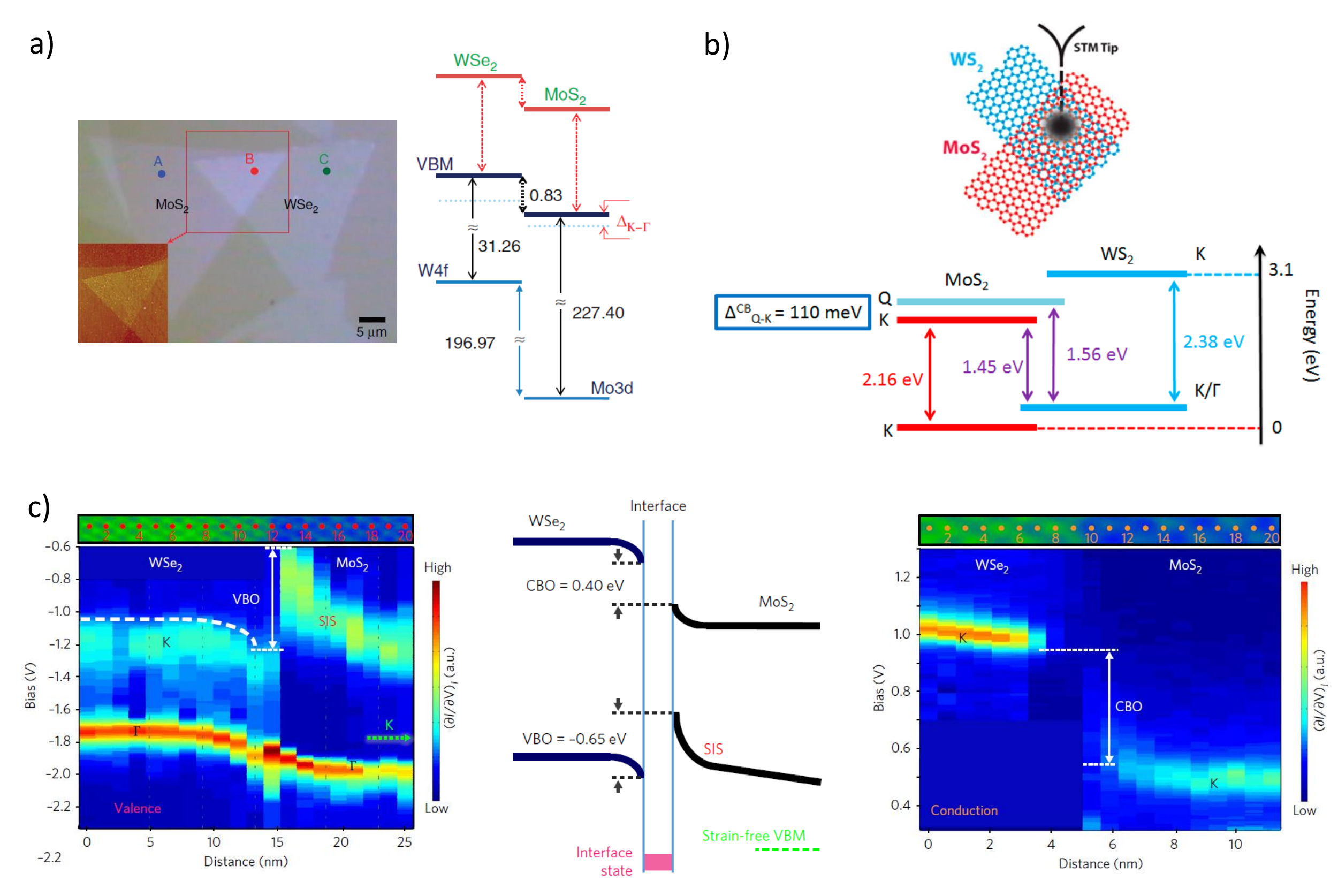}
\caption{Experimentally resolved band alignments (a) type-II for a vertical MoS$_{2}$-WSe$_{2}$ HS \cite{Chiu2015}, (b) type-II for a vertical MoS$_{2}$-WS$_{2}$ \cite{Hill2016}, and (c) type-I for a lateral MoS$_{2}$-WSe$_{2}$ \cite{Zhang2018strain}. In (a) and (b) the schematic setups for the $\mu$-XPS + STM/S measurements are shown, as well as the alignment diagrams. In (c), left and right panels show STS spectra as the tip is approached to the interface (see upper insets showing relative atomic positions), for measuring the VBO and CBO; results schematically shown in the middle panel. Here the levels for the interface states and strain-induced states (SIS) are also shown, with respect to the strain-free MoS$_2$ case (dashed green). (a) Reprinted with permission from \cite{Chiu2015}. Copyright 2015 Creative Commons Attribution 4.0, Nature Communications. (b) Reprinted with permission from \cite{Hill2016}. Copyright 2016 American Chemical Society, Nano Letters. (c) Reprinted with permission from \cite{Zhang2018strain}. Copyright 2018 Springer Nature, Nature Nanotechnology.}
\label{BandAligmentMixed}
\end{figure*}

A type-II band alignment has been inferred in vertical MoTe$_{2}$-MoS$_{2}$ HSs from SKPFM and Raman measurements, and theoretically calculated to be $\simeq 0.66$ eV \cite{Zhang2016ACSnano}. For terrace HSs of the same TMD in a monolayer-bilayer interface, the band alignment for MoS$_{2}$  observed by scanning photocurrent microscopy is found to be type-II \cite{Howell2015}. Later, STS measurements in terraces of WSe$_{2}$ and MoSe$_{2}$ found type-I band alignment \cite{Zhang2016naturecommunications}, with VBO for WSe$_{2}$ (MoSe$_{2}$) of 0.12 eV (0.43 eV) and CBO of 0.15 eV (0.08 eV). This work also reports DFT calculations for vertical HSs of TMDs with different chalcogens. An interesting hybrid bilayer system, with top WS$_2$ layer and bottom WS$_2$-MoS$_2$ lateral HS, is studied by STM/STS and it appears to show type-II band alignment at the HS \cite{Kobayashi2016}.

Chiu \emph{et al.} \cite{Chiu2015} used STS and $\mu$-XPS measurements in vertical MoS$_{2}$-WSe$_{2}$ HSs, finding that the HS bandgap is $1.32\pm0.12$ eV, measured from the VB K-point of WSe$_{2}$ up to the CB K-point of MoS$_{2}$, corresponding to a type-II alignment. They measure the VBO is 0.83 eV, and CBO is 0.76 eV; the quasiparticle gaps of MoS$_{2}$ ($2.15\pm0.01$ eV) and WSe$_{2}$ ($2.08\pm0.01$ eV) are also reported.  The VBO value of $\approx0.8$ eV is supported by DFT calculations (GGA-PBE) \cite{Guo2016}. In other work, Hill \emph{et al.} \cite{Hill2016} studied both vertical MoS$_{2}$-WS$_{2}$ and WS$_{2}$-MoS$_{2}$ HSs by STS, and observed an HS bandgap of $1.45\pm0.06$ eV, measured from the VB K-point of WS$_{2}$ up to the CB K-point of MoS$_{2}$, corresponding to type-II alignment. The quasiparticle gaps of MoS$_{2}$ ($2.16\pm0.04$ eV) and WS$_{2}$ ($2.38\pm0.06$ eV) on the HS setup were determined together with the energy difference between the Q and K-points on the MoS$_{2}$ CB of 110 meV.  The band offset findings are schematically shown in figure\ \ref{BandAligmentMixed}(a)-(b) \cite{Chiu2015,Hill2016}, and a brief summary is given in table \ref{tab:table2}.

\begin{table*}[t]
\caption{\label{tab:table2} Experimental band alignment energy parameters. The parameters are (in order): HS, type of HS, MoS$_{2}$ quasiparticle gap $\Delta_{\textrm{MoS}_{2}}$, WS(Se)$_{2}$ quasiparticle gap $\Delta_{\textrm{WS(Se)}_{2}}$, HS gap $\Delta_{\textrm{HS}}$, valence band offset (VBO), conduction band offset (CBO), and type of alignment. All energies are in eV.}
\begin{indented}
\item[]\begin{tabular}{@{}cccccccc}
\br
HS & HS type &$\Delta_{\textrm{MoS}_{2}}$& $\Delta_{\textrm{WS(Se)}_{2}}$ & $\Delta_{\textrm{HS}}$ & VBO & CBO & band alignment\\
\mr
MoS$_{2}$-WSe$_{2}$ \cite{Chiu2015} & Vertical & 2.15$\pm$0.01 & 2.08$\pm$0.01 & 1.32 & 0.83$\pm$0.07 & 0.76$\pm$0.12 & type-II \\
MoS$_{2}$-WS$_{2}$ \cite{Hill2016} & Vertical & 2.16$\pm$0.04 & 2.38$\pm$0.06 & 1.45 & 0.71 & 0.93 & type-II \\
MoS$_{2}$-WSe$_{2}$ \cite{Zhang2018strain} & Lateral & position-dependent & position-dependent & 0.52 & -0.65$\pm$0.05 & 0.40$\pm$0.05 & type-I \\
\br
\end{tabular}
\end{indented}
\end{table*}

In lateral HSs, two works have addressed band alignment. The study by Gong \emph{et al.} \cite{Gong2014NatMat} finds the DFT alignment in WS$_{2}$-MoS$_{2}$ is type-II, with band offset of 0.07 eV at the VBM, and gaps of 1.59 eV and 1.55 eV for MoS$_{2}$ and WS$_{2}$, respectively. This work also calculates a built-in electric field of over $2\times10^{8}$ N/C at the zigzag interface, which may drive free electrons and holes generated in the vicinity of the interface to recombine preferentially at the interface. More recently, Zhang \emph{et al.} \cite{Zhang2018strain} analyzed the band alignment in a lateral MoS$_{2}$-WSe$_{2}$ HS, finding that the misfit strain induces a type-II to type-I transformation. They used STS mappings of the valence and conduction bands as the tip moves across the interface, as shown in figure\  \ref{BandAligmentMixed}(c).  While a vertical HS of the same materials shows type-II alignment \cite{Chiu2015} (as shown in figure\ \ref{BandAligmentMixed}(a) \cite{Chiu2015} and (b) \cite{Hill2016}), the lateral HS shows type-I alignment, as the MoS$_2$ valence band shows an unexpected spatial variation with respect to WSe$_2$, with CBO=0.4 eV and VBO$=-$0.65 eV, as shown in figure\ \ref{BandAligmentMixed}(c). The strain pushes the VBM of the MoS$_2$ above the $\Gamma$ point, while the CBM behaves more straightforwardly. The band bending is found to start just 5 nm away from the interface on the WSe$_2$ side, while in MoS$_2$ the band bending starts further away. The potential discontinuity is however observed in a window of just 1 nm. To the best of our knowledge, this is the first accurate space-resolved measurement of band alignment in a lateral TMD HS.

\subsubsection{Plasmonics}
\label{subsubsec:PlasmonicsEffectsEXPERIMENT}

Quantum plasmonic effects were recently observed in WS$_2$-MoS$_2$ \cite{Shi2017cascaded} and WSe$_2$-MoSe$_2$ \cite{Tang2018} lateral HSs, measuring photoresponse that suggests these systems might serve as quantum nanodevices with tunable optical response.

Shi \emph{et al.} \cite{Shi2017cascaded} transferred WS$_2$-MoS$_2$ onto a Ag plasmonic plate covered with Al$_2$O$_3$, so as to transfer  the excitonic energy to surface plasmon polaritons. A complex cascade of exciton/surface-plasmon-polariton/exciton conversion in lateral HSs was demonstrated from WS$_2$ to MoS$_2$, as mediated by the plasmonic substrate.  The advantage of having an atomically sharp interface is that the energy transfer has a propagation length of $\sim$40 $\mu$m (2 orders of magnitude larger than in bare TMD), and the pristine interface minimizes energy loss.

The experiments by Tang \emph{et al.} \cite{Tang2018} image near-field tip-enhanced photoluminescence (TEPL) of a lateral 150 nm wide interface (not atomically sharp). They investigate tunneling-assisted hot-electron injection (HEI) at room temperature, observing quenching and enhancement of the PL from the interfacial region due to the attenuation of localized electromagnetic field and hot electron injection. TEPL allowed optical characterization of the HS, showing that the interface PL response can be controlled by varying lateral tip position and picoscale tip-sample distance. For charge tunneling distances of $\sim$20 pm, the electron tunneling facilitates thermionic injection in the quantum regime.

The interface plays a critical role in the enhancement of the TEPL signal: it is the interfacial region that allows the MoSe$_2$ side to accumulate more plasmon-induced hot electrons. This is because of  directional hot electron injection at the interface, due to band alignment. Hot electrons are transferred to the MoSe$_2$ side and when the tip diameter is comparable to the interfacial region (20 to 0.36 nm) the injected hot electrons accumulate in MoSe$_2$, leading to PL enhancement in MoSe$_2$ and quenching on the WSe$_2$ side. Close tip-sample distance favors electron tunneling, leading to extra quenching in the WSe$_2$ PL, while the MoSe$_2$ component is still enhanced.

\subsection{Phase interfaces within the same MX$_{2}$}
\label{subsec:PhaseInterfaces}

Different crystalline phases of the same TMD can also be created by electrostatic potential differences between regions, for example.  Lateral \emph{p-n} junctions within the same TMD \cite{Baugher2014Optoelectronic,Pospischil2014Solar,Ross2014Electrically,Desai2016Mos2} have been studied with a smooth HS profile.

An early report by Eda \emph{et al.} showed the creation of coherent interfaces between semiconducting
H and metallic T phases within MoS$_2$, characterized by STEM \cite{Eda2012Coherent}. In 2014, Lin \emph{et al.} \cite{lin2014atomic} created few-atoms-wide interfaces between MoS$_2$ metallic 1T triangular islands embedded in MoS$_2$ semiconducting 2H phases controlling the growth of triangular 1T regions by electron beam illumination. They observed that the atomic interface shows a dynamic evolution between different H-T phases of MoS$_2$, involving atomic gliding of S and/or Mo planes to achieve the triangular island. Further insight into these experiments is given by DFT calculations \cite{Kretschmer2017}. More recently, Yoo \emph{et al.} \cite{Yoo2017} showed the creation of lateral HSs between MoTe$_2$ 2H-1T' phases, by controlling temperature of the reaction vessel and Te flux (high flux for the 2H phase and low for 1T'). These crystals appear as 2H circular islands, laterally connected to multilayer 1T' regions. SKPFM and Raman show sharp in‐plane interfaces.

\begin{figure}[tbph]
\centering
\includegraphics[width=0.4\textwidth]{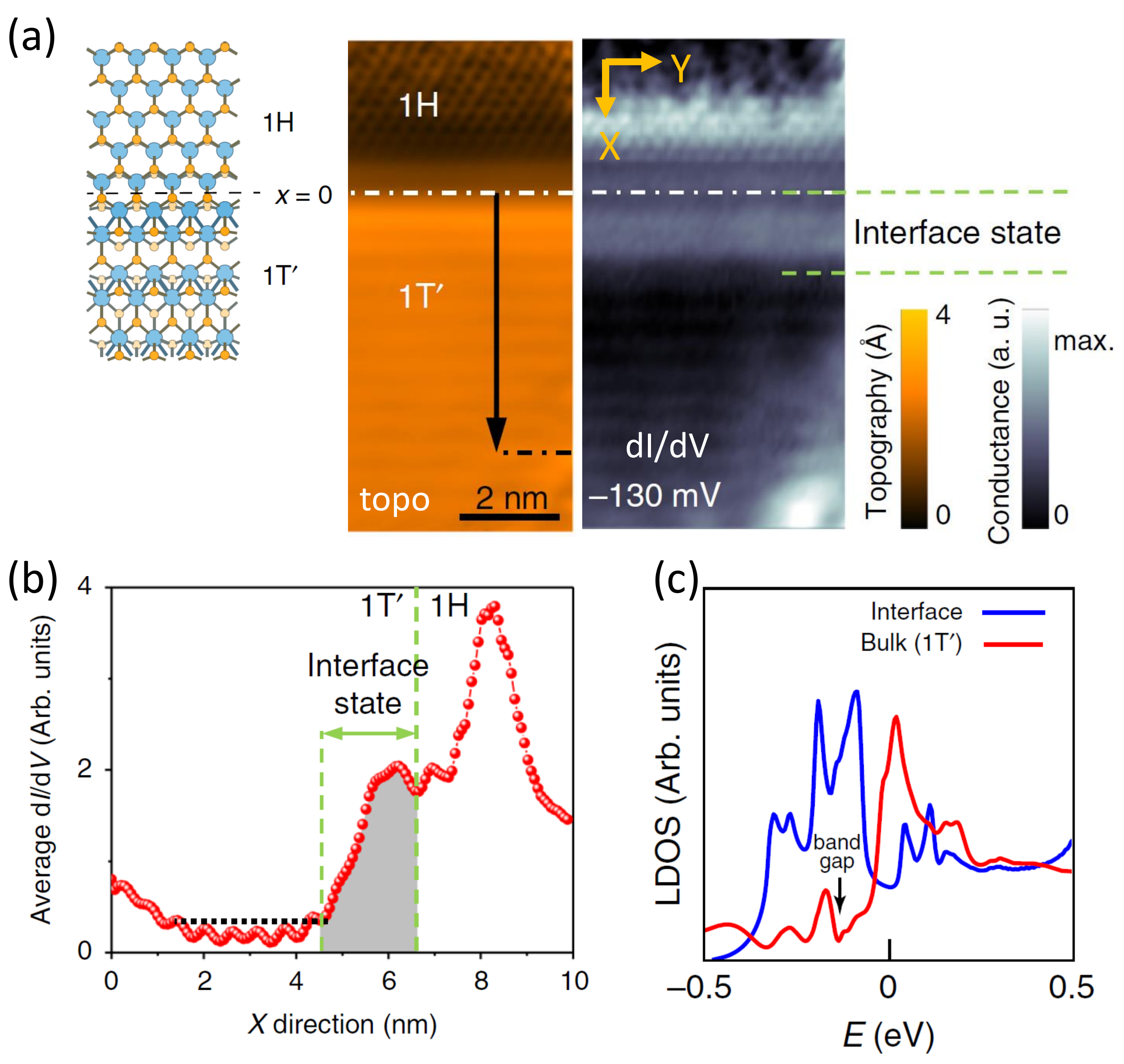}
\caption{1D interface in 1H-1T' WSe$_2$ (a) Model, topography, and dI/dV STS spectra at $-130$ mV. Dashed-dotted white line indicates the interface, and green dashed lines indicate the spatial extent of the interface state. (b) dI/dV along $x$-direction. (c) Comparison between interfacial and bulk LDOS. Reprinted with permission from \cite{Ugeda2018arxiv}. Copyright 2018 Creative Commons Attribution 4.0, Nature Communications.}
\label{Ugeda2018arxiv}
\end{figure}

One fascinating aspect of the 1D 1T'-phase structures is their topological nature. An atomic sharp interface between 1T'- and 1H-WSe$_{2}$ monolayer has been synthesized \cite{Ugeda2018arxiv}, figure\ \ref{Ugeda2018arxiv}(a), to study topological properties at the 1D interface. Topologically protected helical edge states were seen at the interface, showing that such a novel quantum spin Hall insulator platform is possible \cite{Ugeda2018arxiv}. Molecular beam epitaxy (MBE) is used to grow a mixed-phase of monolayer WSe$_2$ and characterized with angle-resolved photoemission spectroscopy (ARPES) and STM/STS, jointly revealing inverted bulk bands, and the existence of topological interface states within the bandgap, figure\ \ref{Ugeda2018arxiv}(c), at crystallographically well-ordered interfaces.  All this in agreement with first principles calculations.
The 1D interfacial states at such atomically sharp interface have characteristic decay penetration length of only 2 nm into the bulk, as shown in figure\ \ref{Ugeda2018arxiv}(b).

\section{Numerical and theoretical descriptions}
\label{sec:TheoreticalAndNumerical}

This section represents the main scope of this review, and focuses on current theoretical advances for the description of lateral HS between TMDs. While experiments have been concentrating on achieving pristine and coherent interfaces with lengths now exceeding  several micrometers, prospective theoretical directions and applications have also been reported.

In section\ \ref{subsec:StructuralAndElectronic} we present advances in numerical and theoretical calculations of different aspects of TMD lateral HSs, and in \ref{subsec:1DNovelPlatform} we review proposals for using these interfaces as an effective platform for unique 1D physics.

\subsection{Structural and electronic properties}
\label{subsec:StructuralAndElectronic}

DFT calculations of lateral HSs have been appearing since 2014, as the first clean interfaces were being grown by several groups. These theoretical works are mostly focused on studying the evolution of the electronic bands and bandgaps, with consideration of band alignment effects. Structural studies focus on geometrical stability and strain, utilizing relatively small unit cells for computation.

The remainder of this section is arranged as follows: first, in \ref{subsubsec:ElectronicStructure} we summarize works on electronic structure for commensurate TMD HSs, highlighting band alignment, stability, and structure. Then, in \ref{subsubsec:IncommensurabilityAndStrain} we provide an overview of how the inherent strain at the interface affects the electronic properties, especially for non-commensurate HSs.

\subsubsection{Electronic structure}
\label{subsubsec:ElectronicStructure}

\emph{Band alignment.- } Let us first review the theoretical works that analyze an important aspect in the electronic structure of TMD HSs, which is the band alignment (or band offset) across the juncture. These offsets are important parameters in material design, as discussed in section\ \ref{subsubsec:BandalignmentExperiment}, and HS modeling requires an accurate knowledge of the alignment. Unfortunately, the band offsets for monolayer materials and their lateral heterostructures are not fully known theoretically.  Depending on the materials involved, some works suggest type-I (nested) alignment, others suggest type-II \cite{Kang2013,Kosmider2013,Wei2014}, and even type-III alignment in some cases \cite{OngunOzcelik2016}.

\begin{figure}[tbph]
\centering
\includegraphics[width=0.5\textwidth]{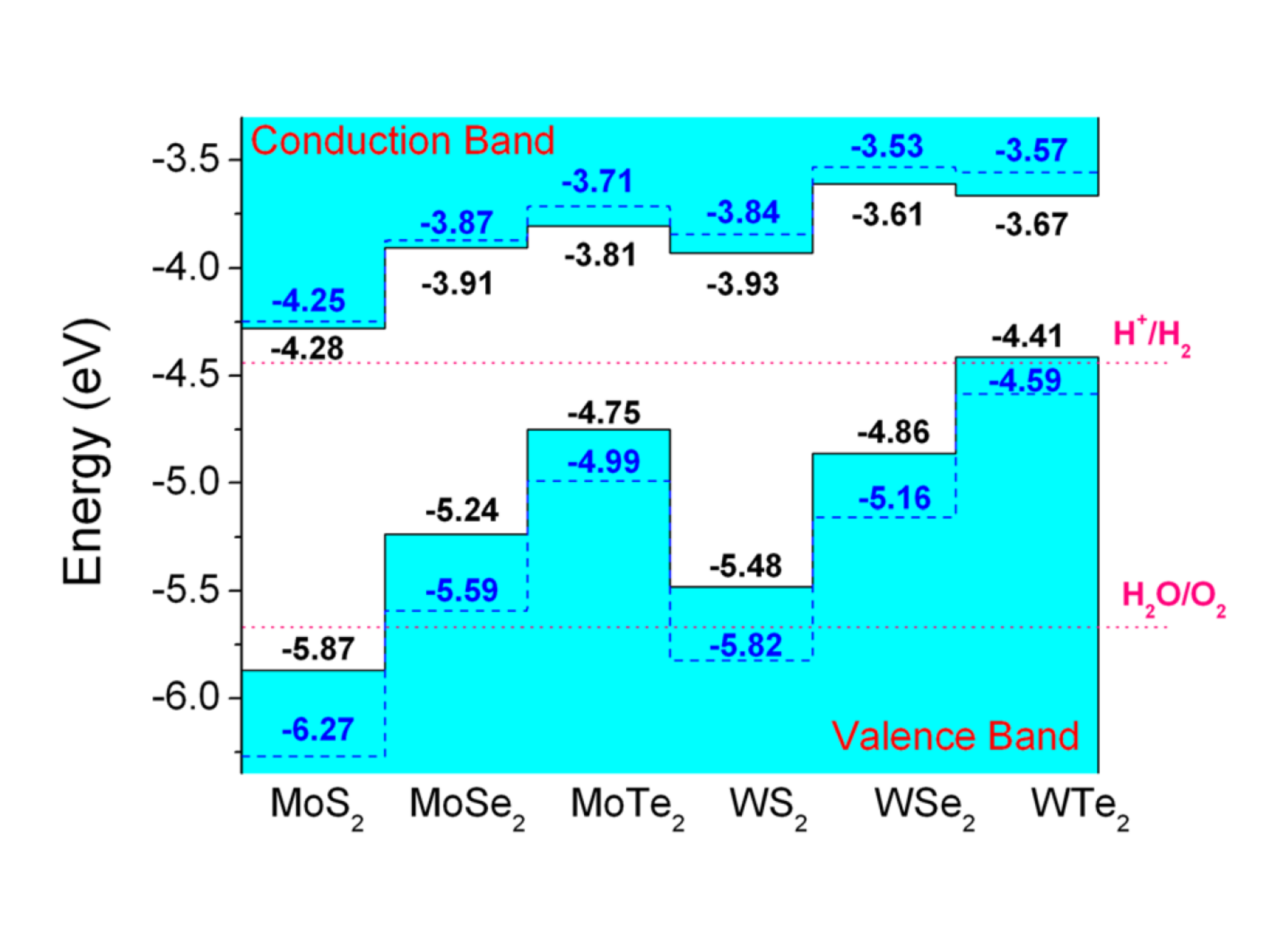}
\caption{Band alignments for six TMDs as shown. Solid (dashed) lines are with PBE functional (HSE06 hybrid functional). Potentials levels for water reduction (H$^+$/H$_2$) and oxidation (H$_2$O/O$_2$) are also shown. Reprinted with permission from \cite{Kang2013}. Copyright 2013 by AIP Publishing, Applied Physics Letters.}
\label{Kang2013}
\end{figure}

Early DFT band alignment studies mostly addressed vertical HSs \cite{Kang2013,Gong2013,Kosmider2013}. The first work to analyze lateral monolayer HSs and the role of band alignment between different TMDs, was reported by Kang \emph{et al.} \cite{Kang2013}. They use the Vienna \emph{ab initio} simulation package VASP \cite{Kresse1996} with projector augmented wave (PAW) \cite{Blochl1994}, in either the generalized gradient approximation of Perdew-Burke-Ernzerhof (GGA-PBE) \cite{PerdewBurkeErnzerhof1996} or the hybrid Heyd-Scuseria-Ernzerhof (HSE06) \cite{Heyd2003} functionals for electronic correlations. They study several lateral TMD combinations, finding similar chemical trends, regardless of the functional used. This suggests a model to establish relative alignment of valence and conduction bands, from the orbital content of the VBM and CBM, which originate from the repulsion between the cation-$d$ and anion-$p$ orbitals. They show that the MoX$_2$-WX$_2$ lateral HS would have a type-II band alignment, as shown in figure \ref{Kang2013}, where each WX$_2$ element is higher in energy than its same-chalcogen MoX$_2$ counterpart. This approach has become popular and remains widely used.
In general PBE (GW$_{0}$) underestimates (overestimates) the bandgaps, and the bandgap accuracy is improved by using hybrid functionals, such as HSE06. Nevertheless, the HSE06 functional overestimates the spin splitting of the valence band \cite{Amin2015,Kormanyos2015}, so that one finds use of PBE and HSE06 as lower and upper bound estimates for gaps, respectively. The band structure of HSs is also very sensitive to the type of atomic stacking \cite{Kormanyos2015,Terrones2013}.

Other DFT calculations have shown that a vertical MoS$_{2}$/WS$_{2}$ HS \cite{Kosmider2013} also shows type-II band alignment with direct bandgap (HSE functional, while PBE predicts indirect bandgap), in contrast to their pristine bilayer counterparts, both of which show indirect bandgaps. Gong \emph{et al.} studied the band alignment between several vertical TMDs \cite{Gong2013}, including the semiconducting group-IVB and metallic group-VB TMDs (IVB: Ti, Zr, and Hf; VB: V, Nb, Ta). They find that tunnel field effect transistors could be built with $p$-$n$ junctions of group VIB $n$-type and group IVB $p$-type HSs. Soon after these predictions, experimental work finds that the lateral WS$_{2}$-MoS$_{2}$ HS alignment is indeed type-II, providing a combined theory-experiment study.  DFT calculations determined a band offset of 0.07 eV at the VBM, and gaps of 1.59 eV and 1.55 eV for MoS$_{2}$ and WS$_{2}$, respectively \cite{Gong2014NatMat}. This work also calculates a strong built-in electric field of over $2\times10^{8}$ N/C at the zigzag MoS$_{2}$-WS$_{2}$ lateral interface.

Interest in the alignment between TMD in lateral HSs has been increasing over the years \cite{Wei2014,OngunOzcelik2016,Amin2015}. Wei \emph{et al.} confirmed the type-II alignment predicted by Kang \emph{et al.}, additionally studying lateral junctions with metallic TMDs \cite{Wei2014}. Other DFT work \cite{Amin2015} looks at vertical and lateral MoX$_{2}$-WX$_{2}$ (X=S, Se, Te) HSs, reporting structural, electronic, optical, and photocatalytic properties. We note, however, that this system is not a single interface between two slabs, but rather an in-plane arrangement of single atomic lines of different transition metals, with zigzag interfaces between each other, which one can describe as a large concentration of parallel grain boundaries. They find all these systems have direct bandgap, with contributions of both the Mo and W to the VBM and CBM.

\begin{figure}[tbph]
\centering
\includegraphics[width=0.5\textwidth]{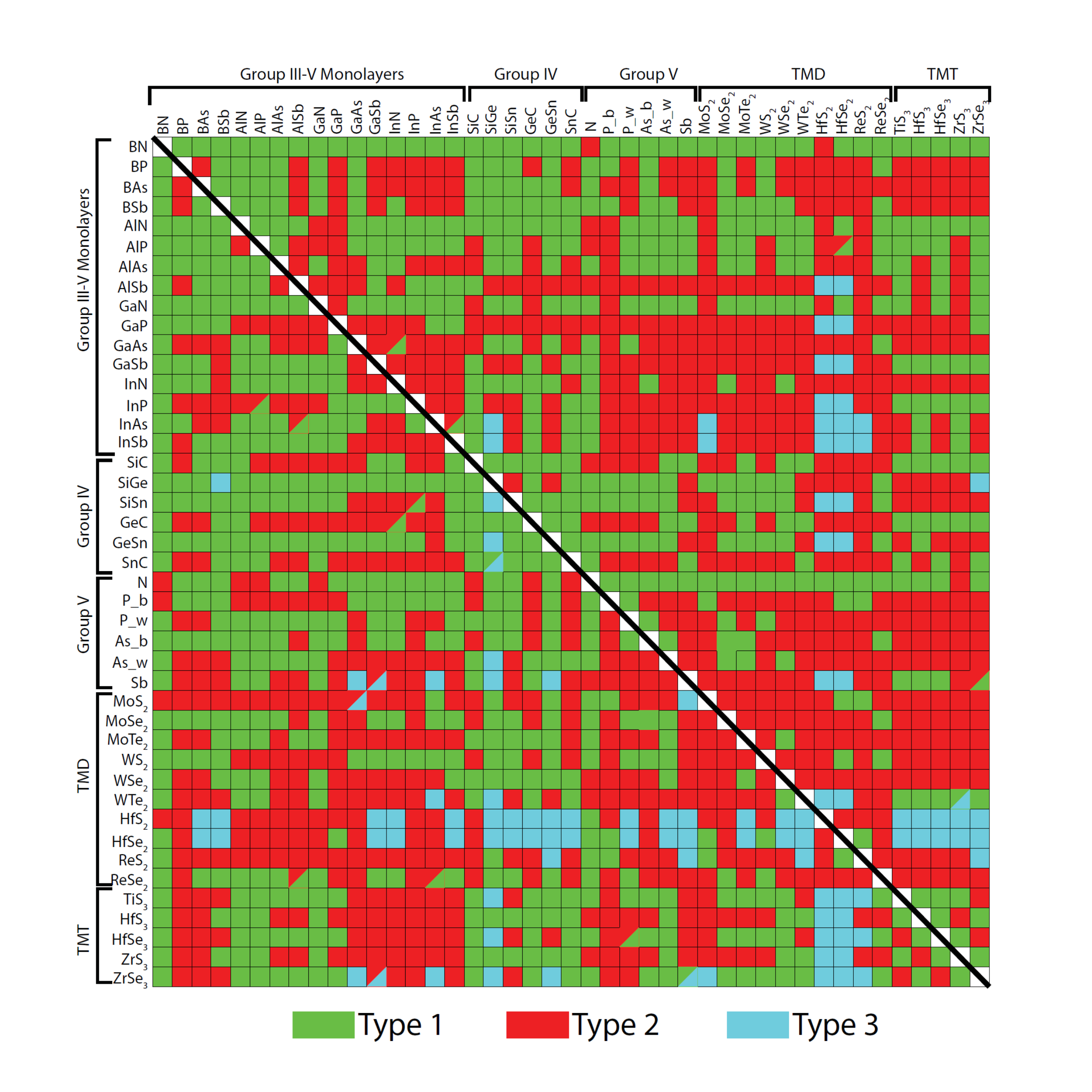}
\caption{Band alignment of different HSs, in the so called \emph{periodic table of heterostructures} \cite{OngunOzcelik2016}. Lower left (upper right) region of the table corresponds to results for the PBE (hybrid HSE06) functional. Two colors for the same HS, indicates accuracy is beyond DFT error range, so the HS can have either type. Reprinted with permission from \cite{OngunOzcelik2016}. Copyright 2016 by the American Physical Society, Physical Review B.}
\label{OngunOzcelik2016}
\end{figure}

More recently, Guo \emph{et al.} have addressed the issue of band alignment for lateral HSs of different TMDs, metallic and semiconducting \cite{Guo2016}. They used the CASTEP plane wave pseudopotential \cite{CASTEP2005} code, with a combination of ultrasoft potentials. For lateral MoS$_2$-WS$_2$ HS, they studied both zigzag and armchair interfaces, finding little difference in the HS projected DOS. This is attributed to the Mo-S bond being relatively non-polar, with only 0.3e charge on each S site. Finally, a comprehensive spin-polarized DFT study of band alignment \cite{OngunOzcelik2016} (PBE and HSE06 functionals) has been carried out for well-known 2D semiconductors, including transition metal di- and trichalcogenides. This results in a useful database, the \emph{periodic table of heterostructures}, including geometries, electronic structure and band offsets, among other properties. This is shown in figure \ref{OngunOzcelik2016}. In this table, the alignment for most group-VIB TMD HSs is proposed to be type-II.

Recent studies have argued for the applicability of the definition of band alignment. Wei \emph{et al}.\ dispute whether the alignment in lateral HSs of different TMDs can be addressed by separately aligning rigid band edges, since the creation of a dipole at the interface \cite{Wei2016}, which has been seen experimentally \cite{Gong2014NatMat}, is an effect that strongly depends on the structure. The size and directionality of the dipole should consider the different TMD crystallite edges, such as zigzag and armchair, as well as terminations involving grain boundaries \cite{Kang2015}, and/or defects \cite{Wei2016,Cao2017}. Such a complete description of the band alignment in lateral TMD HSs is yet to come.

\emph{Band structure of commensurate TMD HSs.- } Most numerical work in lateral TMD HSs has been done in nanoribbons (NRs). Let us first summarize what is known for pristine TMD NRs. DFT studies have shown that zigzag terminated NRs have a magnetic ground state, with metal, half-metal and semiconducting electronic states depending on the NR width \cite{Wei2015,Wen2016}, while larger sizes tend to remain metallic. Armchair-terminated NRs are nonmagnetic and semiconducting \cite{Wei2015,Wen2016}. Zigzag magnetic properties can be enhanced by strain, while bandgaps of armchair NRs decrease with strain. MoS$_2$ tensile (compressive) strain increases (reduces) the bond lengths, so that the bulk bandgap reduces (increases) monotonically and a direct-indirect transition occurs. In contrast, bi-axial tensile strain reduces the gap further \cite{Wei2017}.

Pristine zigzag NRs are found to exhibit magnetic ground states for small NRs (smaller than 8 atomic lines in width) \cite{Wen2016}.
The larger magnetic moment is for the smallest `ribbon' (2 atomic lines, 1 of each atom), while higher MoS$_{2}$ content produces smaller magnetic moment for any width. Pristine armchair NRs have smaller (larger) bandgap for smaller (larger)  width, so that the gap can be tuned by changing the NR width and edge termination. These authors also report a transition from indirect to direct bandgap, when the width increases above 9 atomic lines.

Studies in lateral TMD HSs have been carried out for (nearly) commensurate and incommensurate junctions.  The first ones are built out of TMDs with different transition metal but the same chalcogen (such as MoS$_{2}$-WS$_{2}$), while the latter have typically different chalcogen (such as MoS$_{2}$-MoTe$_{2}$). The latter will be addressed in more detail in section \ref{subsubsec:IncommensurabilityAndStrain}.

An early study by Wang \emph{et al}.\ \cite{Wang2013} tackles structural and electronic properties of commensurate MoS$_{2}$-WS$_{2}$, and incommensurate MoS$_{2}$-MoTe$_{2}$ HSs. They used Quantum Espresso, and PBE-GGA for exchange-correlation, and found that the MoS$_{2}$-WS$_{2}$ HS remains a semiconductor after hybridization, with bandgap of 1.58 eV, smaller than that of the constituents. They also find that the lowest energy superlattice system consists of a MoS$_{2}$ row embedded into a WS$_{2}$ ribbon.
Larger systems were studied with VASP (GGA-PBE for electron exchange correlation) \cite{Wen2016}.

\begin{figure}[tbph]
\centering
\includegraphics[width=0.4\textwidth]{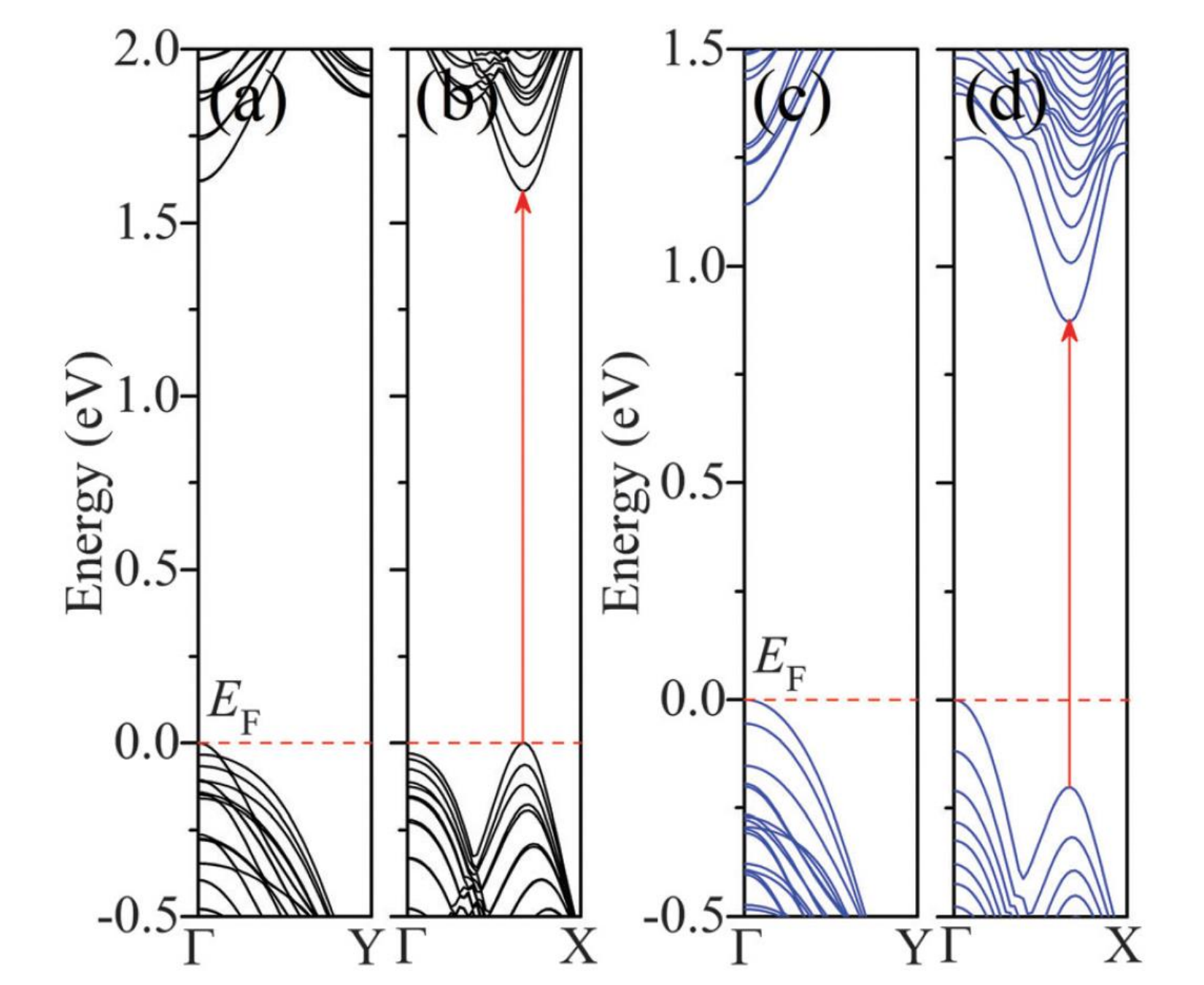}
\caption{Band structures for (a) armchair MoS$_{2}$-WS$_{2}$, (b) zigzag MoS$_{2}$-WS$_{2}$, (c) armchair MoSe$_{2}$-WS$_{2}$, and (d) zigzag MoSe$_{2}$-WS$_{2}$. Red arrows indicate the bandgap. Reprinted with permission from \cite{Wei2015}. Copyright 2015 by Royal Society of Chemistry, Physical chemistry chemical physics.}
\label{Wei2015PCCPFIGURE2}
\end{figure}

Wei \emph{et al}.\ \cite{Wei2015SciRep} have studied the electronic properties of quantum well HSs, a system with two interfaces, such as MoS$_{2}$-WS$_{2}$-MoS$_{2}$, among others. They used PAW+VASP with GGA+PBE for exchange/correlations, and found that the electronic properties of these quantum wells can be engineered by adjusting the strain, resulting in different bandgaps and an indirect-to-direct bandgap transition as the number of unit cells in each HS changes, similar to results by Kang \emph{et al}.\ for single interfaces \cite{Kang2015}. Wei {\em et al}.\ also find type-II alignment in coherent interfaces with strong coupling, suggesting effective separation and collection of excitons as a possible application. The same group studied interface properties in great detail, confirming that excitons should stay confined at opposite sides of the 1D interface due to the type-II band alignment \cite{Wei2015}. Typical band structures for sufficiently large HSs (width $\sim$90\AA) are shown in figure \ref{Wei2015PCCPFIGURE2}. All HSs are found to be semiconducting with direct gaps, at the $A$-point (which is 2/3 of $\Gamma X$) for zigzag HSs, and for the armchair at the $\Gamma$-point.

\begin{figure}[tbph]
\centering
\includegraphics[width=0.45\textwidth]{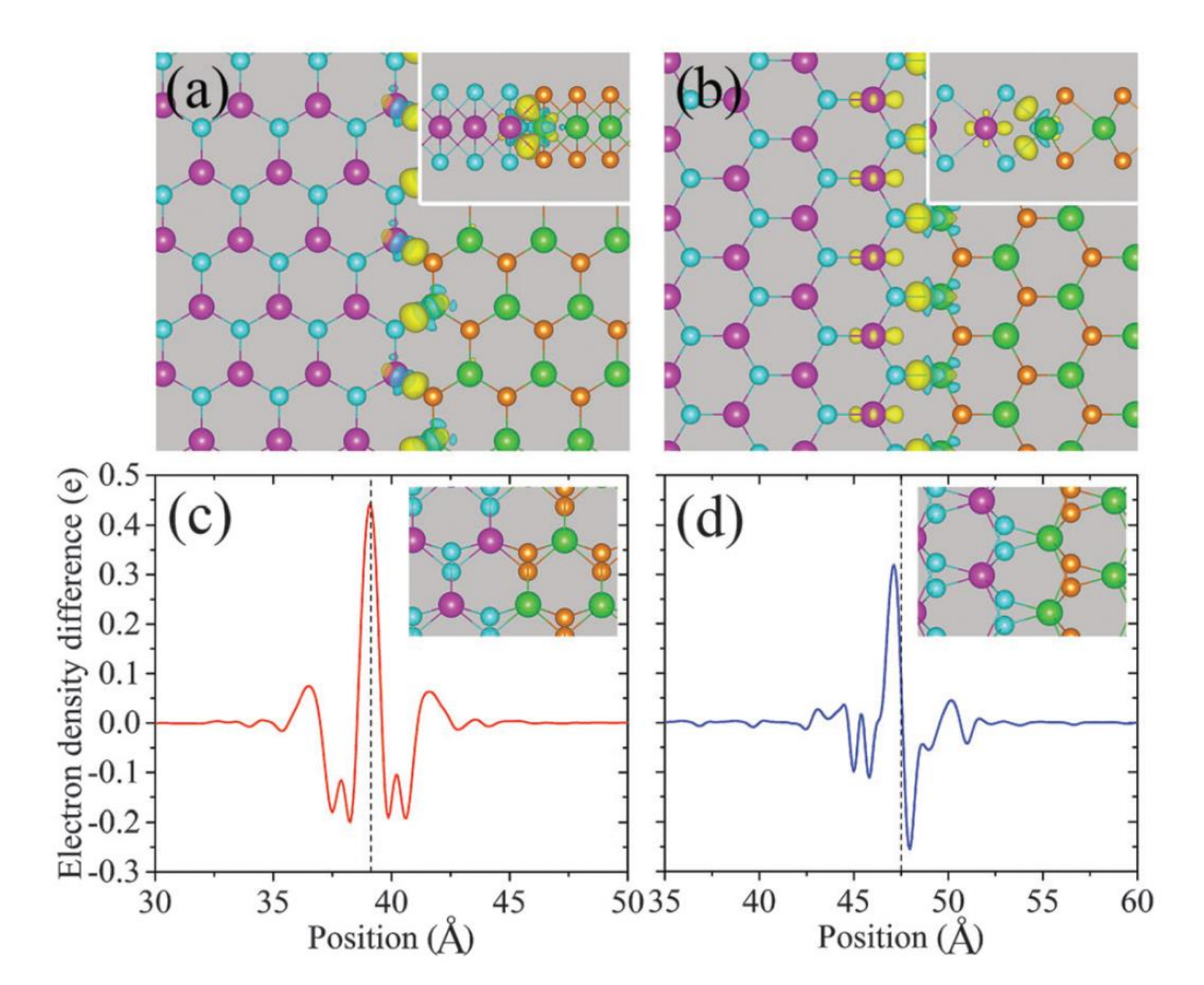}
\caption{For MoS$_{2}$-WS$_{2}$, (a) and (b) show the real-space charge density difference for armchair and zigzag HS, respectively. (c) and (d) show the plane averaged electron density for armchair and zigzag HS. In (a) and (b) yellow (cyan) regions are electron accumulation (depletion). Insets show connection geometries. Reprinted with permission from \cite{Wei2015}. Copyright 2015 by Royal Society of Chemistry, Physical chemistry chemical physics.}
\label{Wei2015PCCPFIGURE3}
\end{figure}

In lateral HSs, no van der Waals forces keep the materials together, rather Mo and W atoms near the interface form competing covalent bondings with the chalcogens. This can be seen for MoS$_{2}$-WS$_{2}$ in figure \ref{Wei2015PCCPFIGURE3} for both armchair and zigzag HSs. Covalent bonding changes can be seen at the interface as electron density probabilities in (a) and (b), and as large electron density difference in (c) and (d), boosting electrical and optical responses exactly at the 1D interface. Note that a net charge transfer does not occur for the armchair HS, where interfacial electrical polarization cancels across the junction, while a net charge accumulation occurs for the zigzag HS.

Analysis of the band-decomposed charge density of the VBM and CBM at the $\Gamma$ point of armchair HSs show localization of states in W and Mo atoms respectively, illustrating a true type-II HS, suggesting monolayer-like optical absorption in these HSs and strong excitonic effects with large binding energies. For zigzag MoS$_{2}$-WS$_{2}$ HS at the $\Gamma$ point, charge is located on opposite sides at the $S$-edges with slight overlap at the interface.

In-plane interfacing effects are further studied, describing charge transfer across the interface, work functions of the different edges connecting at the interface, and the role of defects \cite{Wei2016}. Quantum wells show similar behavior as those in the single interface case of Wei \emph{et al}. \cite{Wei2015}, and include
projections of the wave functions of each material. The VBM and CBM are located at opposite sides of the interface, in WS$_{2}$ and MoS$_{2}$, respectively.  The difference yields the alignment offsets, which for VBM is 0.1 eV and for the CBM is 0.3 eV, different than the core-level alignment values of 0.27 eV and 0.23 eV, respectively. This 0.1 eV offset for the VBM is in agreement with experiments that measure 0.07 eV \cite{Gong2014NatMat}. The HS band structure shows a direct bandgap at the $K$ valley projection. The gaps in quantum wells are found to be lower
than in single-interface HSs, and lower than in the pristine monolayer TMD, for different well widths.
The HS shows type-II band alignment at the interface, with binding energies, $E_b=E_{\mathrm{HS}}-E_{\mathrm{MoS}_2}-E_{\mathrm{WS}_2} \simeq -18$ eV for different well widths, owing to the  metal-S strong covalent bonds formed at the interface. Variations of $E_b$ for large wells are $\simeq 0.02$ eV,
suggesting that the interaction between interfaces is well-screened out for the considered widths.

\begin{figure}[tbph]
\centering
\includegraphics[width=0.4\textwidth]{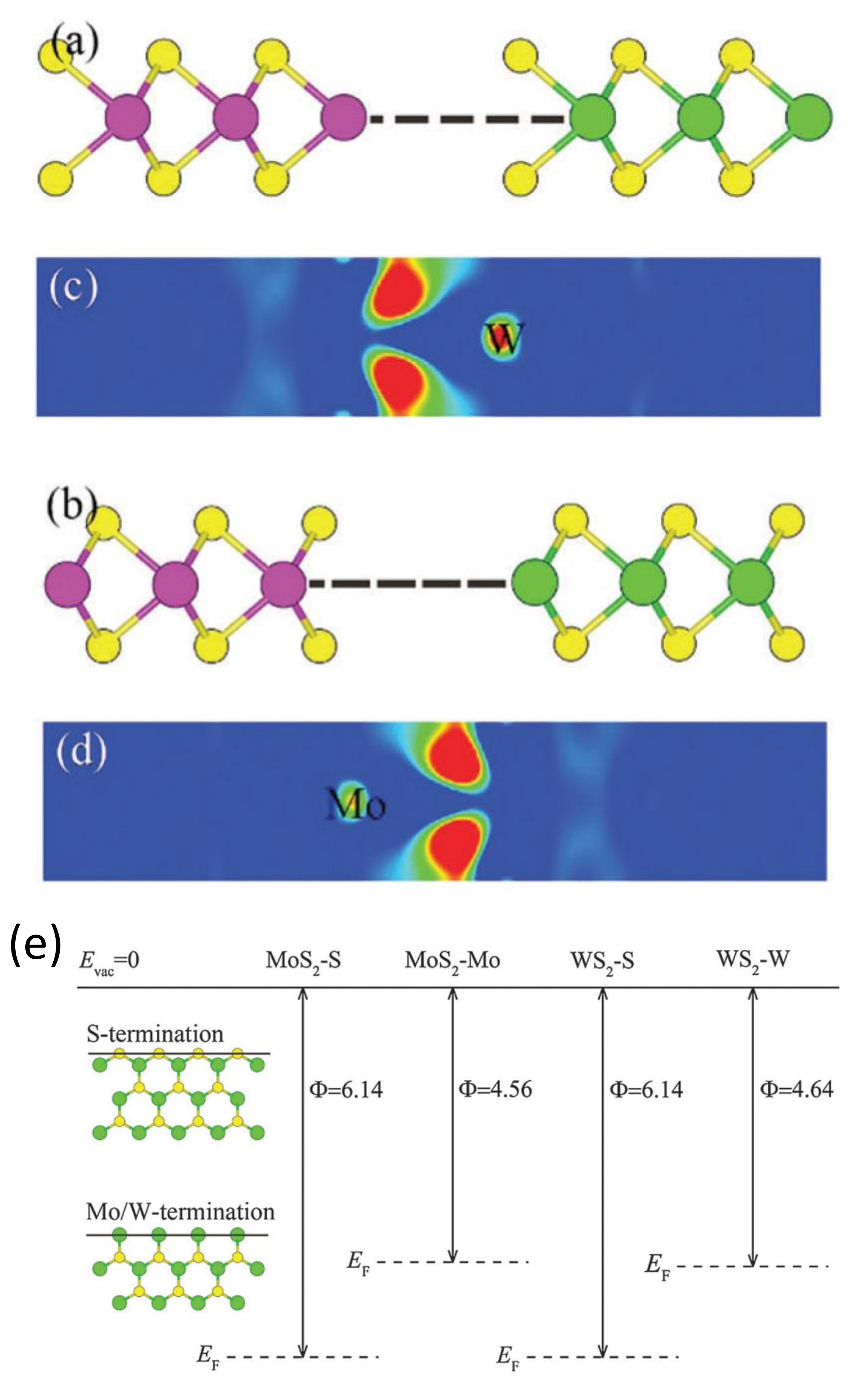}
\caption{For MoS$_{2}$-WS$_{2}$ zigzag interface, (a) and (b) show two in-plane connection patterns between MoS$_{2}$ (purple) and WS$_{2}$ (green), with sulfur (yellow balls). In (a) the Mo-edge connects to the S-edge of the W ribbon, and in (b) it is reversed. (c) and (d) charge density difference for (a) and (b), respectively. (e) Work function $\Phi$ for the four possible zigzag terminations at the interface, as shown. Reprinted with permission from \cite{Wei2016}. Copyright 2016 by Royal Society of Chemistry, Physical chemistry chemical physics.}
\label{Wei2016PCCPFIGURE45}
\end{figure}

\begin{figure}[tbph]
\centering
\includegraphics[width=0.5\textwidth]{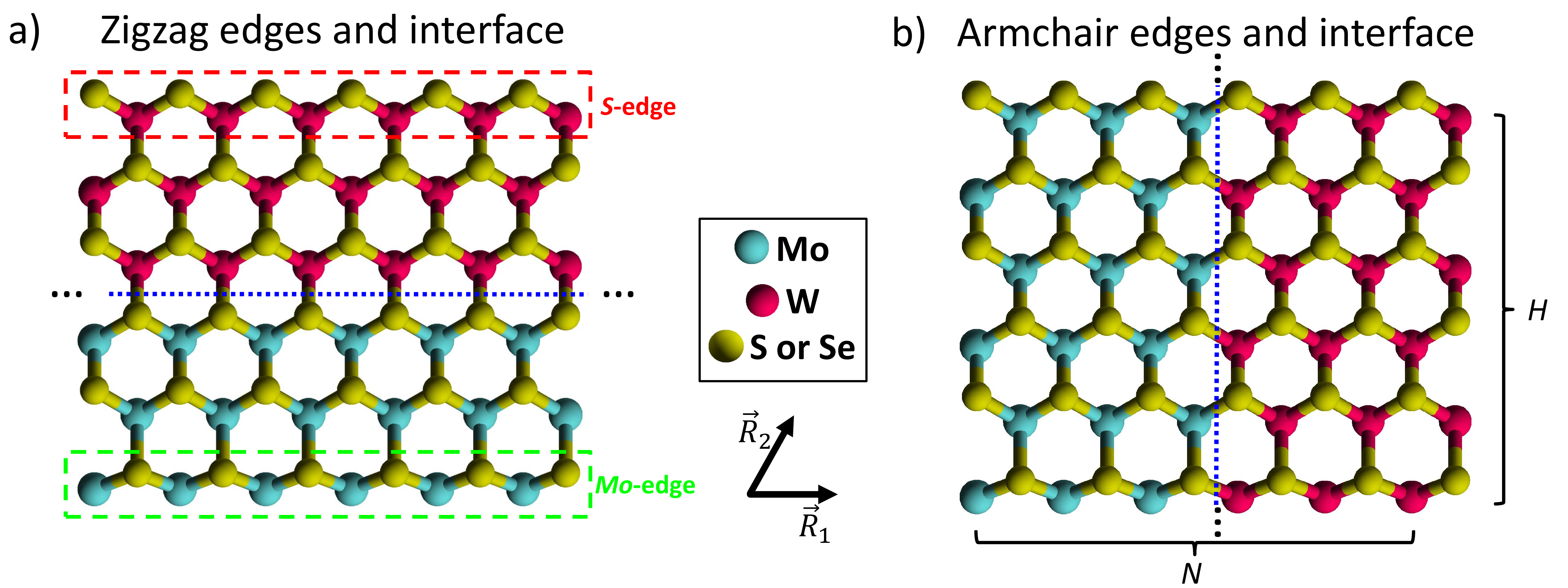}
\caption{Heteroribbons with edges and interfaces for (a) zigzag, and (b) armchair configurations. Metals Mo and W are shown in aqua and red colors, respectively. Chalcogens S or Se are shown in dark yellow. The zigzag (or armchair) heteroribbon is finite along the vertical (horizontal) direction, while periodic boundary conditions are used in the other direction, as indicated by the triple black dots. The interface is shown as a blue dotted line. The zigzag ribbon in (a) has two different edges, the $S$-edge (outermost-atom is a chalcogen) and the $M$-edge (outermost-atom is a transition metal). Reprinted with permission from \cite{AvalosOvando2018Arxiv}. Copyright 2019 by the American Physical Society, Physical Review B.}
\label{AVALOS2018Fig1}
\end{figure}

Most importantly, Wei {\em et al}.\ consider different hybridization geometries at the interface. It is known
that zigzag TMD  terminations can be either chalcogen-- or transition-metal--terminated (see figure\ \ref{AVALOS2018Fig1}), with the S-termination being more stable. Hence, zigzag interfaces can have two patterns: i) Mo-edge with the S-edge of the W ribbon, or ii) W-edge with the S-edge of the Mo ribbon, as shown schematically in figures\ \ref{Wei2016PCCPFIGURE45}(a)--(b).
The hybridization and charge transfer occurring at the interface creates a built-in electric field that leads to a potential gradient  across the interface, and seen in charge density difference maps in figures \ref{Wei2016PCCPFIGURE45}(c)--(d).
The built-in potential is expected to change local work functions; on the S-edge they have the same value for both TMDs, while on the M-edge they are only 0.08 eV apart, as shown schematically in figure \ref{Wei2016PCCPFIGURE45}(e). The charge transfer between MoS$_{2}$ and WS$_{2}$ is hence attributed to the larger work function at the zigzag edges with S-termination. Different zigzag interfacial connections patterns considered in \cite{Wei2016}, have been seen in experiments and are listed in table\ \ref{tab:table1}.

As for defects, chalcogen vacancies are the most recurrent defects in monolayer TMDs. Wei \emph{et al.} \cite{Wei2016} studied S vacancies at the interface and at the two closest S-lines on both the Mo and W sides, in a MoS$_{2}$-WS$_{2}$ HSs. These defects cause localized in-gap states that evolve into overlapping bands in these short period unit cells, and appear below the bottom of the conduction band, contributed mostly by metallic \emph{d}-orbitals. The states closest to the conduction band are contributed by the S-vacancies on the WS$_2$, while the ones lower in energy are on the MoS$_2$ side. When the vacancies are exactly at the interface, both Mo and W sharing the S vacancies are not saturated, creating a two-fold band structure with a gap, where the lower (higher) band is linked to the Mo (W) atom.


\emph{Tight-binding structure of commensurate TMD HSs.- } Other approaches have also been used to describe the electronic structure of HSs. Tight-binding approaches have been among the most common, for either commensurate \cite{Zhang2016SciRep,AvalosOvando2018Arxiv} or incommensurate HSs \cite{Choukroun2018Arxiv}. In this approach, it also becomes straightforward to model vacancies, adatoms and other local defects.

The successful 3-orbital tight-binding (3OTB) model \cite{Liu2013} allows one to build commensurate lateral HS nanoribbons with realistic sharp interfaces, as those seen in experiments \cite{Huang2014NatMat,Sahoo2018Nature}. Different boundary geometries of edges and interfaces (either zigzag or armchair), with periodic boundary conditions (PBC) along the ribbon can be modeled. The NR can be described by a triangular lattice of metal atoms and associated chalcogens, with only three 4\emph{d}-orbitals per metal site. This model exploits the fact that the near-gap (low energy) level structure in TMDs is dominated by the metal 4\emph{d}-orbitals with nearly no contribution from the chalcogen \emph{p}-orbitals \cite{Liu2013}. Other multi-orbital tight-binding models use larger basis sets \cite{Cappelluti2013,Roldan2014,Ridolfi2015,Fang2018}.  These more computationally expensive but powerful formulations validate much of the results seen in the midgap range from the 3OTB approach.
The 3OTB model uses $d_{z^2}$, $d_{xy}$ and $d_{x^2-y^2}$ as basis, and for the HS is given by
\begin{equation}\label{heterolattice1}
  H_{\mathrm{3OTB}} = H^{A}_{\mathrm{pristine}} + H^{B}_{\mathrm{pristine}} + H_{\mathrm{interface}},
\end{equation}
where $H^{A(B)}_{\mathrm{pristine}}$ is the Hamiltonian for each of the two TMDs, and $H_{\mathrm{interface}}$ describes the hoppings at the interface between the two TMD lattices. For TMDs with the same chalcogen atoms, the lattice mismatch is less than 1\% (such as MoS$_{2}$-WS$_{2}$ and MoSe$_{2}$-WSe$_{2}$) \cite{Huang2014NatMat,Gong2014NatMat,Sahoo2018Nature}. This results in corresponding small strain, so that the interface is essentially only compositional. The tight-binding description simply connects the metal atoms across the interface. Differences in real space lattice sizes are translated into different monolayer Brillouin zones (BZ), although the difference is in the m\AA$^{-1}$ range and can be neglected without the necessity of introducing band folding. For each of the pristine TMD lattices (A and B), the 3OTB model is given by \cite{Liu2013}
\begin{equation}\label{lattice1}
  H_{\mathrm{pristine}}^{\mathrm{A(B)}} = H^{\mathrm{A(B)}}_{\mathrm{o}} + H^{\mathrm{A(B)}}_{\mathrm{t}} + H^{\mathrm{A(B)}}_{\mathrm{SOC}},
\end{equation}
where $H^{\mathrm{A(B)}}_{\mathrm{o}}$ is the onsite Hamiltonian and $H^{\mathrm{A(B)}}_{\mathrm{t}}$ has the hopping integrals. $H_{\mathrm o}$ is given by
\begin{equation}\label{lattice2}
  H^{\mathrm{A(B)}}_{\mathrm{o}} = \sum_{ \textbf{l}}^{N_{sites}} \sum_{s=\uparrow,\downarrow}^{\mathrm{spin}} \sum_{\alpha,\alpha'}^{\mathrm{orbitals}} \varepsilon^{\mathrm{A(B)}}_{\alpha\alpha',s}d_{\alpha,\textbf{l},s}^{\dagger\mathrm{A(B)}}d^{\mathrm{A(B)}}_{\alpha',\textbf{l},s},
\end{equation}
where $d^{\mathrm{A(B)}}_{\alpha,\textbf{l},s}$ ($d^{\dagger\mathrm{A(B)}}_{\alpha,\textbf{l},s}$) annihilates (creates) a spin-$s$ electron in orbital $\alpha$, $\in\,\left\{d_{z^2},d_{xy},d_{x^2-y^2}\right\}$ at site $\textbf{l}=l_{1}\textbf{R}_{1}+l_{2}\textbf{R}_{2}$, where \textbf{R}$_{j}$ are the lattice vectors of the triangular lattice for each material, and the onsite energies are given by $\varepsilon^{\mathrm{A(B)}}_{\alpha\alpha',s}$. For a rectangular ribbon, the total number of sites is $N_{sites}=N\times H$, as shown in figure\ \ref{AVALOS2018Fig1}. The nearest-neighbor coupling Hamiltonian is
\begin{equation}
H^{\mathrm{A(B)}}_{\mathrm{t}} = \sum_{\textbf{l,R}_j} \sum_{\alpha,\alpha',s}
t_{\alpha\alpha'}^{(\textbf{R}_{j})\mathrm{A(B)}}d_{\alpha,\textbf{l},s}^{\dagger\mathrm{A(B)}}d^{\mathrm{A(B)}}_{\alpha',\textbf{l}+\textbf{R}_{j},s}+
\mathrm{H.c.},\\
\end{equation}
with different hopping parameters $t_{\alpha\alpha'}^{(\textbf{R}_{j})\mathrm{A(B)}}$.

The SOC in each material is approximated by the metal onsite contributions, $H^{\mathrm{A(B)}}_{\mathrm{SOC}}=\lambda^{\mathrm{A(B)}} L_{z}S_{z}$, where $L_{z}$ and $S_{z}$ are the $z$-components of the orbital and spin operators, respectively, and $\lambda^{\mathrm{A(B)}}$ is the SOC strength for each material. This results in on-site orbital mixings, $\varepsilon_{d_{xy}d_{x^2-y^2},\uparrow}=\varepsilon_{d_{x^2-y^2}d_{xy},\downarrow}=i\lambda^{\mathrm{A(B)}} = -\varepsilon_{d_{xy}d_{x^2-y^2},\downarrow}=-\varepsilon_{d_{x^2-y^2}d_{xy},\uparrow}$, that reproduce well the spin-split valence bands in the 2D crystal and give rise to strong spin-valley locking \cite{Liu2013}.

The interface is also described by nearest neighbor hopping integrals, and needs to take into account two important issues: the band alignment (or offsets) between materials $V_{\mathrm{A-B}}$, and re-scaling of the hoppings across the interface. The band alignment is taken into account through relative shifts of the onsite terms, given by $\varepsilon^{\mathrm{B}'}_{\alpha\alpha',s}=\varepsilon^{\mathrm{B}}_{\alpha\alpha',s}+V_{\mathrm{A-B}}$. These offsets can be taken from DFT results \cite{Kang2013,Guo2016,OngunOzcelik2016}, resulting in either type-I or type-II band alignments, as described above.
The hopping integrals can be written as an arithmetic \cite{AvalosOvando2018Arxiv} or geometric average \cite{Zhang2016SciRep}, with no qualitative difference in results. With the arithmetic average, the Hamiltonian is
\begin{equation}\label{df}
  H_{t}^{\mathrm{A-B}}=\sum_{\textbf{$\gamma$,a}_j} \sum_{s,\alpha,\alpha'}
\delta\left[t_{\alpha\alpha'}^{(\textbf{a}_{j})\mathrm{A}}+t_{\alpha\alpha'}^{(\textbf{a}_{j})\mathrm{B}}\right]d_{\alpha,\textbf{$\gamma$},s}^{\dagger}d_{\alpha',\textbf{$\gamma$}+\textbf{a}_{j},s}
\end{equation}
($+$ H.c.), where $\gamma$ labels the atoms on both sides of the interface. The scaling factor $\delta$ describes the compositional symmetry as well as possible relaxation effects at the interface (it is found that  $\delta=0.1$ is a value consistent with experiments \cite{AvalosOvando2018Arxiv}). The geometric average, with similar results for the state localization at the interface, uses $\sqrt[]{t_{\alpha\alpha'}^{(\mathbf{a}_{j})\mathrm{A}} t_{\alpha\alpha'}^{(\mathbf{a}_{j})\mathrm{B}}}$ as the effective interface hoppings
 \cite{Zhang2016SciRep}.
Zhang \emph{et al.} also consider hopping reconstruction at the ribbon edges, resetting the hopping integrals only for atoms on the borders, since they are connected to fewer atoms than those in the interior. The values used for these edge hopping integrals are inversely proportional to the bond lengths \cite{Zhang2016SciRep}.

\begin{figure}[tbph]
\centering
\includegraphics[width=0.45\textwidth]{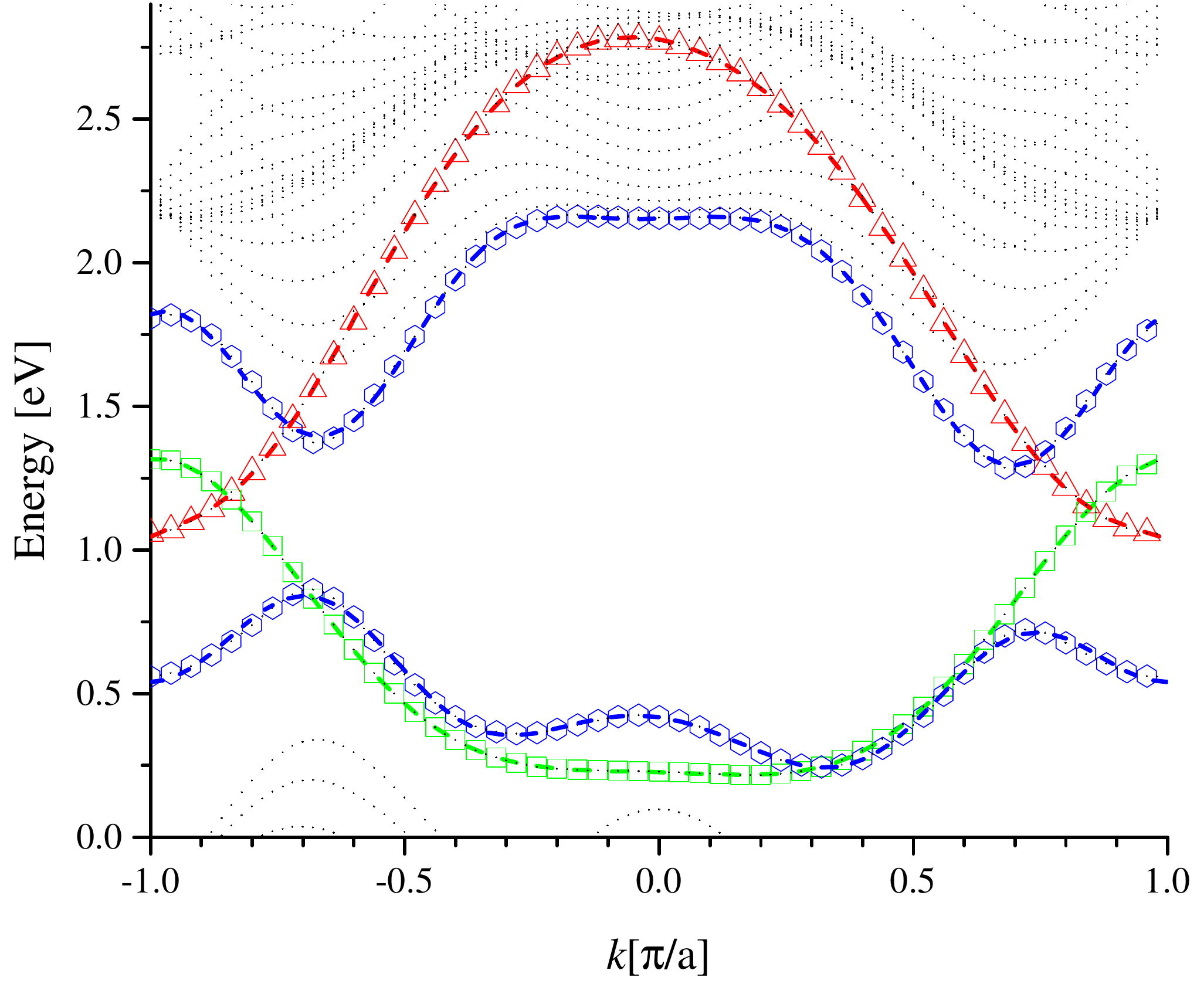}\\
\caption{Fitted bands for the zigzag MoS$_2$-WS$_2$ heteroribbon: The fits of \ref{Avalos20181EffectiveHamiltonianWithPauli} are shown as dashed lines, while symbols indicate the numerical 3OTB bands. We highlight interfacial zigzag bands (blue hexagons), as well as zigzag pristine Mo (green squares) and W (red triangles) edge bands, as shown in figure \ref{AVALOS2018Fig1}. Only spin up states are shown. Reprinted with permission from \cite{AvalosOvando2018Arxiv}. Copyright 2019 by the American Physical Society, Physical Review B.}
\label{AVALOS2018Fig3}
\end{figure}

The band structure of joined nanoribbons at a HS displays bands lying within the bulk gap. These midgap states are located at either the ribbon edges or at the interface of the system. For zigzag HS, all these states cross the gap, as shown in figure\ \ref{AVALOS2018Fig3}. One can identify two interfacial midgap bands, one closer to the conduction band and another to the valence band, with weight in all three orbitals ($d_{z^2,\mathrm{s}}$, $d_{xy,\mathrm{s}}$ and $d_{x^2-y^2,\mathrm{s}}$). The  hybridization across the two materials produces a gap and mixing between the interfacial branches.  This gap is proportional to the hybridization parameter $\delta$, as the chalcogens of one TMD hybridize with the metal atoms of the other TMD.

The interfacial zigzag states can be described analytically by \ref{Avalos20181EffectiveHamiltonianWithPauli}
\begin{eqnarray}\label{Avalos20181EffectiveHamiltonianWithPauli}
H^{\mathrm{interface}}_{\mathrm{eff}} &= \frac{(1-\sigma_{z})}{2} \sum_{n=0}^{\mathcal{N}}\left[t^{(n)}\cos (n k) + s_{z} \ t^{(n)}_{SO}\sin (n k)\right] + \nonumber\\
&\frac{(1+\sigma_{z})}{2}  \sum_{n=0}^{\mathcal{N}}\left[\gamma^{(n)}\cos (n k) + s_{z} \ \gamma^{(n)}_{SO}\sin (n k)\right],
\end{eqnarray}
where $\sigma_{z}$ is the Pauli matrix operating in a two function basis $\left\{ |\phi_{c}\rangle , |\phi_{v}\rangle \right\}$, and $s_{z}$ is the corresponding spin operator. The constants are the $n$th-nearest neighbor hoppings $t^{(n)}$ ($\gamma^{(n)}$) and
spin-orbit interaction $t^{(n)}_{SO}$ ($\gamma^{(n)}_{SO}$) for the lower (upper) interfacial band in the gap, respectively. These are obtained by numerical fitting to the 3OTB results, with excellent agreement, shown in figure \ref{AVALOS2018Fig3} \cite{AvalosOvando2018Arxiv}.

For armchair HSs, the electronic structure is fully semiconducting, with no states crossing the bulk bandgap.  The type-II alignment allows for easy differentiation of two interfacial bands in the gap, one per each material, but displaced to lower energy with respect to the pristine edge band. The interfacial gap also scales with $\delta$, as in the zigzag case, except that for small $\delta$ the gap does not close \cite{AvalosOvando2018Arxiv}.

The 3OTB model was first used for describing lateral MoS$_2$-WS$_2$ HSs by Zhang \emph{et al.} \cite{Zhang2016SciRep}, and transport quantities, as described in section \ref{subsubsec:Transport}. They built an HS of lateral alternating MoS$_2$ and WS$_2$ slabs, and consider different hopping strengths at the edges of the ribbon, to include reconstruction effects.
They find the HS has high-performance thermoelectric response, as the interfaces reduce the thermal conductivity.
Recently, the model was used for describing zigzag and armchair MoS$_2$-WS$_2$ and MoSe$_2$-WSe$_2$ interfaces, to show 1D confinement of states at the interface. Furthermore, it was shown that the interface can act as an unusual effective 1D-host when magnetic impurities are hybridized to it \cite{AvalosOvando2018Arxiv}. Driven by the complex spin and orbital texture of the interfacial states, anisotropic and sizable non-collinear (Dzyaloshinskii-Moriya) effective exchange interactions arising between the impurities. These and other behavior are discussed further in section\ \ref{subsec:1DNovelPlatform}.

\subsubsection{Incommensurability and strain}
\label{subsubsec:IncommensurabilityAndStrain}

The properties of commensurate TMD HSs described previously, can be strongly affected when strain is present. Usually, when TMD lateral HSs of different chalcogen atoms such as MoS$_{2}$-WSe$_{2}$ or MoS$_{2}$-MoTe$_{2}$ are formed, the lattice constant for the heavier chalcogen system is much larger, leading to sizable intrinsic strain at the interface. This mismatch has been experimentally measured to be as large as 4\% \cite{Duan2014NatNano,Xie2018ParkGroup}, introducing strain and requiring consideration of lattice relaxation effects. In the following, we review some of these effects.

\emph{Band alignment.- }
Guo \emph{et al.} \cite{Guo2016} address band alignment in a lateral HS between TMDs with different chalcogen, MoS$_{2}$-WSe$_{2}$, which has a 3.7\% lattice mismatch. This structure has type-II band alignment, with the states of WSe$_2$ lying higher in energy than MoS$_2$, and the charge neutrality point lying close to midgap. The comprehensive spin-polarized DFT study by \"Oz\ifmmode \mbox{\c{c}}\else \c{c}\fi{}elik \emph{et al.} \cite{OngunOzcelik2016} confirmed that in group-VIB TMD HSs the band alignment is mostly type-II, with only a few combination of TMDs being type-I, as strain is considered.

Early studies by Wang \emph{et al.}\ \cite{Wang2013} considered structural and electronic properties of non-commensurate MoS$_{2}$-MoTe$_{2}$ HSs, using Quantum Espresso, and PBE-GGA for exchange-correlation. Their HS is interesting to study because of the different bandgaps of each pristine system: MoS$_{2}$ has the largest and MoTe$_{2}$ the smallest. Due to the large difference in lattice constants (about 10\%), they consider a 10 MoS$_{2}$-9 MoTe$_{2}$ supercell. This HS shows metallic behavior, originating from atoms displaced at the interface. A similar study in MoS$_2$-WS$_2$ zigzag and armchair interfaces found strain-driven type-II to type-I band alignment transition when tensile strain is applied to the WS$_2$ side, as well as localized in-gap states in the presence of grain boundaries \cite{Kang2015}.

The MoS$_{2}$-WSe$_{2}$-MoS$_{2}$ quantum well HSs studies by Wei \emph{et al.} \cite{Wei2015SciRep} show that their electronic properties can be engineered by adjusting the strain.  This leads to different bandgaps and to an indirect-to-direct bandgap transition as the number of unit cells in each HS changes. Typical results for large HSs are shown in figure\ \ref{Wei2015PCCPFIGURE2}(c)--(d). The MoSe$_{2}$-WS$_{2}$ armchair HS is semiconducting with a direct gap in $\Gamma$, but the zigzag HS is indirect (at the VBM in $\Gamma$ and the CBM in $A$), the difference attributed to the intrinsic electric field across the interface and to lattice mismatch effects. For zigzag  MoSe$_{2}$-WS$_{2}$ HSs,
the projected band-decomposed charge density shows that both VBM and CBM are confined to the Mo-side, suggesting a type-I band alignment with a smaller gap than in pristine MoSe$_{2}$, due to the presence of the built-in dipole at the interface.

Defects such as S-vacancies at the interface on incommensurate MoS$_{2}$-WSe$_{2}$ interfaces have also been recently addressed, showing that even non-pristine interfaces show sharp electrostatic potential profile changes at the interface, as in commensurate HSs \cite{Cao2017}.

\begin{figure}[tbph]
\centering
\includegraphics[width=0.45\textwidth]{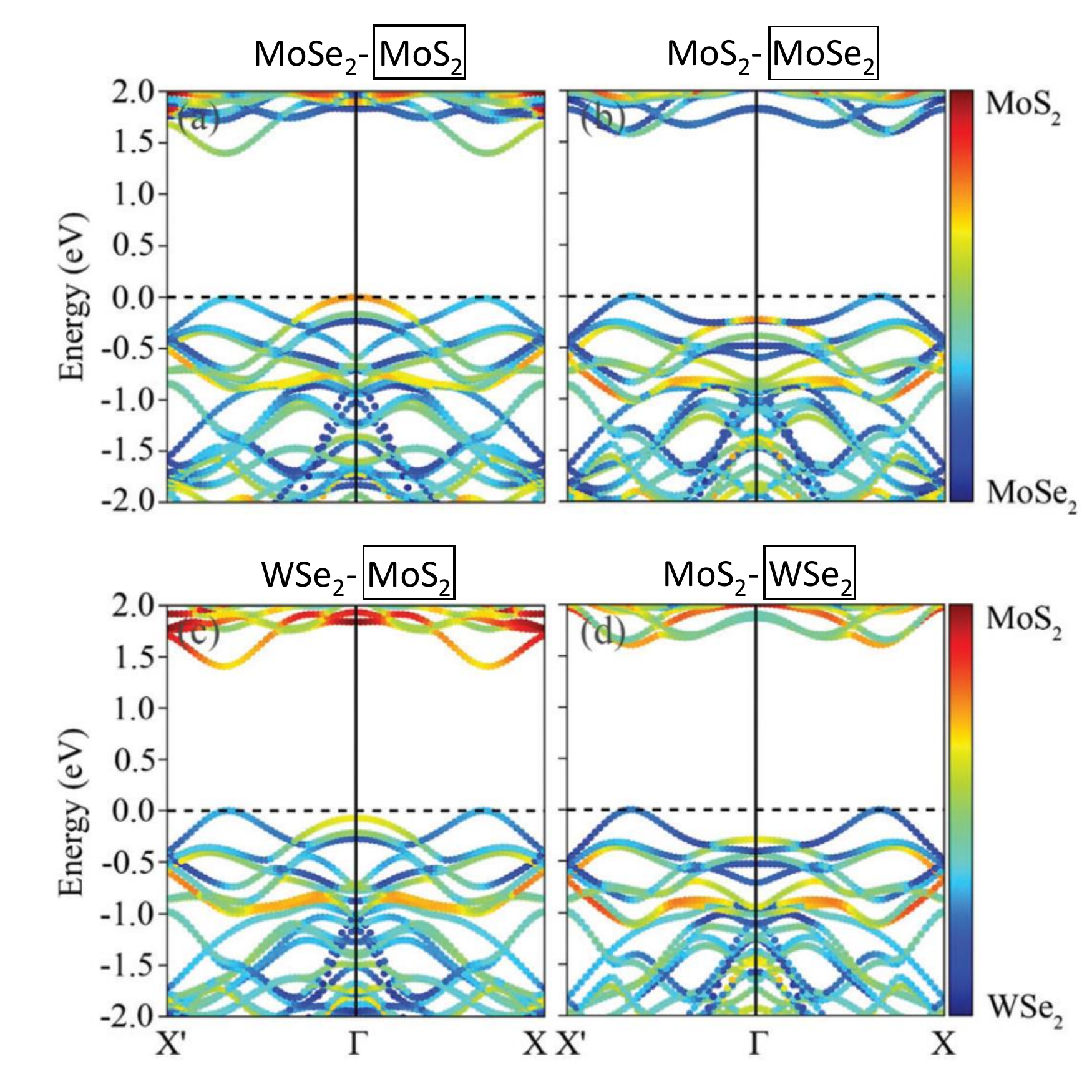}
\caption{Band structures for HSs as shown, with strain applied to the TMD inside the square while the other remains unaffected. Tensile (compressive) strain is applied to the Se (S) based TMD. Reprinted with permission from \cite{Wei2017}. Copyright 2017 by Royal Society of Chemistry, Physical chemistry chemical physics.}
\label{Wei2017PCCPFIGURE3}
\end{figure}

{\em Straintronics} in lateral TMD HSs has also been studied. Strain changes atom bondings, resulting in bandgap changes, and/or indirect-direct bandgap crossover \cite{Kang2015,Wei2017,Lee2017,Mu2018MRExpress}. Electronic effects driven by strain, such as band alignment transition under tensile strain from type-II to type-I in a MoS$_{2}$-WS$_{2}$ HS were characterized in four HSs \cite{Wei2017},
as depicted in figure\ \ref{Wei2017PCCPFIGURE3}.
The band structure calculations find that while cases (b), (c), and (d) in figure \ref{Wei2017PCCPFIGURE3} show direct bandgap, case (a) does not. The projections of the wave function contributions for each material do not change considerably after four atomic lines ($\sim$22 \AA),
signaling strong interfacial behavior. Contribution of states to the VBM and CBM in cases (a) and (b) show type-I alignment, while
cases (c) and (d) show clear type-II, with the VBM localized on WSe$_2$ and the CBM on MoS$_2$. The relative alignments are attributed in part to SOC effects, since the VBM at the K-valleys are shifted upwards, overcoming the tensile strain-induced shift of the VBM at the $\Gamma$-point. These results highlight that intrinsic strain at 1D interfaces gives rise to different electronic properties. Interestingly, systems with  simultaneous strain show direct bandgaps.

Strain can also affect the solar power conversion efficiency in commensurate quantum wells HS.  Lee \emph{et al.} found that type-II band alignment can be preserved with up to 12\% of uniaxial strain \cite{Lee2017}. Straintronics can also manifest in more exotic HSs, such as WS$_{2}$-WSe$_{2}$-MoS$_{2}$ quantum wells \cite{Mu2018MRExpress}. The bandgap can be continuously tuned changing the size of the central quantum well component. Lattice mismatch induces strain, direct-indirect bandgap transitions, and differences in band alignment. They used \emph{ab initio} molecular dynamics to verify thermodynamic stability of the interfaces, finding that room temperature does not break bonds, and that the hexagonal structure holds, supporting interface stability \cite{Mu2018MRExpress}. This finding is also reflected in the phonon dispersion curves, showing only branches with positive frequencies.
An electrostatic potential difference associated with the built-in electric field is seen at the interfaces. No sharp drop in the macroscopic average indicates also strong hybridization at the interfaces.

\emph{Tight binding.- }
Tight binding models have also recently addressed the effects of inconmensurability, studying WTe$_2$-MoS$_2$ and MoTe$_2$-MoS$_2$ HSs \cite{Choukroun2018Arxiv}, using an 11-orbital basis \cite{Cappelluti2013,Roldan2014,Ridolfi2015,Fang2018}. Choukroun \emph{et al.} used  this approach to model tunnel field effect transistors on in-plane heterojunctions, and studied quantum transport with NEGF. The original 11-orbital TB Hamiltonian doubles, as both TMDs  must be considered in the transport simulation cell. The model uses all five metal $d$-orbitals, as well as  the three $p$-orbitals for each of the chalcogen layers. The model describes first neighbor hoppings M-M, M-X, and X-X, and second neighbors X-X, and considers strain between the different TMD lattices. The coupling Hamiltonian between both TMD lattices is taken to be the arithmetic average between hoppings on both sides of the interface (see equation \ref{df}),
\begin{equation}\label{Choukroun2018ArxivHopping}
T_{n+1,m}^{A/B}=\left(T_{n+1,m}^{A}+T_{n+1,m}^{B}\right)/2,
\end{equation}
where A(B) are the TMDs on either side of the interface. This is analogous to the approach in \cite{AvalosOvando2018Arxiv}.

\emph{Strain tensor.- }
Recently, the 2D strain tensor ${\epsilon}$ in lateral WSe$_2$-MoS$_2$ HSs has been characterized as \cite{Zhang2018strain}
\begin{equation}
{\epsilon} = \left[
  \begin{array}{cc}
  \epsilon_{aa} & \epsilon_{ab}  \\
  \epsilon_{ba} & \epsilon_{bb}  \\
  \end{array}
\right],
\end{equation}
in terms of appropriately defined directions  $a$ and $b$. In a strainless case, such as an HS with the same chacolgen, vectors $\textbf{a}$ and $\textbf{b}$ can be defined in terms of a rectangular unit cell, where $\textbf{a}$ is parallel to the zigzag interface, and $\textbf{b}$ is along the perpendicular armchair direction. In the presence of shear strain on the MoS$_2$ side (smaller lattice constant), the unit cell is now a trapezoid, with moir\'e pattern spacings $\lambda_a$ and $\lambda_b$, along the $\textbf{a}$ and $\textbf{b}$ directions given by
\begin{eqnarray}
\lambda_a&=&a'_{\mathrm{Mo}}/\delta_a,\,\,\,\,\,\,\,\,\,\,\mathrm{with}\,\,\,\,\,\delta_a=|a_\mathrm{W}-a'_\mathrm{Mo}|/a_\mathrm{W}, \label{strain1} \nonumber \\
\lambda_b&=&b'_{\mathrm{Mo}}/\delta_b,\,\,\,\,\,\,\,\,\,\,\mathrm{with}\,\,\,\,\,\delta_b=|b_\mathrm{W}-b'_\mathrm{Mo}|/b_\mathrm{W}, \label{strain2}
\end{eqnarray} \label{straintensor1}
where no-prime (prime) values correspond to unstrained (strained) lattices, and $\delta$'s are lattice mismatches. See figure\ \ref{FigInterfaceStrain}(a)--(b) for the schematic representation of these quantities. The shear angle $\beta$ of the moir\'e pattern is related to the atomic lattice shear angle $\alpha$ by
\begin{equation}
\tan{\beta}=A_\beta\tan{\alpha},\,\,\,\,\,\,\,\,\,\,\mathrm{with}\,\,\,\,\,A_\beta=1/\delta_a. \\
\end{equation} \label{straintensor2}
This approach allows one to relate the moir\'e pattern spacing obtained with STM to the atomic lattice spacing, allowing the first one to act as a \emph{magnifying glass} with amplification factor $A_\beta$, inversely proportional to the mismatch: a tensile (compressive) strain in the Mo-side (W-side) will reduce (increase) the mismatch and will increase (reduce) the moir\'e pattern periodicity. Experimental data for $\lambda$'s should allow the determination of $\epsilon$: for a lateral WSe$_2$-MoS$_2$ interface, it is found that $\epsilon_{aa}=1.17\%$, $\epsilon_{bb}=-0.26\%$, and $\epsilon_{ab}=\epsilon_{ba}=0.69\%$ \cite{Zhang2018strain}. These parameters could be introduced into  tight-binding descriptions of atomic lattices to realistically account for strain distributions around interfaces.

\emph{Coarse-grained simulations.- }
Coherent WSe$_2$-WS$_2$ HSs have been recently grown \cite{Xie2018ParkGroup}, where the WS$_2$ (WSe$_2$) lattice constant is stretched (compressed) to achieve an integrated superlattice with almost-no-dislocations, as shown in figure\ \ref{FigInterfaceStrain}(c)--(e).
Lattice constant measurements along the directions parallel ($a_{\parallel}$) and perpendicular ($a_{\perp}$) to the interface, allowed estimates of the corresponding lattice mismatches $\delta_{\parallel}=0$ and $\delta_{\perp}=1.2\%$. A coarse-grained force-field model was used to model this system. The model needs to consider nearest-neighbor bonds and angular interactions to accurately reproduce experimental results. The energy of the HS is given by the sum of the harmonic bond and angular potentials,
\begin{equation}
E_{latt}=\frac{1}{2} \sum_{\rm bonds} k_b (r-r_0)^2 + \frac{1}{2} \sum_{\rm angles} k_{\theta} (\theta-\theta_0)^2.\\
\end{equation} \label{XiePark}

After an initial configuration of atoms is defined, following the scheme presented in figure\ \ref{XiePark}(a)-(b), $E_{latt}$ is minimized using second-order damped dynamics, until convergence is achieved. Atomic bonding is parameterized from 2D Young's moduli for WS$_2$ ($Y_{2D}=140$ N/m) and WSe$_2$ ($Y_{2D}=116$ N/m) DFT calculations, obtaining $k_b$'s and $r_0$'s. Although angular interactions are important, as they reflect the shear stiffness modulus, the moduli for TMDs is yet unknown.  However, it is found that $k_{\theta}=20$ rad$^{2}$ yields reasonable results. The simulations including angular coupling find $\delta_{\parallel} =0$ and $\delta_{\perp}=1.3\%$, as shown in figure\ \ref{XiePark}(c), in excellent agreement with the aforementioned experimental values.

\begin{figure}[tbph]
\centering
\includegraphics[width=0.45\textwidth]{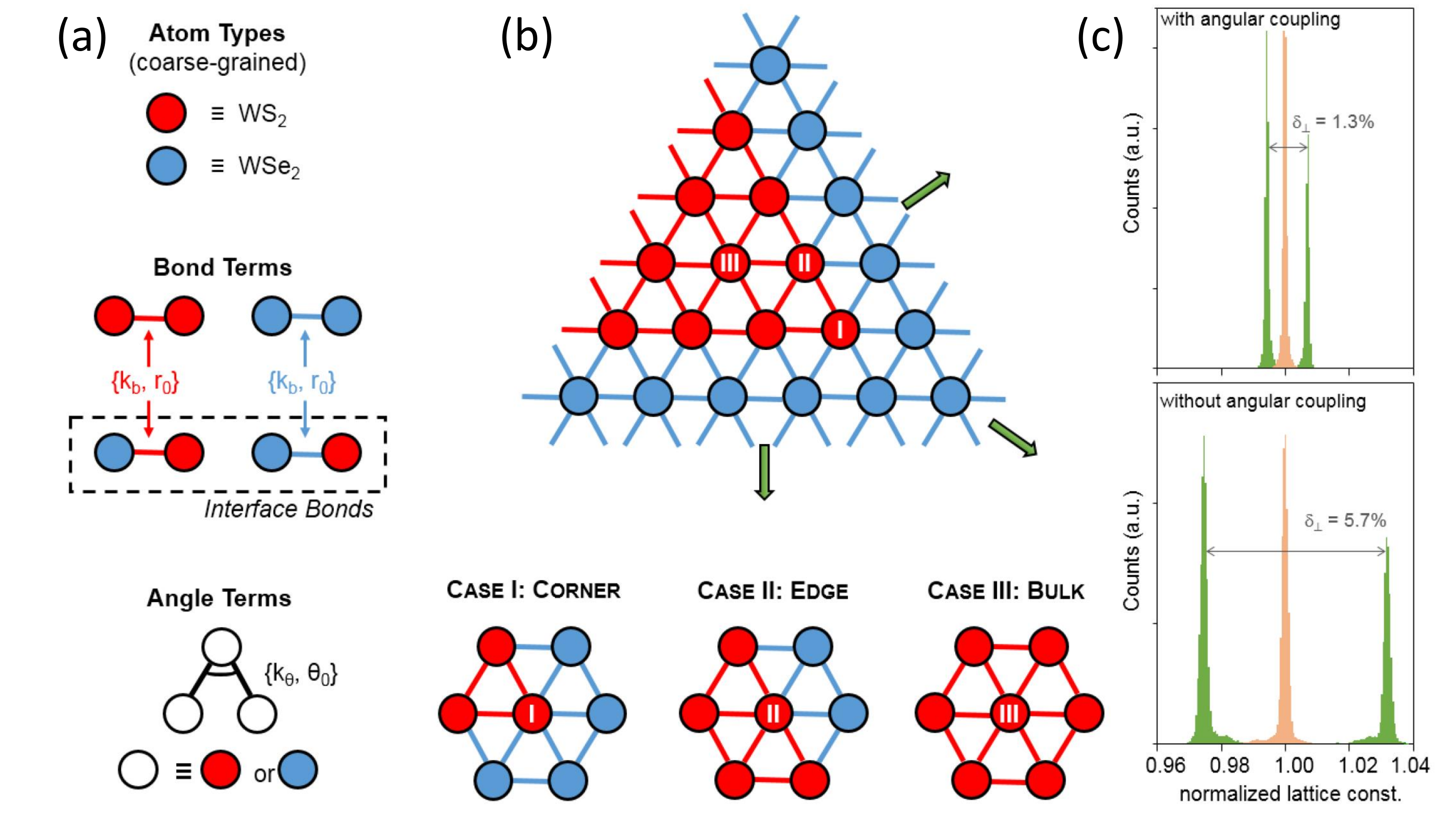}
\caption{Coarse-grained simulations for WS$_2$-WSe$_2$ coherent HS. (a) Atoms types (WS$_2$ in red, WSe$_2$ in blue), four possible bonding among atoms (growth directionality matters), and angular terms. $k_b$ and $r_0$ describe the bond force and equilibrium distance; similarly, $k_{\theta}$ and $\theta_0$ are angular force constant and equilibrium angle. (b) Coherent HS with its growth directions shown as green arrows,  labeling three possible cases for surrounding neighbors shown on the bottom. (c) Simulation results for $\delta_{\parallel}$ (orange) and $\delta_{\perp}$ (green). Reprinted with permission from \cite{Xie2018ParkGroup}. Copyright 2018 The American Association for the Advancement of Science, Science.}
\label{XiePark}
\end{figure}

\emph{Molecular dynamics simulations.- }
Strain effects have also been recently addressed with models based on classical potentials. Jiang \emph{et al.} \cite{Jiang2018misfit} studied the misfit strain-induced buckling for different interfaces in TMD lateral HSs, using molecular dynamics calculations. As more experiments are rapidly appearing, they highlight the need for theoretical methods (other than DFT) that are able to consider properties such as misfit strain, thermal transport, or sharpness of the interface, in systems with larger sizes. In their work they used 50,000 atom simulations and Stillinger-Weber (SW) potentials.  Calculating the strain distribution along the interface yields that misfit strain can induce significant buckling on various TMDs in patterns consistent with experiments.
The incommensurate lattices cause compressive stress in the TMD with largest lattice constant, and a buckling instability may occur.
The SW potential is a nonlinear potential given by two- ($V_2$) and three-body ($V_3$) interaction terms as
\begin{eqnarray}
V_2(r_{ij})&=&\epsilon A(B \sigma^p r_{ij}^{-p}-\sigma^q r_{ij}^{-q})e^{[\sigma(r_{ij}-a\sigma)^{-1}]}, \label{Jiang2018misfitSWpotentialEq1} \nonumber \\
V_3(\vec{r}_i,\vec{r}_j,\vec{r}_k)&=&\epsilon\lambda e^{[\lambda\sigma(r_{ij}-a\sigma)^{-1}+\lambda\sigma(r_{jk}-a\sigma)^{-1}]} \nonumber \\
&&\times (\cos\theta_{jik}-\cos\theta_0)^2, \label{Jiang2018misfitSWpotentialEq2}
\end{eqnarray} \label{Jiang2018misfitSWpotential}
where $r_{ij}$ is the distance between atoms $i$ and $j$, and $\theta_{jik}$ is the angle between the bonds $r_{ij}$ and $r_{jk}$, $\theta_0$ is the equilibrium angle, and the parameters are naturally TMD-dependent. The structure is first relaxed, then thermalized at 4.2 K, using  Large-scale Atomic/Molecular Massively Parallel Simulator (LAMMPS) \cite{LAMMPS}.  Several HSs were studied, including MoS$_2$-WSe$_2$, MoS$_2$-WTe$_2$, MoS$_2$-MoSe$_2$, and MoS$_2$-MoTe$_2$, all exhibiting strain distributions consistent with available experiments \cite{Zhang2018strain,Xie2018ParkGroup}. The TMD with smaller lattice constant shows only small tensile strain, and the edges of the sample not interfaced with another TMD also present compression due to bending of the interface. The TMD with larger lattice constant, however, shows significant compressive strain at the interface, and small tensile strain at the edges. The effect is also seen in triangular lateral heterostructures. They find that both strain, tensile and compressive, decay exponentially as $\propto e^{-x/\xi}$, with a critical length of $\xi  \simeq 15$ \AA.

\subsection{1D novel platform}
\label{subsec:1DNovelPlatform}

Experiments have shown enormous progress in achieving nearly-clean 1D interfaces between TMDs, and theoretical calculations have confirmed the remarkable stability and interesting electronic structure of lateral HSs.  An increasing number of experimental and theoretical efforts have started exploring effective uses for these lateral interfaces, in areas as diverse as optics, magnetism, and transport.

Section\ \ref{subsubsec:Optics} shows studies in optics, which have addressed excitonic effects around the interface \cite{Yang2017,Lau2018ArxivExcitons}, as well as wave guiding and spin-valley selection effects \cite{Ghadiri2018JAP}. A combined low-energy continuum description and tight binding approach, has found that the 1D HS interface exciton has similar binding energy as the 2D excitons in pristine monolayer TMDs, with somewhat larger effective radius.
This finding suggests effective optoelectronics applications involving 0D quantum dot confinement of excitons \cite{Lau2018ArxivExcitons}, associated with the formation of new W-S chemical bonds that favor exciton recombination \cite{Yang2017}.

The interface has moreover been recently proposed to serve as 1D-like host for long-range non-collinear magnetic interactions when magnetic impurities are hybridized to the interface, finding large tunability and stable conditions for the interaction to occur \cite{AvalosOvando2018Arxiv}. This is further described in section\ \ref{subsubsec:MagneticProperties}.

Transport effects have been studied with DFT, tight binding \cite{Zhang2016SciRep,Choukroun2018Arxiv}, and effective mass approximation \cite{Mishra2018oneDimensional} approaches. A 3-orbital tight binding model has been used to describe MoS$_2$-WS$_2$ and found to exhibit efficient thermoelectric characteristics, depending on the number and width of lateral HSs segments \cite{Zhang2016SciRep}. The 11-orbital TB model has been used for modelling MoTe$_2$-MoS$_2$, and found to be a possible system to implement high performance tunnel effect transistors \cite{Choukroun2018Arxiv}. The effective mass approximation which describes electrons at the K-points has been used to study transport properties, finding a one-dimension spin polarized channel at the interface \cite{Mishra2018oneDimensional}.
Similarly, theoretical transport studies have found that HSs can be used as gateless electron waveguides and spin valley filters/splitters \cite{Ghadiri2018JAP}.

In \ref{subsubsec:PhaseInterfacesTHEORY} we review studies of atomically clean interfaces between phases of the same TMD. Finally, in \ref{subsubsec:WithOtherMaterials} we briefly summarize lateral HSs proposed between group VIB semiconducting TMDs and metallic TMDs, and posible uses for them.

\subsubsection{Optical effects}
\label{subsubsec:Optics}

The study of excitons (bound states of an electron and hole) in a TMD, has been a topic of great interest from the outset of TMD monolayer studies, as exciton properties are essential for determining optical response. The attention has focused on pristine monolayers and vertical heterostructures.  In the latter, the exciton may be spatially separated, and with lower binding energy than in pristine TMDs, providing long exciton lifetimes and tunability. More recently, however, lateral HSs excitons have been seen in experiments \cite{Huang2014NatMat,Gong2014NatMat,Duan2014NatNano}, promising exciting properties by the inherent 1D interface of the planar HS \cite{Wei2015,Wei2015SciRep,Mu2018MRExpress,Yang2017,Lau2018ArxivExcitons}.

Early DFT studies suggested excitonic localization on either side of the interface, based on the projected density of states of the band structure, and reflecting the associated type-II band alignment. This alignment would allow hole and electron to be located on different sides of the interface, favoring the selective formation of the exciton right at the interface \cite{Wei2015,Wei2015SciRep,Mu2018MRExpress,Yang2017}. A photoexcitation charge transfer study, using time-domain DFT along with nonadiabatic molecular dynamics, was carried out in lateral (and vertical) MoS$_{2}$-WS$_{2}$ HSs \cite{Yang2017}. They use VASP, with PBE-GGA, along with Grimme DFT-D3 for the molecular dynamics simulations. In the lateral HS case, an exciton-like state is seen to be localized at the interface due to Coulomb interaction, with an exciton recombination factor 3 times faster than in the vertical HS. The coupled electron-hole at the interface enhances electron-phonon coupling, due to the formation of new W-S chemical bonds.

Lau \emph{et al.} \cite{Lau2018ArxivExcitons} have recently theoretically studied excitonic states at the 1D armchair interface between two TMDs, with type-II band alignment. They considered one interface (WSe$_{2}$-MoSe$_{2}$), and two interfaces (WSe$_{2}$-MoSe$_{2}$-WSe$_{2}$), as well as a triangular MoSe$_2$ area enclosed by WSe$_2$, i.e., an heterotriangular quantum dot (QD) with surrounding interface. They analyzed the exciton binding energy $E_b$, effective radius $a_b$, optical dipole $D$ (related to exciton lifetime), and intervalley coupling strength $J$, using two approaches for solving the exciton problem. They find that the exciton radius increases with band offset, to be much larger than the 2D TMD exciton, while the binding energy does not decrease significantly.
The optical transition dipole decreases with band offset, up to one order of magnitude smaller than in pristine 2D TMD. Excitons in triangular QD structures show confinement of one carrier inside the QD, while the other remains close but in the second material, separated by the interface.  They find this effect is tunable, with optical selection rules depending on the QD size.

The exciton is studied in the effective mass approximation with
\begin{equation}
\label{Lau2018Hamiltonian1}
H=-\frac{\hbar^2}{2m_e}\nabla^{2}_{\textbf{r}_e}-\frac{\hbar^2}{2m_h}\nabla^{2}_{\textbf{r}_h}+V_C(|\textbf{r}_e-\textbf{r}_h|)+V_I(\textbf{r}_e,\textbf{r}_h),
\end{equation}
where $m_e$ $(m_h)$ are the electron (hole) effective masses, $\textbf{r}_e$ $(\textbf{r}_h)$ their real space positions, and $V_I(\textbf{r}_e,\textbf{r}_h)$ is the interface potential defined as $V_I(\textbf{r}_e,\textbf{r}_h)=V_e(\textbf{r}_e)+V_h(\textbf{r}_h)$, with contributions of $V_e$ $(V_h)$ for the electron (hole) lattice potentials, that include band offsets at the interface. The electron-hole Coulomb interaction $V_C(|\textbf{r}_e-\textbf{r}_h|)$ is given by the Keldysh 2D potential \cite{Keldysh1979}
\begin{equation}
\label{Lau2018Coulomb2}
V_C(\textbf{r})=-\frac{e^2\pi}{2r_0}\left(H_0\left(\frac{r}{r_0}\right)-Y_0\left(\frac{r}{r_0}\right)\right),
\end{equation}
with $H_0$ and $Y_0$ the Struve and Bessel functions of the second kind, respectively.
The interface potential favors the electron an hole staying on opposite sides of the interface, while the attractive Coulomb interaction opposes that effect.
\begin{figure}[tbph]
\centering
\includegraphics[width=0.5\textwidth]{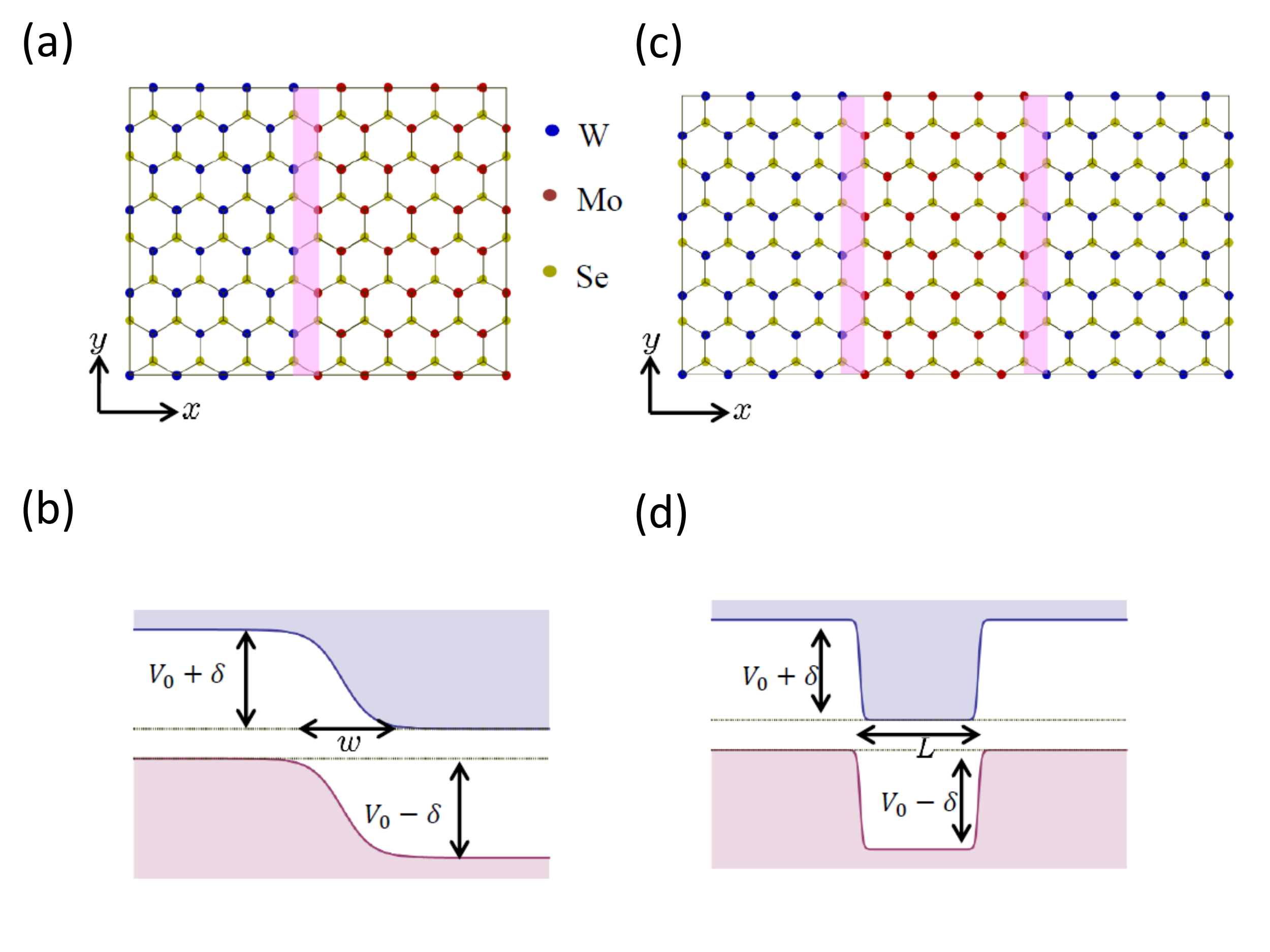}
\caption{Atomically sharp armchair interface (in purple) of lateral HS between (a) WS$_{2}$-MoS$_{2}$ and (c) WS$_{2}$-MoS$_{2}$-WS$_{2}$. (b) and (d) are \emph{p-n} and \emph{p-n-p} junctions, respectively. For HSs between  different TMDs, $V_0$ is the band offset, $\omega$ is the interface width (set to zero for atomically sharp interfaces), $\delta$ is the difference in band offset for electron and hole, and $L$ is central region width in (d). Reprinted with permission from \cite{Lau2018ArxivExcitons}. Copyright 2018 by the American Physical Society, Physical Review B.}
\label{Lau2018ArxivExcitonsFIGURE1}
\end{figure}
Equation \ref{Lau2018Hamiltonian1} can be rewritten in terms of center-of-mass and electron-hole pair relative motion. Assuming the Bohr-Oppenheimer approximation to be valid, the wave function is separable,
\begin{equation}
\label{Lau2018Hamiltonian3}
\Phi(\textbf{R},\textbf{r})=\Psi(\textbf{R})\Theta(\textbf{r}),
\end{equation}
with $\textbf{R}$ as the position of the center of mass $M=m_e+m_h$, and relative coordinate $\textbf{r}$ with reduced mass $\mu=m_e m_h/(m_e+m_h)$. The interface potential is modeled as

\begin{eqnarray}
V_e(x_e)&=&\frac{V_0+\delta}{2}\left(1-\tanh \left(\frac{x_e}{\omega}\right)\right), \label{electronpotential} \\
V_h(x_h)&=&-\frac{V_0-\delta}{2}\left(1-\tanh \left(\frac{x_h}{\omega}\right)\right), \label{holepotential}
\end{eqnarray} \label{interfacepotentials}
where $\omega$ is the interface width, characterizing the sharpness of the band offset $V_0$, as shown in figure\ \ref{Lau2018ArxivExcitonsFIGURE1}. The translational symmetry along $y$ allows the center-of-mass motion to be written as $\Psi(\textbf{R})=\Psi(x)e^{i p_y y}$, and the relative motion and center-of-mass equations, respectively as
\begin{equation}
\label{Lau2018Hamiltonian4}
\left[-\frac{\hbar^2}{2\mu}\nabla^{2}_{\textbf{r}}+V_C(r)+V_I(x,\textbf{r})\right]\Theta(x,\textbf{r})=E(x)\Theta(x,\textbf{r}),
\end{equation}
\begin{equation}
\label{Lau2018Hamiltonian5}
\left[-\frac{\hbar^2}{2M}\frac{\partial^2}{\partial x^2}+E(x)\right]\Psi(x)=E_g \Psi(x).
\end{equation}
In \ref{Lau2018Hamiltonian5}, $E_g$ is the ground state  for the type-II interface exciton. Equation \ref{Lau2018Hamiltonian4} is solved by: i) real-space tight-binding (for small supercells), or ii) perturbation expansion in a hydrogen-like basis (for larger systems).

\begin{figure*}[tbph]
\centering
\includegraphics[width=1.0\textwidth]{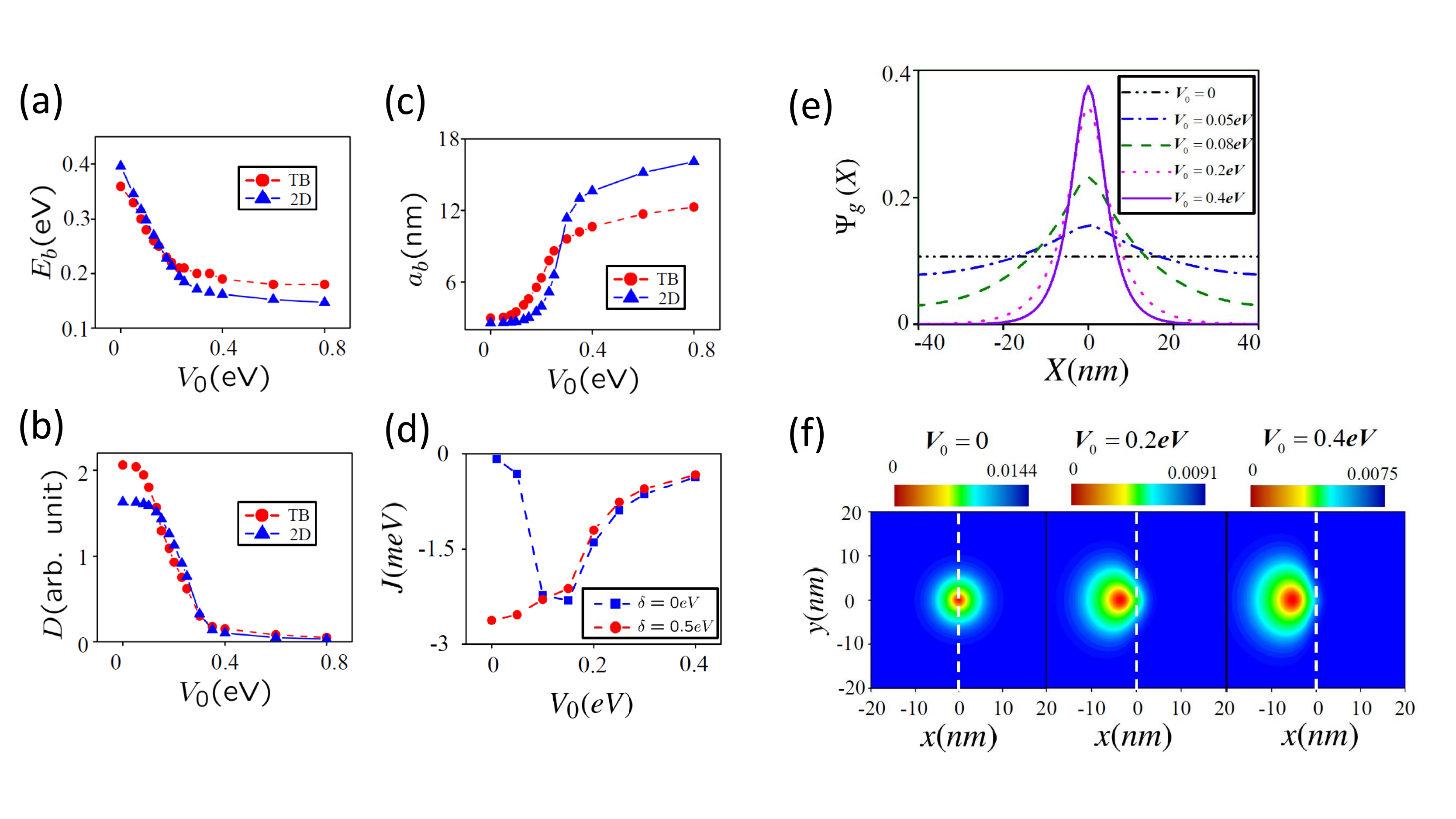}
\caption{Results for exciton at WS$_{2}$-MoS$_{2}$ single interface. (a) Binding energy $E_b$, (b) exciton radius $a_b$, and (c) optical dipole $D$ vs band offset $V_0$, obtained with tight-binding model (red symbols) and the continuum 2D hydrogenic basis model (blue symbols), and $\delta=0$. (d) Valley coupling $J$ vs band offset $V_0$, for symmetric (asymmetric) band offset $\delta=0$ eV  ($\delta=0.5$) in blue (red) symbols. (e) and (f) are wave functions in real space for the center of mass, and relative coordinate, respectively. The interface is at $x=0$, and results are for different band offsets $V_0$. Reprinted with permission from \cite{Lau2018ArxivExcitons}. Copyright 2018 by the American Physical Society, Physical Review B.}
\label{Lau2018ArxivExcitonsFIGURE75c3}
\end{figure*}

Typical results for interface exciton binding energy $E_b$, effective radius $a_b$, and optical dipole $D$ vs band offset $V_0$ are shown in figure\ \ref{Lau2018ArxivExcitonsFIGURE75c3}(a)--(c).
The results fall into two different regimes: small ($V_0<0.1$ eV), and large band offsets ($V_0>0.4$), driven by the competition between Coulomb interaction and interface potential. In the small-$V_0$ regime, $E_b$ is relatively large ($\approx0.35$ eV), $a_b$ small, and $D$ is large, meaning that $V_C$ dominates, leading to an exciton ground state with similar characteristics to the 2D exciton. On the other hand, for the large-$V_0$ regime, $E_b$ is relatively small ($\approx 0.2$ eV), $a_b$ large ($\approx 5$ nm), and $D$ is small, as $V_I$ dominates over $V_C$, yielding long lifetimes. Note that although smaller, $E_b=0.2$ eV, the binding energy is still of the same order of magnitude as for the 2D excitons. Lastly, the intervalley exchange $J$ vs band offset is studied for a symmetrical ($\delta=0$ eV) and asymmetrical ($\delta=0.5$ eV) band offset, as shown in figure\ \ref{Lau2018ArxivExcitonsFIGURE75c3}(d). For the symmetric case, three-fold symmetry with $V_0=0$ does not allow valley-mixing, but as $V_0$ increases $J$ reaches a maximum before decaying. For the asymmetric case, $\delta\neq0$ has already broken the symmetry and $J$ is already maximum, decreasing for larger band offsets. This suggests that the interface exciton has a $(|K\rangle\pm|K'\rangle)$ valley state, and it will couple with linearly polarized light instead of circular polarization. Results for larger supercells based on a continuum model lead to similar results, as shown in figure\ \ref{Lau2018ArxivExcitonsFIGURE75c3}(a)-(c) in blue symbols.

The ground-state solutions for \ref{Lau2018Hamiltonian4} and \ref{Lau2018Hamiltonian5}, $\Theta(x,\textbf{r})$ and $\Psi(x)$, are shown in figure\ \ref{Lau2018ArxivExcitonsFIGURE75c3}(e)-(f), for different band offsets $V_0$. The figure shows that $\Psi(x)$ is spread across the heterostructure width for small $V_0$, with 2D-like exciton behavior. This changes for large $V_0$, as the exciton center of mass is now located at the interface, indicating that electron and hole are separated on opposite sides. The contour maps in figure\ \ref{Lau2018ArxivExcitonsFIGURE75c3}(f), show that the extent of the relative coordinate function increases for larger $V_0$. Although for small $V_0$ the exciton behaves as in a 2D pristine TMD, for $V_0>0.2$ eV the exciton size is much larger, decreasing its overlap and enhancing its lifetime.

\begin{figure}[tbph]
\centering
\includegraphics[width=0.5\textwidth]{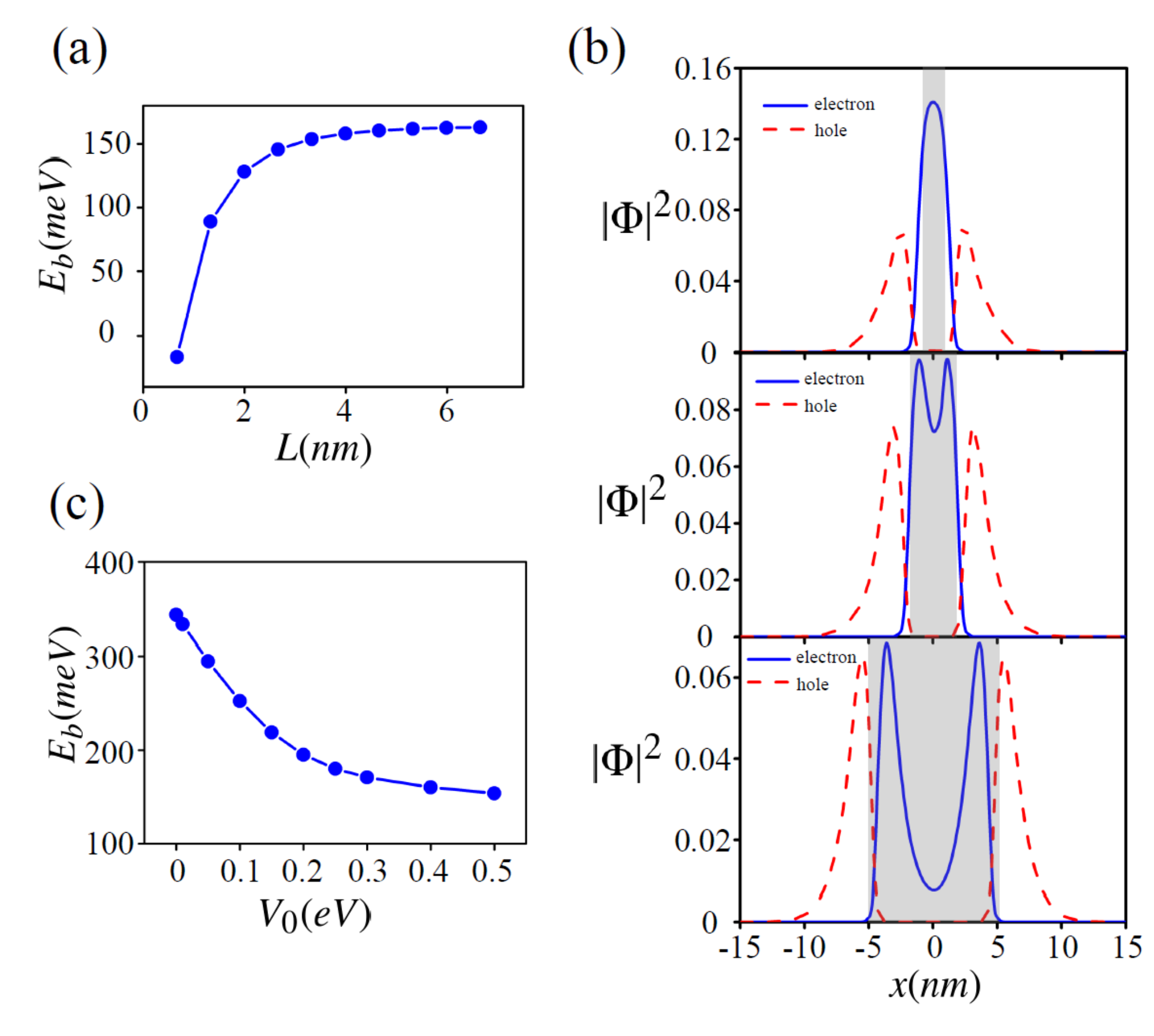}
\caption{(a) Binding energy $E_b$ vs different MoS$_{2}$ widths $L$ for WS$_2$-MoS$_2$-WS$_2$ structure. (b) Magnitude squared of the electron and hole wave functions for three different widths of MoS$_{2}$ ($L=1.3$, 3.3, 10 nm, respectively). MoS$_{2}$ region is shown in gray. (c) Binding energy $E_b$ vs band offset $V_0$, for $L=3.3$ nm. Reprinted with permission from \cite{Lau2018ArxivExcitons}. Copyright 2018 by the American Physical Society, Physical Review B.}
\label{Lau2018ArxivExcitonsFIGURE8}
\end{figure}

These authors consider also a double lateral interface, as in the system WS$_{2}$-MoS$_{2}$-WS$_{2}$, shown in figure\ \ref{Lau2018ArxivExcitonsFIGURE1}(c). Type-II band alignment dictates that the electron should mostly remain in the central MoS$_{2}$ region, while the hole would be in the outer WS$_{2}$ regions, to an extent depending on the competition between attracting Coulomb interaction, width of the central region $L$ and band offset $V_0$. For small $L$, the interface exciton has small binding energy due to the overlap across the central nanoribbon.  However, binding increases for larger $L$ until it saturates to the same value of the single interface case described before ($E_b \simeq 0.16$ eV at $a_b \simeq 5$ nm for $V_0=0.5$ eV), as shown figure\ \ref{Lau2018ArxivExcitonsFIGURE8}(a). This suggests that in WS$_{2}$-MoS$_{2}$-WS$_{2}$ HSs the excitons will be separated at each interface for $L> 5$ nm, as seen in figure\ \ref{Lau2018ArxivExcitonsFIGURE8}(b).

\begin{figure}[tbph]
\centering
\includegraphics[width=0.5\textwidth]{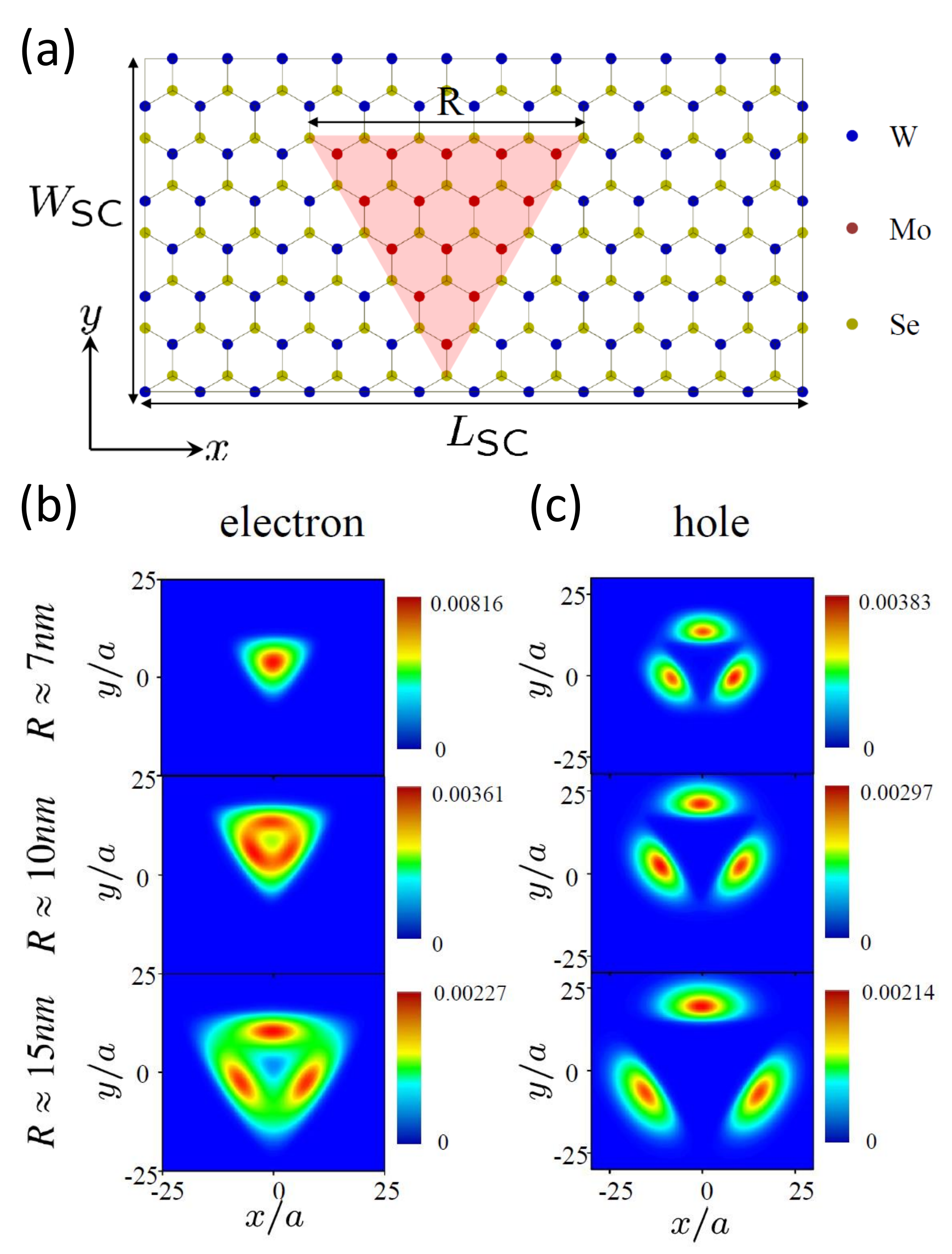}
\caption{Results of the excitons for triangular HS of WS$_{2}$-MoS$_{2}$-WS$_{2}$ triple zigzag interface, with a central triangular MoS$_{2}$ region, enclosed by WS$_{2}$. (a) Schematics of the heterostructure, with $R$ the MoS$_{2}$ triangle side. (b) electron and hole relative motion spatial wave function probability distribution vs the heterostructure real-space super cell dimensions, for $R=7,10,15$ nm, and $V_0=0.3$ eV. Reprinted with permission from \cite{Lau2018ArxivExcitons}. Copyright 2018 by the American Physical Society, Physical Review B.}
\label{Lau2018ArxivExcitonsFIGURE9and12}
\end{figure}

Lastly, excitons in triangular MoS$_{2}$ flakes, QDs enclosed by WS$_{2}$, are studied (see figure\ \ref{Lau2018ArxivExcitonsFIGURE9and12}).
A finite basis representation of n-electron/m-hole states is used to find binding energy and wave function of the interface exciton.
They find maximum exciton binding energy for an optimal flake size $R$, due to competitions of quantum confinement and Coulomb interaction terms, as shown in figure\ \ref{Lau2018ArxivExcitonsFIGURE9and12}(b)-(c). For the smallest central QD region size, the electron is spread over the entire QD, but for the largest size, one can see spatially separated wave functions at each interface. For sufficient large QDs, the wave functions are spatially separated, with a $E_b\approx 0.14$ eV, in agreement with the single interface calculations. They propose a valley-dependent effective model for three-fold symmetric QD systems with overlapping states as

\begin{equation}
\label{Lau2018Hamiltonian6}
    H^{\triangle\textrm{QD}}_{\textrm{eff,valley}}=
    \left(\begin{array}{ccc}
                    E_0 & te^{i\theta_{\tau}} & te^{-i\theta_{\tau}}\\
                    te^{-i\theta_{\tau}} & E_0 & te^{i\theta_{\tau}}\\
                    te^{i\theta_{\tau}} & te^{-i\theta_{\tau}} & E_0\\
                  \end{array}\right),
\end{equation}
with basis $\{|\Phi\rangle,C_3|\Phi\rangle,C_3^2|\Phi\rangle\}$. The wave function of a 1D interface exciton at one edge, $|\Phi\rangle$, is modified by $2\pi/3$ rotation operators $C_3$. $E_0$ is the exciton binding energy, and $te^{i\theta_{\pm1}}$ the transition between wave functions at different edges, with $\tau=\pm1$ indicating the valley index for $K$ and $K'$, respectively. For non-valley mixing, three energy states $E_j=E_0+2t\cos(2j\pi/3-\theta)$ are found, with twofold degeneracy. Numerical results show that the transition coefficient $t$ depends on the overlap of the quasi-1D excitonic wave functions at the corners of the triangular MoS$_2$ QD, with an interesting sign change: $t<0$ for small QD (large Coulomb interaction, large exciton overlap), and $t>0$ for large QD (small Coulomb interaction, small overlap). One of the excitonic states is bright, and the other two are dark under circularly polarized light excitation.

For the more general case of valley-mixing,
\begin{equation}
\label{Lau2018Hamiltonian7}
    H^{\triangle\textrm{QD}}_{\textrm{eff-intervalley}}=
    \left(\begin{array}{cc}
                    H^{\triangle\textrm{QD}}_{\textrm{eff,+1}} & H^{\triangle\textrm{QD}}_{\textrm{eff,pq}}\\
                    H^{\triangle\textrm{QD}}_{\textrm{eff,pq}} & H^{\triangle\textrm{QD}}_{\textrm{eff,-1}}
                  \end{array}\right),
\end{equation}
the intervalley mixing is reduced by symmetry to two independent matrix elements, and
Lau {\em et al.} provide estimates for them \cite{Lau2018ArxivExcitons}.

Other theoretical treatments of excitons in lateral HSs include Wei \emph{et al.} \cite{Wei2015SciRep}, who predict a coherent lattice and strong coupling at the interface with type-II alignment.  They suggest this as a possible mechanism for effective separation and collection of excitons at the HS. This expectation was confirmed by a detailed interfacial study that finds excitons pinned to the HS with carriers on opposite sides of the 1D interface \cite{Wei2015}. Other DFT studies have also agreed \cite{Yang2017,Mu2018MRExpress}. On the other hand, when defects are distributed along/near the interface, resulting in non-sharp junctions, a smooth electrostatic potential profile is expected, reducing the HS exciton localization and weak overlap properties \cite{Cao2017}.

\begin{figure}[tbph!]
\centering
\includegraphics[width=0.45\textwidth]{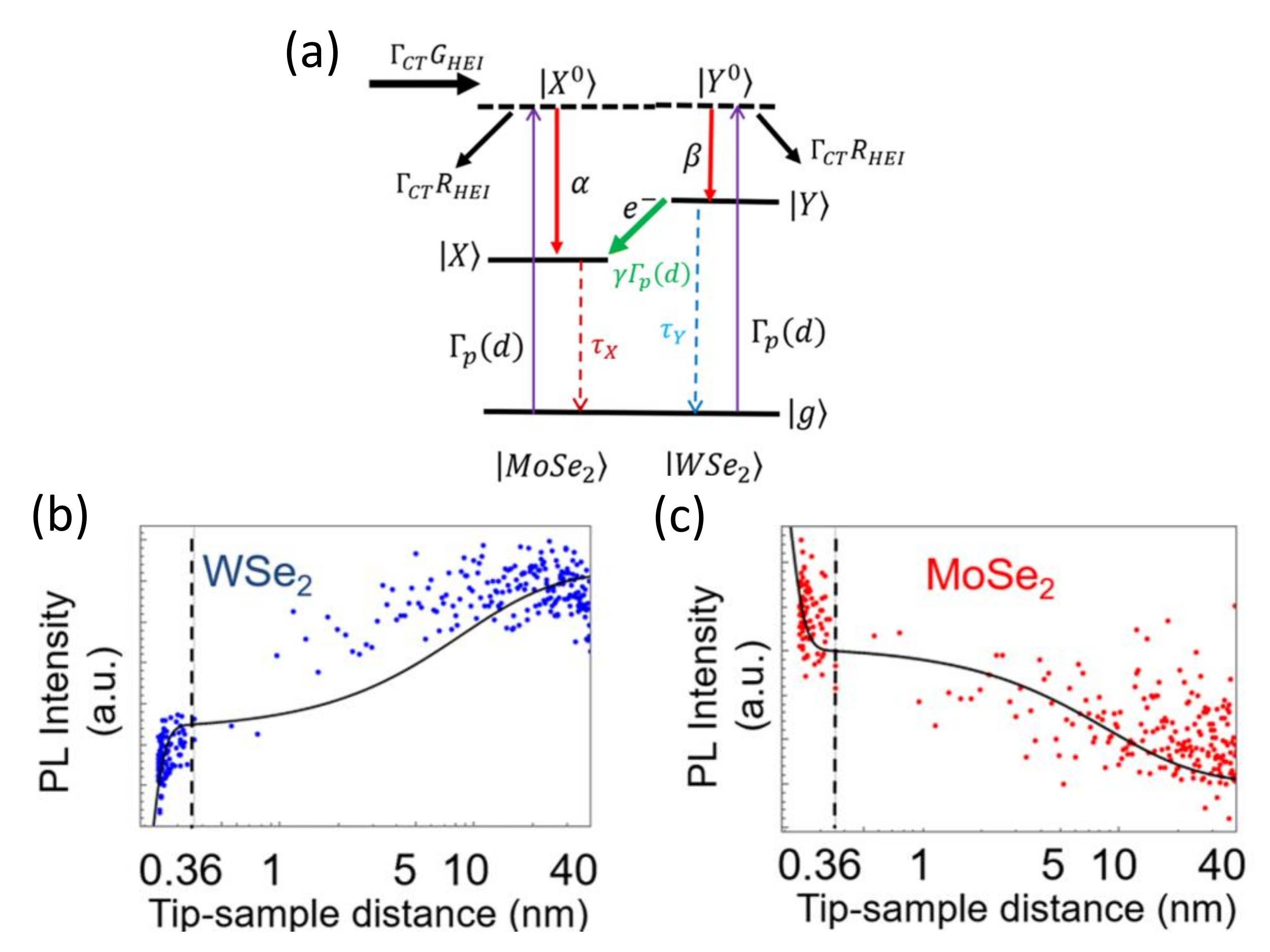}
\caption{(a) Schematics of WSe$_2$-MoSe$_2$ lateral HS energy diagram showing arrows for hot-electron injection (HEI) caused by the interface: HEI (black), plasmon-induced charge transfer (green), and tip-enhanced PL TEPL (purple). These lead to tunable quenching/enhancement of PL (dashed arrows). $\Gamma_{\mathrm{CT}}G_{\mathrm{HEI}}$ ($\Gamma_{\mathrm{CT}}R_{\mathrm{HEI}}$) is the hot-electron injection rate (HE decay rate) for states $|X^{0}\rangle$ and $|Y^{0}\rangle$. $\Gamma_{\mathrm{CT}}$ is the tunneling between tip and sample; $\Gamma_{p}$ the local optical excitation by the tip near field; $\alpha$ ($\beta$) the exciton generation rate of MoSe$_2$ (WSe$_2$); $\gamma\Gamma_{p}$ the photoinduced charger transfer across the interface; and $\tau_X$ ($\tau_{Y}$) the exciton relaxation time in MoSe$_2$ (WSe$_2$). (b) and (c) Experiment PL intensities (dots) at either side of the interface (WSe$_2$ and MoSe$_2$, respectively) vs tip-sample distance, with theoretical model (lines). $d \simeq 0.36$ nm signals the transition from the classical to the quantum tunneling regime. Reprinted with permission from \cite{Tang2018}. Copyright 2018 by the American Physical Society, Physical Review B.}
\label{Tang2018}
\end{figure}

We want to call attention to quantum plasmonic effects recently observed in WSe$_2$-MoSe$_2$ lateral HSs, by Tang \emph{et al.} \cite{Tang2018}, mentioned in section \ref{subsubsec:PlasmonicsEffectsEXPERIMENT}.
These authors suggested that hot electron injection enhances the PL signal of the MoSe$_2$ side, since there is an increased recombination rate, while hot electron injection in WSe$_2$ quenches its PL due to charge transfer across the interface. This competition is tunable with tip position away from the interface, and by the tip-sample distance. The model shown schematically in figure\ \ref{Tang2018}(a) accounts for the competition between hot electron injection (HEI) and the tip-enhanced PL.
Rate equations relate the initial excited state populations $N_{X0}$ and $N_{Y0}$ with that of exciton $|X\rangle$ in MoSe$_2$, and $|Y\rangle$ in WSe$_2$ \cite{Tang2018}.

The model fits well both the classical ($d>0.36$ nm) and the quantum ($d<0.36$ nm) regimes, as shown in figures\ \ref{Tang2018}(b)-(c). As the tip approaches the surface, the photoinduced charge transfer $\gamma\Gamma_{p}$ suppresses PL in WSe$_2$ and enhances it in MoSe$_2$, as seen in experiments.

\subsubsection{Magnetic interactions}
\label{subsubsec:MagneticProperties}

The tight binding models presented in \ref{subsubsec:ElectronicStructure} and \ref{subsubsec:IncommensurabilityAndStrain}, for commensurate and incommensurate lateral HSs, respectively, have one great advantage: they can reliably simulate large systems that may also include defects such as vacancies and/or adatoms.  Of particular interest are the effective interactions between magnetic adatoms hybridized at or near the lateral interface between MoS$_{2}$-WS$_{2}$ and MoSe$_{2}$-WSe$_{2}$ HSs. As the interfacial states are highly localized, the HS acts as a 1D effective host interaction between the magnetic impurities. This Ruderman-Kittel-Kasuya-Yosida (RKKY) indirect exchange interaction between magnetic impurities is expected to be drastically different, due to the strong spin-orbit coupling these materials, and the effective 1D dimensionality of the states at the HS. The impurities are modeled by an additional term in the Hamiltonian given by
\begin{equation}\label{impurities1}
  H_{\mathrm{I}}={\cal J} \sum_{i=1,2} \textbf{S}_{i}\cdot\textbf{s}_{\alpha_i}(\textbf{l}_i),
\end{equation}
which describes the local exchange coupling ${\cal J}$ between the impurity spin $\textbf{S}_{i}$ and electrons in orbital $\alpha_i$ at the location of the impurity $\textbf{l}_{i}$ at/near the interface \cite{AvalosOvando2018Arxiv}. $\textbf{s}_{\alpha_i}(\textbf{l}_i)$ is the electron spin density at the impurity location. After integration of the electronic degrees of freedom, one obtains the inter-impurity effective exchange interaction as
\begin{eqnarray}\label{jeffective1}
H_{RKKY} &=& J_{XX}\left(S_{1}^{x}S_{2}^{x}+S_{1}^{y}S_{2}^{y}\right)+J_{ZZ}S_{1}^{z}S_{2}^{z}\nonumber\\
 & &+J_{DM}\left(\textbf{S}_{1}\times \textbf{S}_{2}\right)_{z},
\end{eqnarray}
where  $J_{XX} = J_{YY}$ (in-plane), $J_{ZZ}$ (Ising), and $J_{DM}$ (Dzyaloshinskii-Moriya), as mediated by the TMD HS host are proportional to the static spin susceptibility tensor of the electron system \cite{RudermanKittel1954,Kasuya1956,Yosida1957,Imamura2004}. These $J$ parameters (jointly called $J_{\mathrm{eff}}$ for simplicity below) control the impurity interaction, and can be calculated by different approaches, including: i) consideration of the energy difference between triplet and singlet impurity configurations after diagonalization of the full Hamiltonian $H=H_{\rm HS} + H_{\rm I}$, and ii) second order perturbation theory \cite{AvalosOvando2018Arxiv}.

\begin{figure}[tbph]
\centering
\includegraphics[width=0.45\textwidth]{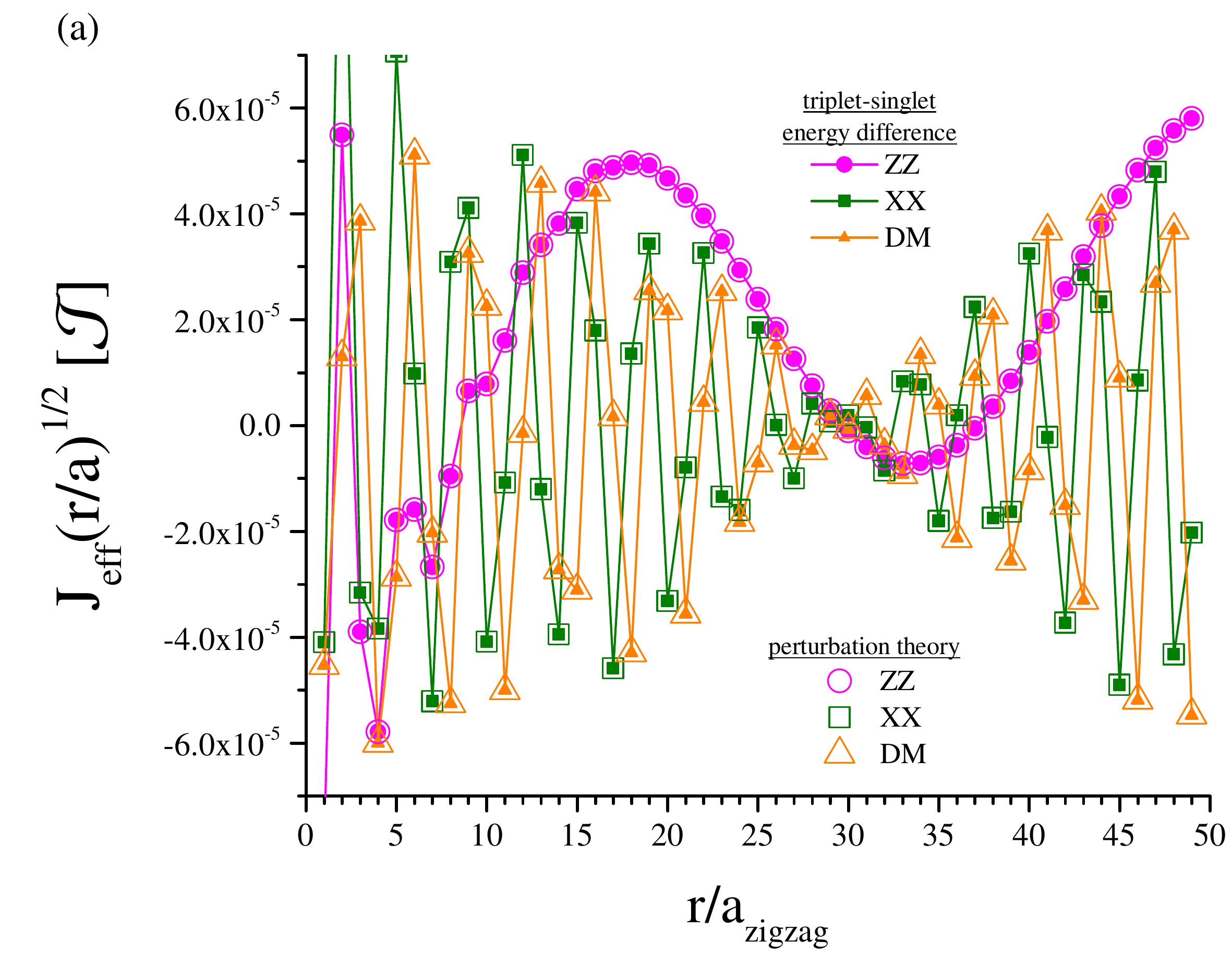}
\caption{RKKY interaction for impurities on a zigzag lateral HS interface vs impurity separation. The interaction is shown in units of a typical TMD-impurity hybridization magnitude ${\cal J} (=10$ meV), and scaled by the impurity separation $r^{1/2}$, normalized by the zigzag lattice constant. Full (empty) symbols indicate triplet-singlet energy difference (perturbation theory) results: magenta for Ising $J_{ZZ}$, dark green for $J_{XX}$ and orange for $J_{DM}$ terms. Reprinted with permission from \cite{AvalosOvando2018Arxiv}. Copyright 2019 by the American Physical Society, Physical Review B.}
\label{AVALOS2018Fig4a}
\end{figure}

The resulting set of $J_{\rm eff}$ exhibit long-range sub-1D behavior, as well as strong DM interactions.  This proves that the interface states behave indeed as unusual 1D hosts in the p-doped regime. This is illustrated in figure\ \ref{AVALOS2018Fig4a}, where typical RKKY interactions at a HS are shown vs separation $r/a$ between the two magnetic impurities; the impurities are hybridized at the interface and the Fermi level of the system is assumed to be in the bulk midgap. The various interactions are seen to oscillate with a decaying envelope, as one would expect. The oscillations in $J_{\rm eff}$ values describe how the lowest energy impurity alignment changes between ferromagnetic ($J_{\mathrm{eff}}<0$) and antiferromagnetic ($J_{\mathrm{eff}}>0$), depending on their separation $r/a$. More importantly, the interaction is seen to decay with distance as $J_{\mathrm{eff}}\propto1/r^{1/2}$, much slower that the expected $r^{-1}$ for a simple 1D system.  This suggests that the HS interface hosts a long range interaction between impurities with rather unusual features.
Results for armchair interfaces are similar in magnitude and decay envelope but exhibit different oscillation patterns \cite{AvalosOvando2018Arxiv}.

It would be interesting to explore magnetic ensembles with long range and helical magnetic interactions in lateral HSs.  Such systems are highly desirable in many different contexts in condensed matter.  For example,  transferring quantum information in spin chains \cite{Menzel2012} or studying the possible emergence of Majorana bound states when in proximity to superconductors \cite{Kim2014}.  Lateral interfaces between TMDs could provide a new platform for future studies with unique doping or gating tunability.

\subsubsection{Transport properties}
\label{subsubsec:Transport}

Heterostructures should exhibit electronic transport features absent in their pristine counterparts. In TMDs, some of these properties include spin filtering, rectification, and enhanced energy conversion in thermoelectric transport.

Electronic transport in commensurate MoS$_2$-WS$_2$ HSs has been studied using non-equilibrium Green functions (NEGF) \cite{zhou2015RSCAdv,zhou2016PCCP,an2016JMCC}, within the Atomistix ToolKit package \cite{atomistix}. The currents through the device are calculated by the Landauer-B\"uttiker with the Fisher-Lee relation for transmission. These authors find significant negative differential resistance (NDR) in different systems, arising from the level structure differences and offset across the HS, which produces transmission resonances.

Zhou \emph{et al.} \cite{zhou2015RSCAdv} studied several perpendicular and parallel ribbon geometries of MoS$_2$-WS$_2$ HSs, finding that an armchair interface shows rectifying behavior, which is suppressed as the number of WS$_2$ slabs decreases.  These HSs are proposed  for spintronics due to spin filtering and NDR capabilities \cite{zhou2016PCCP}. The rectification ratio, $RR(V)=|I(V)/I(-V)|$, which characterizes the asymmetry current-voltage, is predicted to be as high as 18.3. The rectifying behavior is analyzed in terms of transmission and DOS profiles, which show higher amplitude peaks for positive than for negative $V$, and produce large $RR(V)$ values. For zigzag (magnetic) edges there is no asymmetry between negative and positive bias, although NDR can be obtained but with smaller efficiency than in armchair, and vanishes with increasing WS$_2$ content. The spin content can also be analyzed through the spin-filtering efficiency (SFE),
\begin{equation}
    SFE=\frac{T_{up}(E_F)-T_{down}(E_F)}{T_{up}(E_F)+T_{down}(E_F)} \, ,
    \label{SFEzhou2016PCCP}
\end{equation}
where $T_s(E_F)$ is the transmission coefficient for spin $s$ at the Fermi level. The SFE efficiency is found to reach 60\%, and attributed to the larger contribution to the DOS from the \emph{p}-orbital of S atoms on the ribbon edges. This effect is independent of the ribbon width, and shows different behavior at different edges.

A similar study \cite{an2016JMCC} finds NRD as well, with electrons propagating along the M-edge, and never along the X-edge. The transmission channels appear indeed as contributed mostly by the metal $d$-orbitals at the Mo-edge. Local current calculations show two types of current channels: the predominant Mo-Mo hopping current, and Mo-S-Mo hopping current (via Mo-S bonds). These results suggest that different widths and metal-vacancies at the sulfur-edge will have little impact on the transport features.

Motivated by nanostructured thermoelectric materials that can efficiently convert wasted heat into electricity (and vice versa), and that a modified thermoelectric material could be made more efficient by interfacial  effects, Zhang \emph{et al.} \cite{Zhang2016SciRep} studied thermoelectric properties of MoS$_2$-WS$_2$ HSs.  They show that this system is expected to have higher figures of merit than those of the counterparts, as the interface reduces the lattice thermal conductivity more than electron transport.

\begin{figure*}[tbph]
\centering
\includegraphics[width=0.9\textwidth]{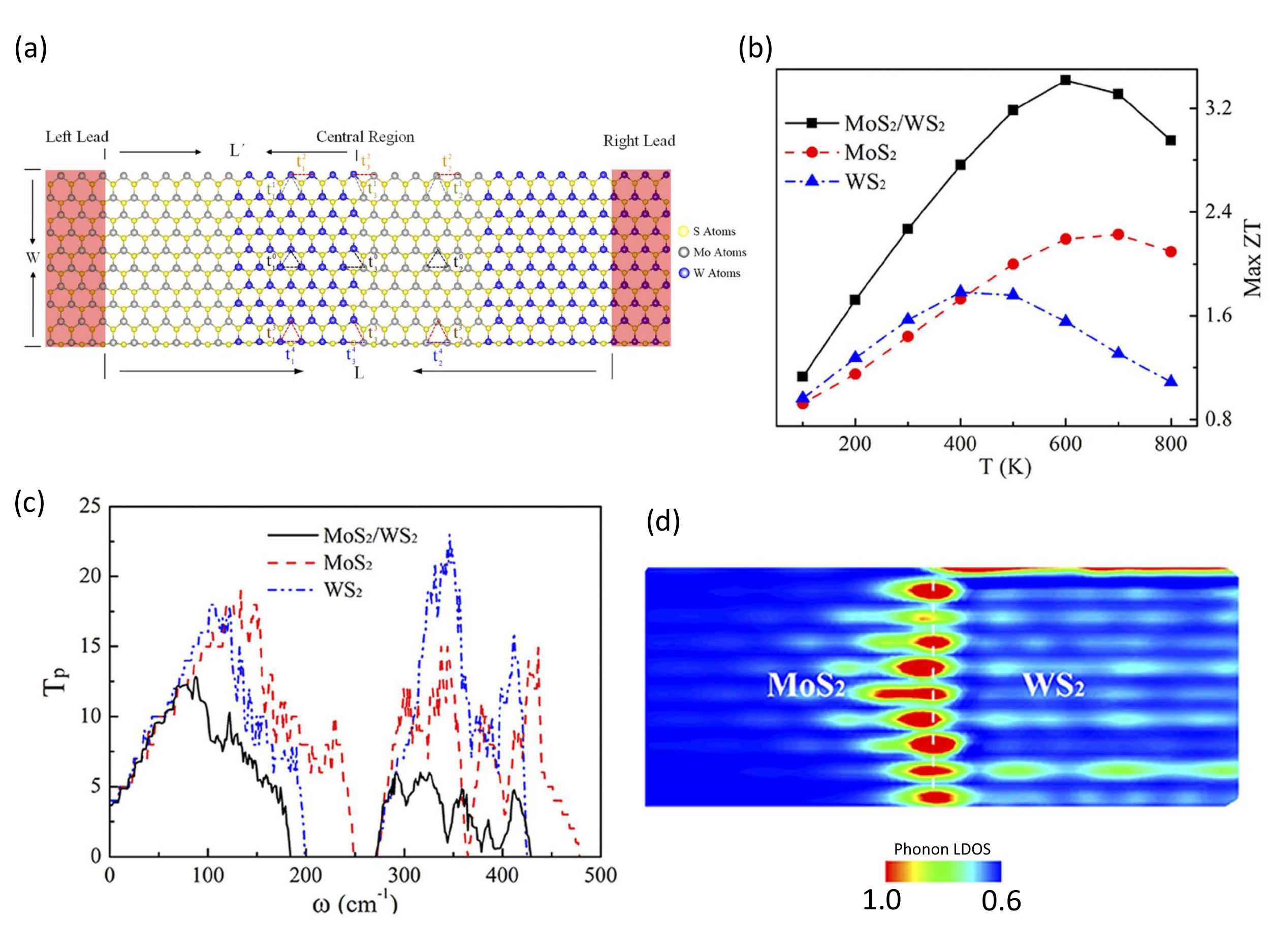}
\caption{(a) Atomic structure of the zigzag terminated MoS$_{2}$-WS$_{2}$ hybrid nanoribbons. The central scattering region is composed by $N_{\mathrm{Mo/W}}$ periodic Mo-W slabs, in this example $N_{\mathrm{Mo/W}}=2$, defined as $N_{\mathrm{Mo/W}}=L/L'$. Atomic hoppings are expressed by $t^{m}_{\alpha}$, where $m$ indicates transversal location of the hopping and $\alpha$ indicates in which material is it. (b) Maximum $ZT$ as a function of temperature $T$, for the hybrid nanoribbon with just one interface ($N_{\mathrm{Mo/W}}=1$), and their respective pristines counterparts. (c) Phonon transmission $T_{p}$  as a function of phonon frequency $\omega$, for $N_{\mathrm{Mo/W}}=1$. (d) Phonon local density of states for $N_{\mathrm{Mo/W}}=1$, where the interface is shown in dashed white and the colors represent the strength of phonon localization. Reprinted with permission from \cite{Zhang2016SciRep}. Copyright 2016 Creative Commons Attribution 4.0, Scientific Reports.}
\label{Zhang2016SciRep}
\end{figure*}

The thermoelectric energy conversion efficiency of zigzag-edge MoS$_2$-WS$_2$ NRs is calculated using a system of $N_{\mathrm{Mo/W}}$ dual slabs as the active scattering region, connected to metallic leads, as shown in figure\ \ref{Zhang2016SciRep}(a). The thermoelectric efficiency of the system is characterized by the figure of merit $ZT=S^2\sigma T/\kappa$, where $T$ is the system temperature, $S$ the Seebeck coefficient, the electronic conductance $\sigma$, and $\kappa=\kappa_{e}+\kappa_{p}$ is the total thermal conductance with contributions of electrons ($\kappa_{e}$) and phonons ($\kappa_{p}$). Systems with $ZT>1$ are considered efficient energy converters. These various quantities are calculated using non-equilibrium Green functions, except for $\kappa_{p}$ which is obtained in a harmonic approximation.

The highest thermoelectric efficiencies are achieved by tunning the number of MoS$_2$-WS$_2$ dual slabs, $N_{\mathrm{Mo/W}}$, i.e. the number of interfaces. For a single interface ($N_{\mathrm{Mo/W}}=1$), $ZT=2.3$ at room temperature, while $ZT=1.6(1.5)$ for pristine WS$_{2}$(MoS$_{2}$), as seen in figure \ref{Zhang2016SciRep}(b). Higher efficiencies are reached with more interfaces, so that for $N_{\mathrm{Mo/W}}=6$ it reaches $ZT=5.5$ at $T=600$ K or $ZT=4$ at $T=300$ K, nearly 3 times as large as in the pristine components. Higher $N_{\mathrm{Mo/W}}$ reduces efficiency, however.
These results are attributed to a sharp decrease of phononic thermal conductance $\kappa_{p}$ at the interfaces, especially as the effects on $\sigma$, $S$ and $\kappa_{e}$ are small. The phonon transmissions $T_{p}$ for the hybrid and the pristine systems reveal two main reasons for the $\kappa_{p}$ drop: a decrease of  the spectral range, as well as the reduction in $T_{p}$ magnitude itself. The first is shown in figure\ \ref{Zhang2016SciRep}(c), where phonon gaps for WS$_{2}$ (MoS$_{2}$) are 73 (24) cm$^{-1}$, while for the MoS$_2$-WS$_2$ hybrid nanoribbon is 87 cm$^{-1}$, hence a larger gap for the phonons to overcome. Also shown in figure\ \ref{Zhang2016SciRep}(c) is the clearly smaller $T_{p}$ from 80 cm$^{-1}$ onward.  The interface then acts as a potential barrier, localizing phonons and suppressing their transmission from left to right in figure\ \ref{Zhang2016SciRep}(d).

Different theoretical studies have looked at the stability and transport quantities of HSs when a variety of gas molecules is adsorbed; in MoS$_{2}$-WS$_{2}$ HS, adsorbed molecules studied include CO, H$_2$O, NH$_3$, NO, and NO$_2$ \cite{Sun2016}. The calculations are performed with VASP and PBE-GGA, while the vdW correction for molecule adsorption includes the Grimme long-range correction. Transport properties are calculated with the Landauer-B\"uttiker formula, while adsorption stability is analyzed by considering $E_{\mathrm{ads}}=E_{gas/HS}-E_{gas}-E_{HS}$, where a negative $E_{\mathrm{ads}}$ indicates adsorption being energetically favorable.
NH$_3$ is reported to act as a charge donor, while all other studied molecules act as acceptors. The largest device sensitivity is found for CO and NO$_2$, due to their larger binding energy, which deeply modifies the TMD HS electronic structure.

Other approaches have also been used to calculate electronic transport properties through TMD lateral HSs. A recent study on WTe$_2$-MoS$_2$ and MoTe$_2$-MoS$_2$ HSs study was reported \cite{Choukroun2018Arxiv}, using an 11-orbital tight binding model \cite{Fang2018}. This work models tunnel field effect transistors on lateral HSs, and study quantum transport with NEGF. Analyzing DOS and characteristic I-V curves of several systems as the channel length and backgate voltages are changed, finds that the MoTe$_2$-MoS$_2$ HS can serve as ultra-low power consumption device, with low sub-threshold swings and high $I_{\mathrm{on}}/I_{\mathrm{off}}$ ratios.

A different study on transport by Ghadiri \emph{et al.} \cite{Ghadiri2018JAP} considers a possible Goos-H{\"a}nchen lateral shift of valley electrons arriving at the interfacial scattering region, in lateral HSs of normal MoS$_{2}$ and `ferromagnetic' WS$_{2}$, in WS$_{2}$-MoS$_{2}$-WS$_{2}$ and MoS$_{2}$-WS$_{2}$-MoS$_{2}$ quantum well systems.  The magnetic TMD is achieved by deposition on a magnetic insulator system.

The Goos-H{\"a}nchen (GH) shift occurs in optics when an incident beam of light is fully reflected off an interface and displaced laterally from the anticipated geometrical path.  The shift occurs as the incident wave packet is reshaped by the interface due to each plane wave component experiencing a different phase shift. Similarly, Goos-H{\"a}nchen-like (GHL) shifts of electrons transmitted through an interface are also observed.
Such GH and GHL shifts are expected in 2D materials due to local strains, as predicted on graphene \cite{WuPRL2011}. It is natural that the inherent local strain and band alignment at the interface between WS$_{2}$ and MoS$_{2}$ would cause GH and/or GHL shifts. The ferromagnetic proximity induced on WS$_2$ should also result in strong valley-dependence for incident waves at the interface.

\begin{figure}[tbph!]
\centering
\includegraphics[width=0.4\textwidth]{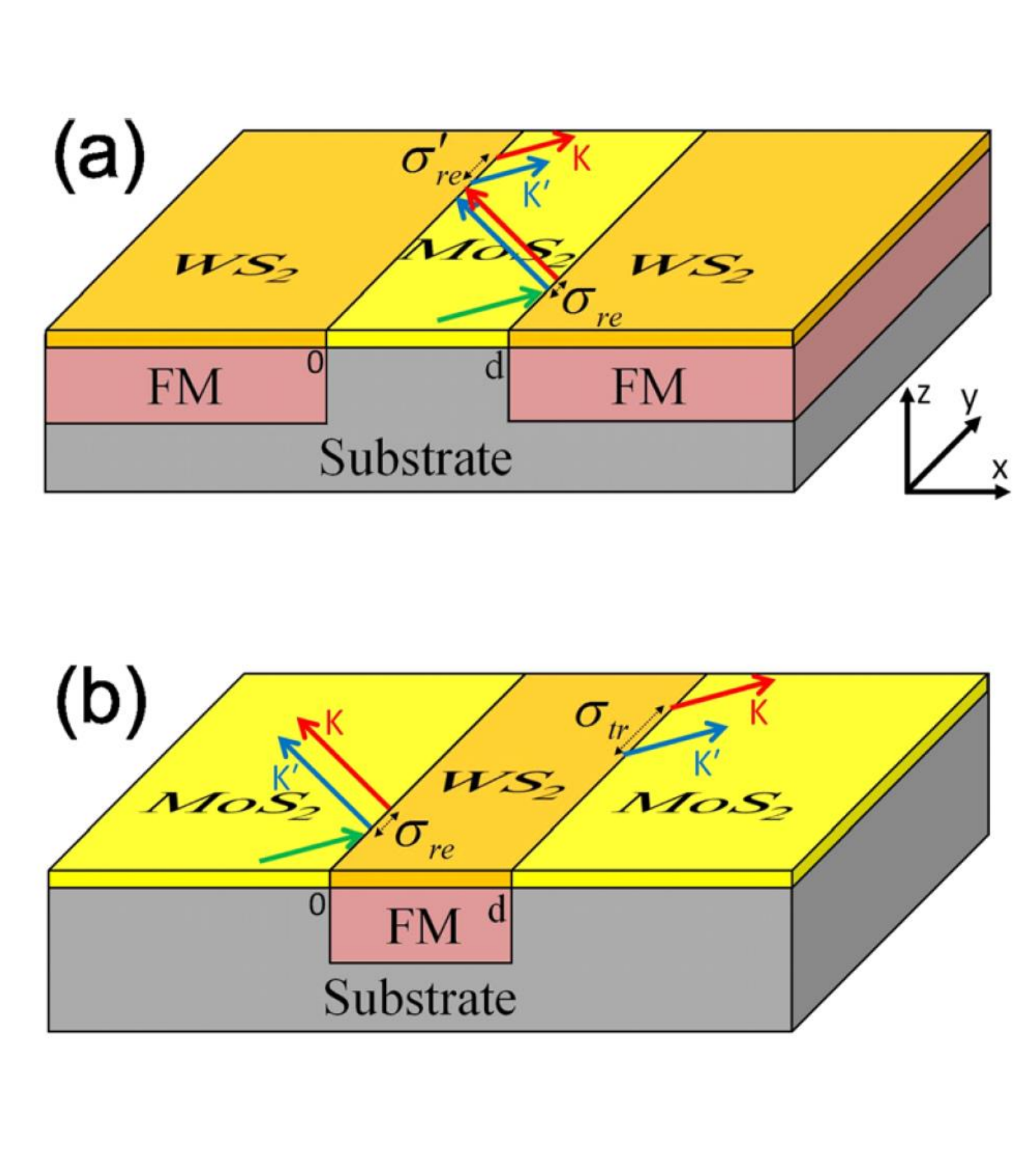}
\caption{Lateral quantum well HSs, (a) WS$_{2}$-MoS$_{2}$-WS$_{2}$ and (b) MoS$_{2}$-WS$_{2}$-MoS$_{2}$. The WS$_2$ regions exerience proximity-induced exchange bias. The main results are indicated by arrows: the initial incident beam (green), and the $K$ (red) and $K'$ (blue) valleys. In (a) the incident beam is in the central region, and the reflected $K$ and $K'$ are spatially separated. In (b) the incident beam on the left WS$_{2}$ region, reflect back and transmit spatially separated $K$ and $K'$ valleys into the MoS$_{2}$. In both cases, the incident electron comes from MoS$_{2}$, since its CBM is lower than in WS$_{2}$. Reprinted with permission from \cite{Ghadiri2018JAP}. Copyright 2018 AIP Publishing, Journal of Applied Physics.}
\label{Ghadiri2018JAP}
\end{figure}

To obtain the transport properties of the system they used wave function matching, modifying the low-energy two-orbital representation from \cite{Xiao2012}, including substrate-induced exchange bias on WS$_{2}$, to write the effective Hamiltonian as

\begin{equation}
\label{GhadiriHamiltonian}
H_j =    \left(\begin{array}{cc}
                    E_{jc}-h_{j}s_{z} & \tau a_{j} t_{j} k_{j} e^{-i\tau\theta_{j}} \\
                    \tau a_{j} t_{j} k_{j} e^{i\tau\theta_{j}} & E_{jv}+\tau s_{z}\lambda_{j} -h_{j}s_{z}
                  \end{array}\right) \, ,
\end{equation}
where $j$ indicates each region $j=1,2,3$ in the structure, and $k_{j}$ and $\theta_{j}$ are the $x$-component of the momentum and angle of the electron wave vector $\mathbf{k}_{j}$, respectively. The pristine material parameters include $E_{jc}$ $(E_{jv})$ the conduction (valence) band minimum (maximum), lattice constant $a_{j}$,  hopping integrals $t_{j}$, induced SOC splitting at the valence band 2$\lambda_{j}$, valley index $\tau=1(-1)$ for K (K'), as well as the substrate-induced exchange $h_{j}$, and $s_{z}=+1(-1)$ is the electron spin. The energy dispersion and corresponding pseudospinors allow them to find
transmission $T_{A(B)}$ and reflection amplitudes, as well as associated GH (transmitted electrons GHL) shifts $\sigma_{re(tr),s_{z}}^{\tau}$ for different quantum well systems, shown schematically in figure \ref{Ghadiri2018JAP}.  We should comment that it is not clear if the required wave function matching in this work has considered the non-hermiticity of the effective Hamiltonian for non-uniform hopping integrals.  This requires consideration of different matching conditions \cite{Kolesnikov1997,Silin1998,Ratnikov2012}.

When MoS$_{2}$ is the quantum well middle region, it can act as a waveguide due to multiple internal reflections at both interfaces, leading to effective confinement in the middle MoS$_{2}$ region. The $\sigma_{re,s_{z}}^{\tau}$ GH shifts are calculated to be of the order of the Fermi wavelength $\lambda_{F}$, leading to the spatial separation of electrons with distinct valley indexes as shown in figure\ \ref{Ghadiri2018JAP}(a). Multiple internal reflections increase $\sigma_{re,s_{z}}^{\tau}$, creating an efficient valley splitter. Additionally, a spin splitter can be achieved by selecting specific angles of incidence.

When the magnetized WS$_{2}$ is the middle region, the transport across the structure shows transmission resonances and GHL spin/valley dependent lateral shifts of the reflected and transmitted beams, as shown in figure\ \ref{Ghadiri2018JAP}(b). GHL shifts can be enhanced near transmission resonances, achieving full spin-valley beam splitter as structure and current injection parameters are tuned.

In an effective mass approximation, Mishra \emph{et al.} \cite{Mishra2018oneDimensional} have found that a 1D spin channel exists at the lateral in plane interface between two TDMs, produced by the Rashba electric field perpendicular to the HS \cite{Sinova2015}.
Lateral metal-metal HSs with strong interfacial SOC have also been suggested to host spin currents \cite{Borge2017}. In the case of the TMD HS, the electric field is provided by the band offset at the interface. As electrons  propagate parallel to the interface, they experience an effective magnetic field pointing out of plane which polarizes them. The interface channel is modeled using effective mass approximation for the conduction band minimum of the K-point. The Hamiltonian describing longitudinal (L) and transverse (T) motion at the interface is given by
\begin{equation}
\label{MishraHamiltonian1}
H=\frac{p^2_x}{2m_L}+\frac{p^2_y}{2m_T}+\frac{\alpha}{\hbar}|E_y| p_x \sigma_z  + V_{\rm conf}(x,y),
\end{equation}
which includes spin orbit coupling $\alpha$, and confining potential $V_{\rm conf}$.
The one-dimensionality of the spin generation becomes evident, showing a peak exactly at the interface.

Although the pseudo magnetic field at the interface does not produce a transverse force on the carriers, there could be a flow out of the interface by diffusion, leading the polarization to depend on the spin diffusion length. If the diffusion length is larger or similar to the device width, the 1D spin channel is lost.

When the device is p-doped, and sectors of the valence band are reached, there is cooperation between the interface spin polarization and the intrinsic spin orbit. The spin polarization then reaches 0.75\%, much larger than the 0.1\% seen in the conduction band.

\subsubsection{Interfaces between different phases: topological, structural and transport effects}
\label{subsubsec:PhaseInterfacesTHEORY}

A few works have addressed the coexistence between phases separated by an atomic sharp interface, as those already experimentally available--see section\ \ref{subsec:PhaseInterfaces}.

In 2016, Olsen \cite{Olsen2016} suggested the design of lateral TMD HS across different regions within the same 1T'-MoS$_2$ monolayer. The first phase is the natural quantum spin Hall insulator of 1T'-MoS$_2$, while a second region is made into a trivial insulating phase by adsorption of different atomic species, including O, F, and Cl. DFT calculations show that a topological 1D metallic state arises at the sharp interface between these phases. The interfacial state is further tested against boundary reconstruction and disorder, and seen to persist as a single level crossing the gap. Hence this platform is suggested to study topologically protected conductivity in 1D.

A monolayer triangular island of T'-MoS$_2$ phase surrounded by H-MoS$_2$ phase has been synthesized in experiments upon electron irradiation, as mentioned in section\ \ref{subsec:PhaseInterfaces}. With this system in mind, Kretschmer \emph{et al.} \cite{Kretschmer2017} studied structural transitions and effects of strain and vacancies. They find that charge redistribution promotes phase transitions, driven by electronic excitations and formation of vacancies while a monolayer is illuminated. The interface between T' and H phases is found to be atomically sharp with some S-deficiency.

The effects of the type and crystallographic orientation of the interface between 1T and 1H phases of MoS$_2$, have been recently studied in transport calculations, with DFT and non-equilibrium Green functions \cite{Aierken2018}. The interface geometry between phases is found to be decisive, as an armchair interface is more conductive than a zigzag.  This occurs because in the first, the Mo zigzag  chains are placed along the transport direction. Electron doping and Mo-substitutional doping of Re or Ta atoms is suggested for stabilization of the 1T metallic phase and reduction of the Schottky barrier.

Other phase-based planar HSs have been also proposed, creating phases by top/bottom gating potentials. Lateral field-effect transistors composed from adjacent regions of T' and H phases of MoSe$_2$, appear
to have better performance and lower-power consumption than conventional CMOS \cite{Marian2017}. Nevertheless, ideal contacts are assumed and no atomic geometry interface effects are further analyzed.

\subsubsection{Lateral heterostructures with other materials}
\label{subsubsec:WithOtherMaterials}

Semiconducting group VIB TMDs have also been proposed to form lateral HSs with metallic NbS$_2$ \cite{Wu2015NbS2,Liu2017NbS2,Yang2017NbS2}, and NiTe$_2$ \cite{Aras2018}. Moreover, as NbS$_2$ (also NbSe$_2$ and TaS$_2$) exhibits superconductivity and charge density waves at low temperatures \cite{Heil2017,Xi2015NbSe2,Xi2016Ising,Sohn2018NbSe2,Barrera2018tuning},
while NiTe$_2$ has been shown to be a type-II Dirac semimetal \cite{Xu2018NiTe2}, this suggests new and interesting lateral HS properties with other TMDs. We briefly summarize results in this rapidly developing area.

Armchair-terminated (zigzag interfaces) of MoS$_2$-NbS$_2$ quantum wells have been predicted to exhibit semiconducting or metallic behavior, depending on which material is the quantum well, respectively \cite{Wu2015NbS2}. Zigzag-terminated structures were seen to exhibit resonance tunneling transport \cite{Liu2017NbS2}, while armchair-terminated are predicted to be excellent ambipolar devices \cite{Liu2017NbS2,Yang2017NbS2}.
The bandgaps of these NRs appear insensitive to their lateral dimension \cite{Wu2015NbS2, Liu2017NbS2}.

Periodic arrays of alternating strips of zigzag-terminated MoTe$_2$-NiTe$_2$ ribbons have been modeled \cite{Aras2018}. They show metallic behavior for narrow strips, while large strips serve as metal-semiconductor junctions with tunable Schottky barriers, due to the confinement of electronic states in different TMDs.

Other examples of lateral HSs of group VIB semiconducting TMDs and metallic CrX$_2$ and VX$_2$ materials have also been suggested for solar energy and photocatalytic water-splitting applications, aided by the corresponding band alignments \cite{Wei2014,Zhao2018Designing}.

\section{Applications}
\label{sec:Applications}

Materials growth and device design continue to improve, allowing for  deeper tunability and functionality in experiments, while more  theoretical proposals continue to appear.  One open question now is how feasible and effective are some of these proposed applications. In this section we discuss devices and applications already achieved in experiments, as well as some proposed by theoretical studies.

\subsection{Experimentally achieved}
\label{subsec:ExperimentalApplications}

Lateral TMD HSs have suggested effective alternatives for producing in-plane \emph{p-n} junctions, critical components in electronic and optoelectronic applications.
Of particular interest is the ability for enhanced exciton trapping at interfacial regions, due to the built-in potential at the interface.

The strong PL enhancement observed at the interface would ideally be the result of enhanced recombination at the HS \cite{Gong2014NatMat,Huang2014NatMat,Duan2014NatNano}, even if in many samples it is likely assisted by exciton trapping by defects.
As discussed previously, the band alignments shown in figure\ \ref{Kang2013} help determine which material may act as \emph{n}-type or \emph{p}-type across the interface, such as WS$_2$ being \emph{n}-type and WSe$_2$ \emph{p}-type \cite{Duan2014NatNano}, or WS$_2$ serving as \emph{p}-type and MoS$_2$ as \emph{n}-type  \cite{Chen2015Electronic}. This has allowed the fabrication of lateral \emph{p-n} diodes \cite{Gong2014NatMat,Duan2014NatNano,Li2015,Gong2015TwoStep,Chen2015Electronic,Chen2016Lateral,Son2016Observation,Zhang2017Robust,Wu2018SelfPowered}, as well as \emph{n-n} \cite{Mahjouri2015patterned} junctions. The \emph{p-n} diodes have been shown to serve as inverters with high voltage gain \cite{Duan2014NatNano}, large rectification ratios ($10^{5}$ \cite{Zhang2017Robust}, 10$^6$ \cite{Xie2018ParkGroup}), and high electroluminescence \cite{Xie2018ParkGroup}.  Also, MoS$_2$-graphene lateral HSs have been built and tested as transistors, with good rectifying behavior and switching ratios as high as 10$^{6}$ \cite{Ling2016}.

Devices with large photoresponse and photovoltaic effects have also been studied.  The solar cell efficiency of lateral WSe$_2$-MoS$_2$ HSs \cite{Tsai2017SingleAtomically} has been found to be high, with excellent power conversion efficiency under illumination, and  omnidirectional light harvesting capability, not seen in vertical TMD  solar cells. The \emph{cheetah spots} mosaic configurations in MoS$_2$-MoSe$_2$ have been used as photodetectors, showing enhanced  performance with respect to the individual components, as band alignments appear to suppress photoexcited electron-hole recombination, leading to effective \emph{n}-doping of MoS$_2$ and \emph{p}-doping of MoSe$_2$ \cite{Chen2017InPlaneMosaic} . Self-powered photovoltaic light sensors based on MoS$_2$-WS$_2$ exhibit large spectral responsivity and detectivity coefficients \cite{Wu2018SelfPowered}.
LED designs have also been built with double HS transistors between WS$_2$ (\emph{n}-type) and WSe$_2$ (\emph{p}-type) \cite{Xie2018ParkGroup} monolayers. The luminescence in these devices originates from the interface, suggesting electrons (from WS$_2$ side) and holes (from WSe$_2$ side) recombine in its vicinity.

\subsection{Theoretically proposed}
\label{subsec:TheoreticalApplications}

Theoretical proposals highlight the powerful features of the lateral TMD HSs. Optoelectronics applications are perhaps the most developed, including excitonic solar cells, and photocatalysis, as effective separation and collection of photo-induced excitons improves efficiency.  However, applications based on electronic transport and magnetic properties have also been proposed.

Excitonic effects have been studied by DFT \cite{Wei2015,Yang2017} and other calculations \cite{Lau2018ArxivExcitons}, as discussed above.   The effective band bending, relative band alignment and associated barriers confine photogenerated carriers to opposite sides of the interface and suppress recombination, which yields higher solar conversion efficiency \cite{Wei2015,Wei2015SciRep,Mu2018MRExpress}. The electron-hole separation was shown to persist for up to 12\% of uniaxial strain.  Moreover, strain can significantly increase the power conversion efficiency of lateral HSs. A 4\% uniaxial tensile strain could increase the efficiency of the lateral MoS$_2$-WS$_2$ (MoSe$_2$-WSe$_2$) heterostructure by about 35\% (15\%), when compared to the pristine system \cite{Lee2017}.

The use of vacancies has also been proposed. Localized arrays of S-vacancies at  interfaces, along with type-II band alignment and built-in electric field could improve the energy conversion efficiency in photocatalysis, given that the ingap states caused by vacancies can activate and optimize hydrogen evolution reactions \cite{Li2016activating}. Strain effects at the interface can also be utilized, as WSe$_2$-MoS$_2$ and MoS$_2$-WSe$_2$ HSs may exhibit photon-induced Coulomb drag over the interface region \cite{Wei2017}.

Lastly, the effect of adsorption of different gas molecules near the HS region has been studied with DFT in MoS$_{2}$-WS$_{2}$ and gases such as CO, H$_2$O, NH$_3$, NO, and NO$_2$ \cite{Sun2016}. The rectification behavior and value of the passing current can be altered by adsorption, and this sensitivity promises HSs as  superior gas sensors in practical applications \cite{Sun2016}.

Transport studies have suggested that the zigzag interface in MoS$_2$-WS$_2$ nanoribbons can be used as high-performance thermoelectric materials, promising applications with high values of the ZT figure of merit \cite{Zhang2016SciRep}. Moreover, WS$_2$-MoS$_2$-WS$_2$ quantum wells have been proposed as spin-valley  filters and splitters without external gating \cite{Ghadiri2018JAP}. Electrons with different spins and valleys can be well separated spatially by tuning the Fermi energy and current incident angle to the interface.
Similar ideas suggest the WS$_2$-MoS$_2$ interface as an electron waveguide, useful in spintronic applications.

Enhanced magnetic exchange driven by the interface electrons has also been predicted at these 1D HSs \cite{AvalosOvando2018Arxiv}. Here the hybridization of magnetic impurities and the tunability of the Fermi level through interface midgap states can serve to implement tunable spin chain systems for information transfer or storage \cite{Menzel2012}. Interactions between magnetic impurities can be tuned by gating and/or separation.


\section{Prospective directions}
\label{sec:ProspectiveDirections}

A great deal of attention has been focused towards obtaining long interfacial regions, with several groups already achieving $\mu$m-lengths.  Many of these are however not fully pristine or sharp, containing defects and a diffusive interfacial region, which in the best cases is only 4 atomic rows \cite{Sahoo2018Nature,Xie2018ParkGroup}. This is an excellent development if compared to the diffusive/alloy interfacial regions that existed only 4 years ago and that could extend from several nm to a few $\mu$m away from the interface. Both zigzag and armchair interfaces can be obtained, the first most often obtained with CVD processes. The cleanest and sharpest pristine zigzag interfaces are now longer than $\sim$50 nm long, while armchair HSs are no more than a few nm long.

Other important experimental issues to be addressed include scalability and HS degradation, as well as selectivity in overall crystal shape and dimensions.
Protection from the environment has seen progress in promising combinations or encapsulation with materials such as graphene and hBN \cite{Ling2016,Murthy2018Intrinsic}.  This should facilitate exploring the real interface features, as well as avoiding HS degradation. State-of-the-art sub-\AA\,microscopy could also be used to further explore these interfaces \cite{Winkler2018Absolute,Jiang2018ElectronPtychography}.

We have also described notable advances in strain control at interfaces providing `predictable wrinkling' and coherent crystals over large scales \cite{Xie2018ParkGroup}.
This is likely the beginning of `kirigami' efforts using TMDs \cite{kirigami}, which would bring additional optical functionalities to that field.

Theoretical efforts to accompany experimental realizations must take into account realistic effects, including atomic relaxation at the interface, dislocations, impurities, vacancies, and strain fields.  Recent DFT studies have started to consider such realistic features of HSs and their effect on electronic, structural and dynamical properties. A first experimental study has tackled the change in band alignment by analyzing atomic lines at the interface \cite{Zhang2018strain}. Nevertheless, more studies are needed, although the intrinsic exposed nature of these HSs gives unprecedented access to study materials interfaces.

Multiscale models are also being used to studying HSs systems where DFT approaches  cannot account for simulations of thousands of atoms, random distribution of atomic vacancies or impurities, and large-scale strains, among others. These studies have been able to predict features not yet seen in experiments or anticipated from DFT calculations.  Further improvements of treatments within effective mass or tight binding treatments or molecular dynamics models must  address realistic systems, supported by analysis of experimental or DFT results. We trust that this review would serve as a starting point where the interplay  experiment-theory-numerics can be readily accessed to motivate further advances.

Doubtlessly, efforts in this field are increasing and experiments becomimg more sophisticated. We anticipate further improvements in length and quality of the interfaces are coming, and they would be important for realistic/practical device geometries and for the implementation of HSs as novel physical environments in which to test tantalizing theoretical proposals.

\section{Summary}
\label{sec:SummaryFinal}

We have presented advances in group-VIB semiconducting TMDs lateral heterostructures. These materials have been on the spotlight in the last few years, given their attractive properties at the monolayer level. The promise of direct bandgaps, large spin-orbit coupling, and controlable interface features makes them promising in optoelectronics, spintronics, and valleytronics applications. Moreover, the combinations of materials have shown different and often enhanced response functions with respect to their pristine counterparts.

Our focus on the lateral connection between distinct TMD monolayers, has analyzed the unusual one-dimensional interfaces with many beautiful examples already seen in experiments. Theoretical studies have started to appear, led by numerical DFT as well as other more recent approaches with complementary emphasis. We have summarized ongoing trends and developments in numerical and theoretical studies, as well as experimental milestones.  Available theoretical studies suggesting possible uses for these unique 1D states at the interface have also been discussed. It is clear that the interface provides an interesting platform for achieving 1D physical systems with unique features.

We look forward to more theoretical studies and experiments in this growing field. Lateral HSs provide exciting opportunities for monolayer systems with tailored properties and the ultimate tunable 2D/1D interchangeable system.


\ack We acknowledge support from NSF grant DMR 1508325.

\section*{References}
\bibliographystyle{iopart-num}
\bibliography{AvalosOvandoEtAl}

\end{document}